\RequirePackage{fix-cm}
\documentclass[twocolumn,epjc3]{svjour3}  
\smartqed  % flush right qed marks, e.g. at end of proof 
\RequirePackage{graphicx} 
\usepackage{color}
\usepackage{amsfonts,amssymb,hyperref,slashed}
\journalname{Eur. Phys. J. C}

\renewcommand{\l}{{\lambda}}
\newcommand{\eqref}[1]{{(\ref{#1})}}

\def\be{\begin{equation}}
\def\ee{\end{equation}}

\begin{document}

\title{
Holographic modeling of nuclear matter and neutron stars\thanksref{t1}
}


\author{Matti J\"arvinen\thanksref{e1,addr1,addr2}
%        \and
%        Second Author\thanksref{e2,addr2,addr3} %etc.
}

\thankstext{t1}{Preprint number: APCTP Pre2021 - 024}
\thankstext{e1}{e-mail: matti.jarvinen@apctp.org}


\institute{Asia Pacific Center for Theoretical Physics, Pohang 37673, Republic of Korea \label{addr1}
           \and
           Department of Physics, Pohang University of Science and Technology, Pohang 37673, Republic of Korea \label{addr2}
%           \emph{Present Address:} if needed\label{addr3}
}

\date{Received: date / Accepted: date}

\maketitle

\begin{abstract}
I review holographic models for (dense and cold) nuclear matter, neutron stars, and their mergers. I start by a brief general discussion on current knowledge of cold QCD matter and neutron stars, and go on discussing various approaches to model cold nuclear and quark matter by using gau\-ge/gra\-vity duality, pointing out their strengths and weaknesses. Then I focus on recent results for a complex bottom-up holographic framework (V-QCD), 
which also takes input from lattice QCD results, effective field theory, and perturbative QCD. Dense nuclear matter is modeled in V-QCD through a homogeneous non-Abelian bulk gauge field. 
Feasible ``hybrid'' equations of state for cold nuclear (and quark) matter can be constructed by using traditional methods (e.g., effective field theory)  at low densities and the holographic V-QCD model at higher densities. I discuss the constraints from this approach to the properties of the nuclear to quark matter transition as well as to properties of neutron stars. Using such hybrid equations of state as an input for numerical simulations of neutron star mergers, I also derive predictions for the spectrum of produced gravitational waves. 

%Insert your abstract here. Include keywords, PACS and mathematical
%subject classification numbers as needed.
\end{abstract}

\tableofcontents

%%%%%%%%%%%%%%%%%%%%%%%%%%%%%%%%%%%%%%%%%%%%%
\section{Introduction}
%%%%%%%%%%%%%%%%%%%%%%%%%%%%%%%%%%%%%%%%%%%%%

Recent observations of binary neutron star mergers by the LIGO and Virgo interferometers have boosted the interest in QCD at finite density. This activity is complemented by ongoing and future heavy ion collisions experiments which will explore quark gluon plasma at higher densities than earlier, and by theoretical advances in the understanding of hot and dense QCD matter.

However, determining even the phase diagram of QCD is a challenging task~\cite{Brambilla:2014jmp}. Producing hot and/or dense QCD matter in the laboratory is extremely complicated, and employing theoretical as well as computational methods is challenging due to the strongly interacting nature of QCD. Analysis of QCD is particularly hard in the regime of high (but not asymptotically high) baryon number density. This is due to various reasons: The main computational tool, lattice QCD~\cite{Ding:2015ona}, only works at low densities due to the well-known sign problem~\cite{deForcrand:2010ys}. Perturbative QCD gives definite predictions only at extremely high densities~\cite{Kurkela:2014vha}. Moreover, effective field theory and other related methods are only reliable at low densities and temperatures. For example, chiral effective theory expansions for nuclear matter work only up to densities comparable to nuclear saturation density~\cite{Tews:2018iwm}. Therefore there is a gap for cool QCD matter at intermediate densities where direct reliable theoretical predictions are not available.

In the absence of first principles calculations, the uncertainties of theoretical predictions for QCD matter at intermediate densities and low temperatures are large. This is true even for basic observables such as the equation of state (EOS), i.e., the relation between the pressure and energy density of QCD matter. Definite constraints for the zero temperature EOS can be obtained by interpolating between reliable results at low and high densities~\cite{Kurkela:2014vha,Annala:2017llu,Most:2018hfd}. Even less is known about the equation of state at nonzero temperature in the same density range, and on other observables such as transport coefficients. 

The properties of the QCD EOS at intermediate densities are interesting among other things because it is known that the densities in neutron star cores lie in this region. Their temperatures are typically small compared to the characteristic QCD energy scale, so that they can be treated as cold objects in QCD analyses. This also means that measurements of neutron stars give information of the properties of cold and dense  QCD (see, e.g.,~\cite{Oertel:2016bki,Annala:2017llu}). There are already plenty of such data available, with additional and more accurate results expected in near future. Measurements of neutron star masses and radii give direct constraints for the QCD EOS~\cite{Lattimer:2012nd}, and measurements of gravitational waves from neutron star mergers from LIGO/Virgo give complementary information about the EOS~\cite{TheLIGOScientific:2017qsa,GBM:2017lvd,Abbott:2018exr,Abbott:2018wiz}. Additional events and improvements in precision are likely to lead to severe constraints to the EOS in near future. It is therefore timely to improve the theoretical status of the predictions for the EOS and other observables of cold QCD matter.

The difficulty of theoretically predicting the behavior of cold QCD matter reflects the fact that the interactions are strongly coupled. It is therefore natural to ask whether AdS/CFT, or gau\-ge/gra\-vity duality in more general, can help to improve the status of theoretical predictions. Namely, gau\-ge/gra\-vity duality (or holography for short) can map strongly interacting field theory to a higher dimensional classical gravity. It is however not obvious that the duality is applicable to QCD. The original formulation~\cite{Maldacena:1997re,Gubser:1998bc,Witten:1998qj} states that the $\mathcal{N}=4$ super Yang-Mills theory is dual to type IIB string theory in 10 dimensions. This field theory is superconformal and nonconfining, that is, significantly different from QCD. Moreover, the duality in its most useful form requires both taking the number of colors $N_c$ and the 't Hooft coupling to infinity, whereas regular QCD has $N_c=3$ and finite coupling. 

Despite these potential issues, gau\-ge/gra\-vity duality has proved out to be a useful tool in studies of various aspects of QCD. Simple models give surprisingly good description of the spectrum of QCD: approaches include the simple five dimensional actions of the hard~\cite{Erlich:2005qh,DaRold:2005mxj} and soft wall~\cite{Karch:2006pv} models, the light front holography framework which is motivated both by gau\-ge/gra\-vity duality and the light-front wave function description of hadrons~\cite{deTeramond:2005su,Brodsky:2014yha}, a bit more advanced dynamic AdS/QCD models~\cite{Alho:2013dka,Erdmenger:2020flu}, and more stringy models such as the Witten-Sakai-Sugimoto model~\cite{Sakai:2004cn,Sakai:2005yt} and the holography inspired stringy hadron framework~\cite{Sonnenschein:2014jwa,Sonnenschein:2018fph}. Moreover, gau\-ge/gra\-vity duality has, among other things, been helpful in the analysis of transport and hydrodynamics of the quark gluon plasma produced in heavy ion collisions (see, e.g.,~\cite{CasalderreySolana:2011us,DeWolfe:2013cua,Janik:2010we}). Examples of important results are the predictions for the shear viscosity of the plasma~\cite{Policastro:2001yc,Kovtun:2004de} and for the behavior of the plasma in the out-of-equilibrium phase right after the collision (see, e.g.,~\cite{Heller:2011ju,Heller:2013fn}).  So given the earlier success, one may expect that gau\-ge/gra\-vity duality works also in the case of dense QCD matter. And one of the goals of this review is to demonstrate that this is indeed the case. 

The phase diagram of QCD has been studied by using several holographic ``top-down'' models, i.e., models directly based on string theory, as well as ``bottom-up'' models, i.e., models motivated by string theory but adjusted by hand. The former class includes the D3-D7 models~\cite{Karch:2002sh,Kruczenski:2003be,Mateos:2007vn} as well as the Witten-Sakai-Sugimoto model~\cite{Witten:1998zw,Sakai:2004cn,Sakai:2005yt,Aharony:2006da}, and the latter class includes the hard and soft wall models, and models based on Einstein-Maxwell actions. In this review I will focus on the V-QCD bottom-up model~\cite{Jarvinen:2011qe}, which is an extension of improved holographic QCD~\cite{Gursoy:2007cb,Gursoy:2007er} with dynamical flavors, i.e., a quark sector with full backreaction to the glue. This class of models is defined through relatively rich five-dimensional actions, which are inspired by noncritical five dimensional string theory, but generalized to include a large number of parameters that need to be determined by comparing to QCD data. This is the main strength of the model: it is rich enough so that it can be matched with QCD data from various sources and in various phases, and it can then be used to extrapolate these results to regimes which are challenging to analyze by other means.

This review is organized as follows. In Sec.~\ref{sec:qcd} I review the status of the QCD phase diagram, with the stress on the region of cold and dense matter. In Sec.~\ref{sec:holoqcd} I continue the review, discussing various holographic approaches to QCD and its phase diagram at finite temperature and density. In particular, I discuss various approaches to baryons and nuclear matter in Sec.~\ref{sec:baryons}. I introduce the V-QCD model in Sec.~\ref{sec:vqcd}, and discuss implementing nuclear matter in this model in Sec.~\ref{sec:VQCDNM} by using a simple, homogeneous approximation scheme. Sec.~\ref{sec:ns} is devoted to applications to neutron stars. I concentrate on the results derived by using the V-QCD model. Finally, I conclude and discuss future directions in Sec.~\ref{sec:concl}.

%%%%%%%%%%%%%%%%%%%%%%%%%%%%%%%%%%%%%
%%%%%%%%%%%%%%%%%%%%%%%%%%%%%%%%%%%%%
%%%%%%%%%%%%%%%%%%%%%%%%%%%%%%%%%%%%%
%%%%%%%%%%%%%%%%%%%%%%%%%%%%%%%%%%%%%
%%%%%%%%%%%%%%%%%%%%%%%%%%%%%%%%%%%%%
%%%%%%%%%%%%%%%%%%%%%%%%%%%%%%%%%%%%%
%%%%%%%%%%%%%%%%%%%%%%%%%%%%%%%%%%%%%
%%%%%%%%%%%%%%%%%%%%%%%%%%%%%%%%%%%%%
%%%%%%%%%%%%%%%%%%%%%%%%%%%%%%%%%%%%%
%%%%%%%%%%%%%%%%%%%%%%%%%%%%%%%%%%%%%
%%%%%%%%%%%%%%%%%%%%%%%%%%%%%%%%%%%%%
%%%%%%%%%%%%%%%%%%%%%%%%%%%%%%%%%%%%%
%%%%%%%%%%%%%%%%%%%%%%%%%%%%%%%%%%%%%
%%%%%%%%%%%%%%%%%%%%%%%%%%%%%%%%%%%%%
%%%%%%%%%%%%%%%%%%%%%%%%%%%%%%%%%%%%%
%%%%%%%%%%%%%%%%%%%%%%%%%%%%%%%%%%%%%
%%%%%%%%%%%%%%%%%%%%%%%%%%%%%%%%%%%%%
%%%%%%%%%%%%%%%%%%%%%%%%%%%%%%%%%%%%%
%%%%%%%%%%%%%%%%%%%%%%%%%%%%%%%%%%%%%
%%%%%%%%%%%%%%%%%%%%%%%%%%%%%%%%%%%%%
%%%%%%%%%%%%%%%%%%%%%%%%%%%%%%%%%%%%%
%%%%%%%%%%%%%%%%%%%%%%%%%%%%%%%%%%%%%
%%%%%%%%%%%%%%%%%%%%%%%%%%%%%%%%%%%%%
%%%%%%%%%%%%%%%%%%%%%%%%%%%%%%%%%%%%%
%%%%%%%%%%%%%%%%%%%%%%%%%%%%%%%%%%%%%
%%%%%%%%%%%%%%%%%%%%%%%%%%%%%%%%%%%%%
\section{Dense QCD and neutron stars} \label{sec:qcd}
%%%%%%%%%%%%%%%%%%%%%%%%%%%%%%%%%%%%%

In this section, I will briefly review the current status of the QCD phase diagram, various tools to probe it, and the predictions for the EOS of cold QCD matter. In particular I will discuss how the current neutron star data constrain the EOS.

\begin{figure}
\centering  \includegraphics{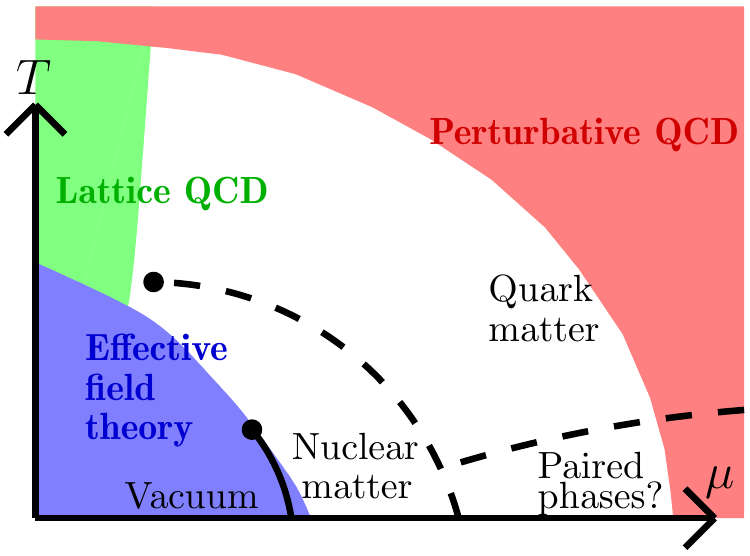}
\caption{A simple sketch of the (possible) QCD phase diagram as a function of quark chemical potential and temperature. The black curves are first order phase transitions ending at critical points. The colored regions show roughly the ranges of applicability of various theoretical and computational methods. We stress the fact that the existence of the nuclear to quark matter transition and the transition to ``exotic'' phases are still open questions by marking them by dashed curves. }
\label{fig:phases}       % Give a unique label
\end{figure}

%%%%%%%%%%%%%%%%%%%%%%%%%%%%%%%%%%%%%
\subsection{Theoretical methods to study the phase diagram} \label{ssec:methods}
%%%%%%%%%%%%%%%%%%%%%%%%%%%%%%%%%%%%%

A sketch of the QCD phase diagram is given in Fig.~\ref{fig:phases}, where the black curves are first order phase transitions. I also show the regions where various theoretical and computational methods for the analysis of the phase diagram work; these will be discussed in more detail below. Main classes of ``standard'' theoretical tools include
%\begin{enumerate}
% \item 
 lattice QCD simulations,
% \item 
 effective field theory, and 
% \item 
QCD perturbation theory.
%\end{enumerate}

\emph{Lattice QCD} is the main tool to obtain genuinely non-perturbative information about the phase structure of QCD. However, as I pointed out in the introduction, at finite chemical potential lattice QCD analysis suffers from the well known ``sign problem''~\cite{deForcrand:2010ys,Gattringer:2016kco}. That is, the Euclidean path integral becomes complex at finite $\mu$, while it is real at $\mu=0$. At large values of $\mu$, the path integral develops a rapidly oscillating phase, so that the integral involves precise cancellations between contributions from nearby regions in field space, which are extremely difficult to handle numerically. As the chemical potential grows, the severity of the issue increases exponentially. 

At small values of the chemical potential, however, the oscillations can be handled by using, e.g., reweighting methods or Taylor expansion. Consequently, the QCD equation of state, among other things, can be analyzed in this region. Near the critical crossover temperature of about 155~MeV, this means that the simulations are reliable up to $\mu/T \approx 1$~\cite{deForcrand:2010ys} (with $\mu$ begin the quark chemical potential). 
The dependence of the EOS on $\mu$ at small $\mu$ can be conveniently described in term of the dimensionless cumulants
\begin{equation}
 \chi_n(T) =  T^{n-4}\frac{\partial^n p(T,\mu)}{\partial \mu^n}\bigg|_{\mu=0} 
\end{equation}
which have been computed on the lattice up to $n=10$~\cite{Borsanyi:2011sw,HotQCD:2012fhj,Borsanyi:2018grb,Bazavov:2020bjn} (see also~\cite{Borsanyi:2021sxv}). Notice that the pressure of QCD is even under the change of sign of $\mu$ due to charge conjugation invariance. Therefore only the cumulants with even $n$ are nonzero.

By \emph{effective field theory} I refer to a wide class of methods in hadron and nuclear theory, which make use of the description of QCD matter in terms of hadronic degrees of freedom. These include systematic chiral perturbation theory (typically with neutrons and pions only)~\cite{Tews:2012fj,Keller:2020qhx}, other effective Lagrangians with modeled nucleon-nucleon potentials~\cite{Serot:1984ey,Akmal:1998cf}, statistical methods for light nuclei, baryons, and mesons~\cite{Hempel:2009mc}, Skyrme models for baryons and meson interactions between them \cite{Stone:2006fn,Adam:2010fg,Ma:2016gdd}, extended liquid drop models~\cite{Lattimer:1991nc}, as well as mean field theory descriptions~\cite{Shen:1998by}. I do not attempt to review all these models here. See~\cite{FiorellaBurgio:2018dga}  for a recent review on the EOS using this kind of approaches. Quite in general, these descriptions rely on modeling nuclear matter through interactions between individual nucleons, which turns out to be reliable only up to densities around the nuclear saturation density. Potential models could in principle be made better if we knew precisely the interactions for the neutron rich matter appearing in neutron star cores, but scattering experiments can only be made with existing nuclei for which the neutron to proton ratio is not high enough.

\emph{QCD perturbation theory} works at asymptotically high energies where the coupling of QCD becomes small thanks to asymptotic freedom (see~\cite{Ghiglieri:2020dpq} for a review). In the context of the phase diagram at finite temperature and density, this means the region of asymptotically high temperatures and chemical potentials~\cite{Vuorinen:2003fs,Kurkela:2009gj,Kurkela:2016was}. However, for good convergence of these methods temperatures or (quark) chemical potentials well above 1~GeV are required. Convergence can be improved by using resummation (of hard thermal loops), see, e.g.,~\cite{Andersen:1999sf,Blaizot:1999ip,Baier:1999db,Mogliacci:2013mca,Fujimoto:2020tjc}, but reliable predictions at low temperatures and neutron star core densities still cannot be achieved.  

In the remaining white region of Fig.~\ref{fig:phases}, none of the methods described above work reliably, so that no (reliable) first principle results are available. Our knowledge of the phase diagram in this region relies on modeling of QCD. There is a vast literature on such models, including various modifications of the Nambu-Jona-Lasinio model~\cite{Fukushima:2003fw,Ratti:2005jh,Ciminale:2007sr,Kashiwa:2007hw,Costa:2008yh}, quasiparticle extensions of the (resummed) perturbative QCD~\cite{Peshier:1999ww,Rebhan:2003wn,Peshier:2005pp}, and recently the functional renormalization group methods~\cite{Drews:2014spa,Drews:2016wpi,Otto:2019zjy} which is expected to capture some nonperturbative features of QCD. Another possibility is to use gau\-ge/gra\-vity duality, which I will discuss in this review.

Notice that not even the details of the phase diagram are known in the white region. In particular, even the existence of the nuclear to quark matter transition remains a conjecture, and this is why I marked it as a dashed line in Fig.~\ref{fig:phases}.
Moreover, there is the possibility that various ``exotic'' phases appear in the diagram. These include paired phases such as the color-flavor-locked phase, which is actually expected to extend up to asymptotically high chemical potentials, different kind of color superconducting phases with various configurations~\cite{Alford:1997zt,Alford:2007xm}, and quarkyonic phases which share features with both the nuclear and quark matter phases~\cite{McLerran:2007qj}.

\begin{figure}
\centering  \includegraphics[width=0.48\textwidth]{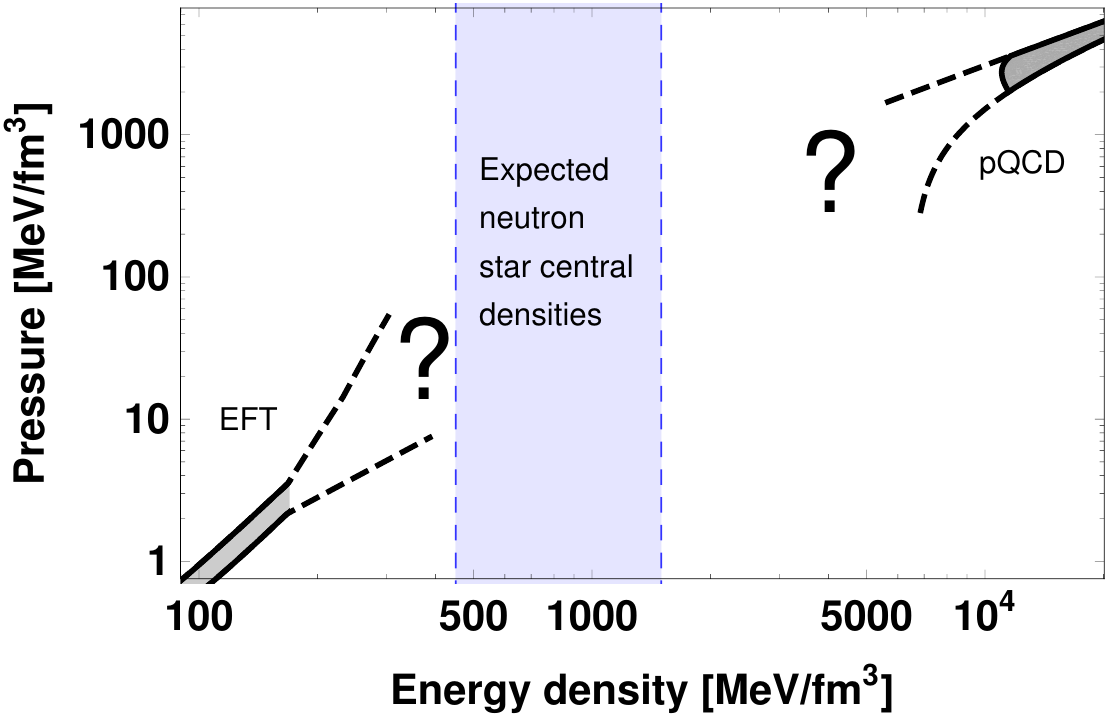}
\caption{Predictions for the QCD EOS from systematic theoretical methods at low~\cite{Tews:2012fj,Hebeler:2013nza} and high~\cite{Kurkela:2009gj} densities compared to the estimated range of central densities in neutron stars.}
\label{fig:eossketch}       % Give a unique label
\end{figure}

Improving the knowledge of QCD in the region of intermediate densities would therefore be essential to pin down the phase structure and EOS of QCD in this region. And this region is not only of academic interests but has applications in real world: Neutron star central densities are known to lie in this regime of intermediate densities. Such high densities are also probed in supernova explosions, even in the case where no neutron star is formed (e.g. due to the core collapsing into a black hole). See 
Fig.~\ref{fig:eossketch} where I show simple estimate for the validity of chiral perturbation theory results at low density~\cite{Tews:2012fj,Hebeler:2013nza} and the validity of perturbative results at high density~\cite{Kurkela:2009gj}, compared to the estimated densities appearing in the cores of most massive neutron stars.

%%%%%%%%%%%%%%%%%%%%%%%%%%%%%%%%%%%%%
\subsection{Experimental efforts: heavy-ion collisions} \label{ssec:exp}
%%%%%%%%%%%%%%%%%%%%%%%%%%%%%%%%%%%%%

Apart from experimental data for the hadron spectrum, decay widths, and cross sections, which are properties of QCD at zero temperature, there is plenty of data which directly probe the high temperature deconfined phase and the crossover region at low density
from 
heavy-ion collisions carried out at RHIC and LHC~\cite{Busza:2018rrf}. 
There are also substantial efforts to extend the experimentally probed region towards higher densities. The most important ongoing program is the beam energy scan at RHIC at Brookhaven which aims at probing the regime where the QCD critical point is expected to lie. The basic idea is to vary the collision energy of Au+Au collisions at around 10~GeV (much lower than the maximum 200~GeV collision energy) and search for evidence of the critical point: non-monotonicity of moments of the net-baryon number distribution as a function of the energy. The first results from the phase II measurements at the STAR detector already report such  non-monotonicity at 3.1$\sigma$ level~\cite{STAR:2020tga}. 

Regions with even higher densities, towards densities appearing in neutron star cores and in neutron star mergers, will be probed in planned future experiments. They will be carried out at the FAIR facility at Darmstadt, Germany (including in particular the CBM experiment), J-PARC at Tokai, Japan as well as at NICA at JINR, Dubna, Russia.

Apart from heavy-ion collisions, there is also plenty of experimental information about dense QCD coming from measurements of neutron stars. But before discussing them, we should first recall a few basic facts about neutron stars.

%%%%%%%%%%%%%%%%%%%%%%%%%%%%%%%%%%%%% 
\subsection{Neutron stars from a QCD viewpoint} \label{ssec:QCDNS}
%%%%%%%%%%%%%%%%%%%%%%%%%%%%%%%%%%%%%

From theoretical viewpoint, neutron stars are (in simplest approximation) large blobs of static cold dense QCD (nuclear or quark) matter. They are self-gravi\-tating and prevented from collapse by a combination of Fermi pressure and repulsive interactions of the constituents, leading to extremely dense and compact stars with radii around 10 to 15 km. See~\cite{Baym:2017whm} for a recent review.

A static spherically symmetric body in general relativity is described in terms of the Tolman–Oppen\-heimer–Vol\-koff (TOV) equations,
\begin{eqnarray}
 p'(r) &=& - \frac{(\epsilon(p(r))+p(r))(m(r)+4\pi r^3p(r))}{r^2\left(1-\frac{2m(r)}{r}\right)} \ , \\
 m(r) &=& 4\pi \int_0^r d\hat r\, \hat r^2\, \epsilon(p(\hat r)) \ .
\end{eqnarray}
Here the first equation is the equivalent of hydrostatic equilibrium in general relativity, and the second equation defines the total mass within the radius $r$. In order to properly define the radial coordinate $r$, one also needs to specify the metric of the star, which is
\begin{eqnarray}
 ds^2 &=& e^{\nu(r)}dt^2 - \left(1-\frac{2m(r)}{r}\right)^{-1} dr^2 -r^2 d\Omega_2^2 \ ,\\
 \nu'(r) &=& - \frac{2}{\epsilon(p(r))+p(r)} p'(r)
\end{eqnarray}
with the boundary condition that outside the star the Schwarzschild metric is obtained.

\begin{figure}
\centering  \includegraphics[width=0.4\textwidth]{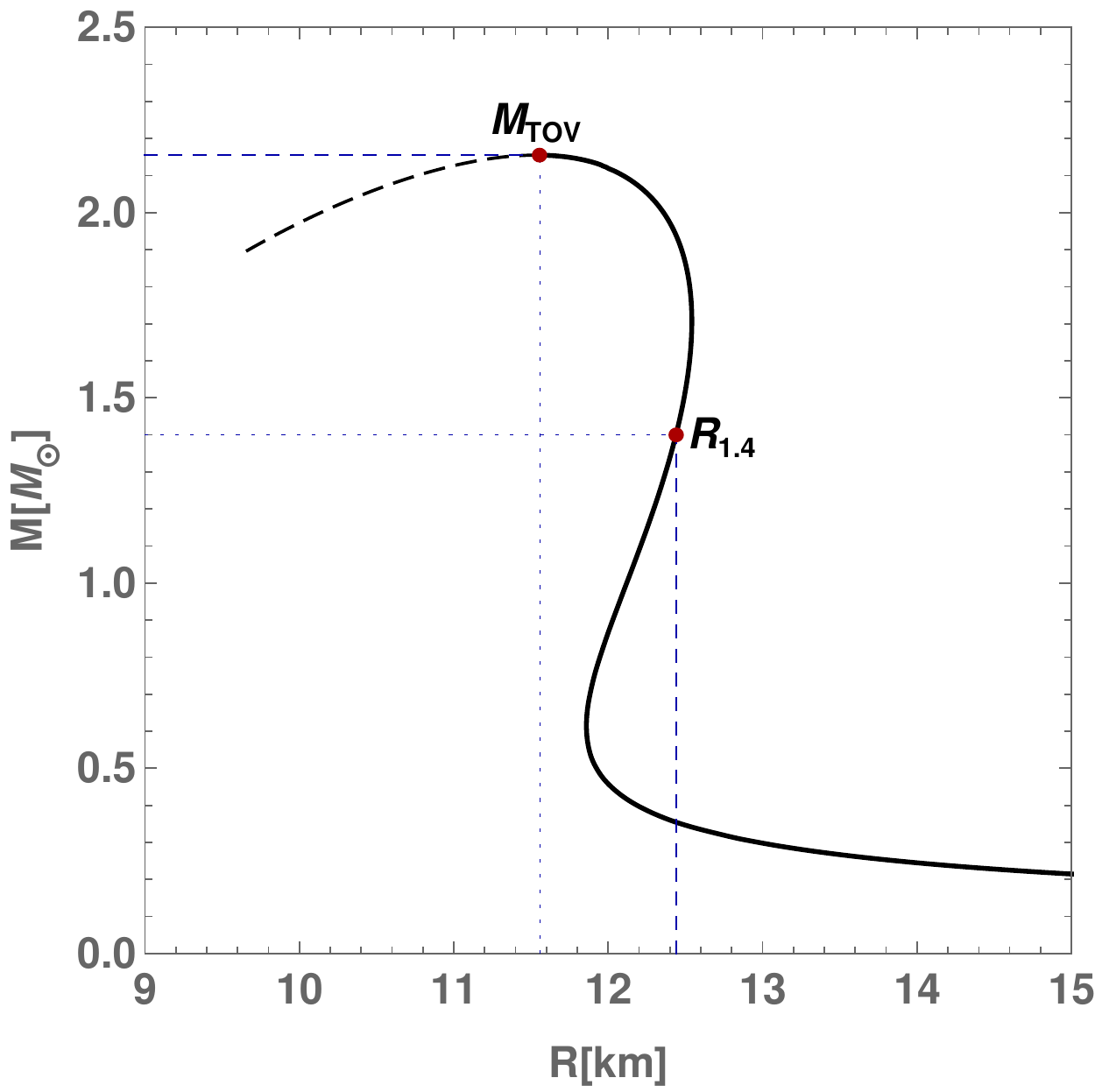}
\caption{An example of a mass-radius curve for neutron stars. The solid (dashed) black curve is the stable (unstable) branch. I also mark the maximum mass $M_\mathrm{TOV}$ of (nonrotating) stars and the radius $R_{1.4}$ at $M=1.4 M_\odot$.}
\label{fig:MRcurve}       % Give a unique label
\end{figure}

A key observation is that the TOV equations only depend on the underlying theory of the matter through the equation of state, i.e., the function $\epsilon(p)$. In practice, the EOS for the bulk of the star is determined by QCD: electroweak interactions only provide small corrections (compared to the uncertainty arising from the uncertainty of the QCD EOS). Solving the TOV equations gives the mass - radius relation for neutron stars. One can show that the mapping between the $M(R)$ curve and the EOS is one-to-one. Therefore measuring the masses and radii of neutron stars provides definite information about the EOS. I show an example of a mass-radius relation in Fig.~\ref{fig:MRcurve}. The central baryon number density of the star $\epsilon_c$ is a monotonic function along the curve, and increases towards top left on the plot. On the part of the curve where
\be
 \frac{dM}{d\epsilon_c}<0 \ ,
\ee
i.e., on the dashed section of the curve, the stars are unstable towards a gravitational collapse leading to  black hole formation. The maximum mass $M_\mathrm{TOV}$ is reached at the onset of the instability. I also mark the radius of the star $R_{1.4}$ at $M=1.4 M_\odot$, which is a typical mass observed in neutron star binaries. It is also possible that there are two separate stable branches of the curve at high masses, in this case there is a range of masses with two stable solutions (``twin stars'')~\cite{Glendenning:1998ag}.

Simply solving the TOV equation does not give the complete picture for neutron stars. First, the neutron stars typically rotate, and the rotation can be extremely fast in the case of millisecond pulsars. However even for the highest measured rotation frequencies, the deformation due to rotation is rather mild, and can be viewed as a relatively small correction to the above picture. For most of the known pulsars, the rotation is much slower so that the deformation is tiny. Second, the temperature of neutron stars is not exactly zero. For newly formed stars the temperature is expected to be relatively high, i.e., almost comparable to the QCD scale, so that the finite temperature corrections to the EOS are nonnegligible. However the star cools down rapidly due to neutrino emission so that the observed temperatures of old neutron stars are suppressed with respect to the QCD scale by orders of magnitude and temperature effects can be safely neglected. Third, neutron stars are known to contain high magnetic fields. For some stars (magnetars)~\cite{Kaspi:2017fwg}, which have particularly high magnetic fields, the strength of the field can be from $10^9$ to $10^{11}$ T. But even such enormously high magnetic fields are still way below the QCD scale: the pion mass squared corresponds to over $10^{13}$~T. Therefore the effect of the magnetic fields on the EOS can be safely neglected even for magnetars, let alone regular neutron stars. 

One should also recall (as I already remarked above) that there are contributions to the EOS from other sectors of the standard model than QCD. Most importantly, there is the pressure of electrons (and other leptons): not even the most massive neutron stars are completely made of neutrons, but also include a fraction of protons and electrons whose numbers must balance for the star to be charge neutral. But these contributions are both easier to compute than the QCD EOS, and are small with respect to the QCD contribution.

Apart from the mass and radius, neutron stars are characterized by a number of other observables. These include the moment of inertia $I$,  quadrupole moment to quadratic order in spin $Q$, and the tidal Love number $\Lambda$ (see, e.g.,~\cite{Yagi:2013awa}). They can all be computed by considering (slowly rotating) star solutions in general relativity and the only input from QCD is still the EOS. Perhaps the most interesting observable for our purposes is the tidal deformability, which measures how much the neutron star is deformed by tidal forces. It can be constrained by measurements of gravitational waves from neutron star mergers, as I will discuss below.

%%%%%%%%%%%%%%%%%%%%%%%%%%%%%%%%%%%%%
\subsection{Experimental efforts: neutron stars and their mergers} \label{ssec:expNS}
%%%%%%%%%%%%%%%%%%%%%%%%%%%%%%%%%%%%%

There is already a fair amount of data for neutron star properties from measurements of isolated pulsars, neutron stars in binaries, and from neutron star mergers. From the point of view of constraining the EOS, perhaps the most interesting are the measurements of neutron star masses and radii (see the review~\cite{Lattimer:2012nd}). 

The masses of a few dozens of stars have been measured. Among these masses, some of the heaviest accurately measured neutron stars are found in pulsar -- white dwarf binaries, and can be measured through the Shapiro time delay of the pulses when the pulsar passes behind its companion (see the review~\cite{Ozel:2016oaf}). Such results include (in units of the solar mass $M_\odot$ and at 1$\sigma$ confidence level)
\begin{itemize}
 \item The (millisecond) pulsar \textbf{J1614-2230} with the mass $M =  1.908\pm 0.016 M_{\odot}$~\cite{Demorest:2010bx,NANOGRAV:2018hou}. 
 \item The pulsar \textbf{J0348+0432} with the mass $M =  2.01\pm 0.04 M_{\odot}$~\cite{Antoniadis:2013pzd}. 
 \item The pulsar \textbf{J0740+6620} with the mass $M =  2.08\pm 0.07 M_{\odot}$~\cite{Cromartie:2019kug,Fonseca:2021wxt}. 
\end{itemize}
These measurements set a stringent lower bound to the maximum mass $M_\mathrm{TOV}$ of neutron stars at around two times the solar mass. This bound requires the EOS of neutron stars to be stiff, i.e., to have high speed of sound $c_s^2 = dp/d\epsilon$, in particular at around core densities. Otherwise such high masses cannot be reached. Several soft EOSs proposed in the literature are already ruled out by these measurements.

The radii of some neutron stars has also been measured but the radius measurements are more difficult than the mass measurements and have usually much larger relative uncertainties. Typical results for the neutron star radii lie between 10 and 15 km.
The radius measurements can be done by using different methods, including spectroscopic measurements of accreting neutron stars, studies of thermonuclear X-ray bursts, and timing observations of signals due to inhomogeneities as the star rotates~\cite{Ozel:2016oaf}.  The ongoing NICER experiment uses the latter method (pulse profile modeling) where long term observation of the the X-rays emitted by the star as well as its modulation as the star rotates are used to estimate the mass and the radius of the star. The results from these measurements have been published in~\cite{Miller:2019cac,Riley:2019yda,Miller:2021qha,Riley:2021pdl}. The analysis is complemented by using data from the XMM-Newton X-ray telescope. In X-ray bursts, matter falls from a companion star to the surface of the neutron star causing an explosive thermonuclear reaction.  Some of the X-ray burst measurements, obtained by analyzing the cooling after the bursts, report relatively accurate results (see, e.g.,~\cite{Nattila:2017wtj}) but these results depend on the modeling of the neutron star ``atmosphere'', i.e., the thin layer at the surface of the star having low density, which brings in additional uncertainty~\cite{Salmi:2020iwq}. 

Another way to study neutron stars is the observation of pulsar ``glitches'', i.e., sudden changes in the rotational frequency of the star~\cite{Haskell:2015jra}. The mechanisms causing glitches are still mostly unknown. They can be analyzed via precise timing measurements, e.g., the SKA and UTMOST programmes.

Apart from properties of a single star, recently binary merger events involving neutron stars have been observed. The first was GW170817 in 2017, which was observed both through gravitational waves by advanced LIGO/Virgo and thereafter by telescopes and observatories ranging basically over the whole spectrum of electromagnetic waves~\cite{TheLIGOScientific:2017qsa,GBM:2017lvd}. Later, another likely merger event (GW190425) was observed by LIGO/Virgo observatory, but in this case the observed gravitational wave signal was weaker and  the electromagnetic counterpart could not be detected~\cite{LIGOScientific:2020aai}. LIGO/Virgo has also observed two events that were most likely mergers of a black hole with a neutron star~\cite{LIGOScientific:2021qlt}, and an additional event (GW190814) where a black hole merged with a $2.6 M_\odot$ object that could be a neutron star or a black hole~\cite{Abbott:2020khf}. 

The first and cleanest observation of a neutron star merger, GW170817, sets also bounds for the EOS. These come from the measurement of the gravitational wave signal, which actually only contains the inspiral phase before the merger. It is likely that gravitational waves were also emitted after the merger, but their frequency was higher, and the sensitivity of the detectors decreases with frequency in the relevant range, so that the aftermerger signal was not detectable. The inspiral signal carries information about the tidal deformability $\Lambda$ of neutron stars. A weak signal of deformation was detected:  the best fit to the data prefers mild deformation. Therefore the data sets a strong upper bound (and a weak lower bound) to the tidal deformability. The analysis by LIGO/Virgo collaboration, which assumed that both merging neutron stars are described by the same EOS, concluded that $\Lambda_{1.4} = 190^{+390}_{-120}$ at 90\% confidence level, where the subscript $1.4$ refers to the mass $M \approx 1.4 M_\odot$ of each of the stars~\cite{Abbott:2018exr}. The upper bound for $\Lambda_{1.4}$ is particularly interesting because it is complementary to the constraint from the mass measurements (i.e., $M_\mathrm{TOV} \gtrsim 2.0 M_\odot$): it excludes EOSs which are too stiff. That is, for an EOS to meet both bounds, it needs to be stiff but not too stiff.

The GW170817 may also set a different kind of bound for the EOS. Namely, the analysis of the electromagnetic signal from the merger suggests that a supramassive neutron star was formed in the merger, which later, within one second or so, collapsed into a black hole (see, e.g.,~\cite{Gill:2019bvq}). If this was the case, the mass of the remnant was above the maximum mass of stable star. This sets an upper bound for $M_\mathrm{TOV}$. Computing the exact bound is, however, involved because only the total mass before the merger is known precisely. Consequently the mass bound depends on the details of the event. Depending on the assumptions, estimates for the bound vary between 2.15 and 2.3 solar masses~\cite{Margalit:2017dij,Rezzolla:2017aly,Shibata:2019ctb}.

Additional and more precise measurement of neutron stars are expected in near future. The radius measurement are becoming more precise due to progress with the methods and new experiments. At the same time, LIGO and Virgo are continuing observations with improved sensitivity, and will soon be accompanied by other gravitational wave observatories such as LIGO-India. Eventually, third generation experiments such as the Einstein telescope will provide detailed information on the gravitational wave signals from neutron star mergers.

%%%%%%%%%%%%%%%%%%%%%%%%%%%%%%%%%%%%%
\subsection{State-of-the-art for cold QCD EOS} \label{ssec:QCDeos}
%%%%%%%%%%%%%%%%%%%%%%%%%%%%%%%%%%%%%

Let me then discuss the state-of-the-art of the EOS of (cold and) dense QCD matter and in particular the effect of the neutron star measurements on it. A model-independent method for studying this is to use parameterized families of EOSs which extrapolate from known results at low and/or high densities, or interpolate between them. 

A popular parametrization is polytropic EOSs where one joins continuously pieces of EOSs which each have constant adiabatic index $\gamma = dp/dn$, where $n$ is the baryon number density. The intervals with constant $\gamma$ can have variable widths (in $n$) and they are typically joined such that the joints are second order transitions, which are artificial in the sense that there is change in the underlying physics which would cause these transitions.
Interpolations between nuclear and quark matter were considered using polytropes in~\cite{Kurkela:2014vha,Annala:2017llu,Most:2018hfd}, whereas~\cite{Hebeler:2013nza,Bedaque:2014sqa,Tews:2018kmu} use only data for nuclear matter, in practice assuming that the quark matter EOS can be matched through a first order phase transition of arbitrary strength. Also other kinds of interpolations have been considered in the literature, for example piecewise continuous Ans\"atze~\cite{Annala:2019puf} or other continuous parametrizations~\cite{Bedaque:2014sqa,Tews:2018kmu} for the speed of sound.

\begin{figure}
\centering  \includegraphics[width=0.48\textwidth]{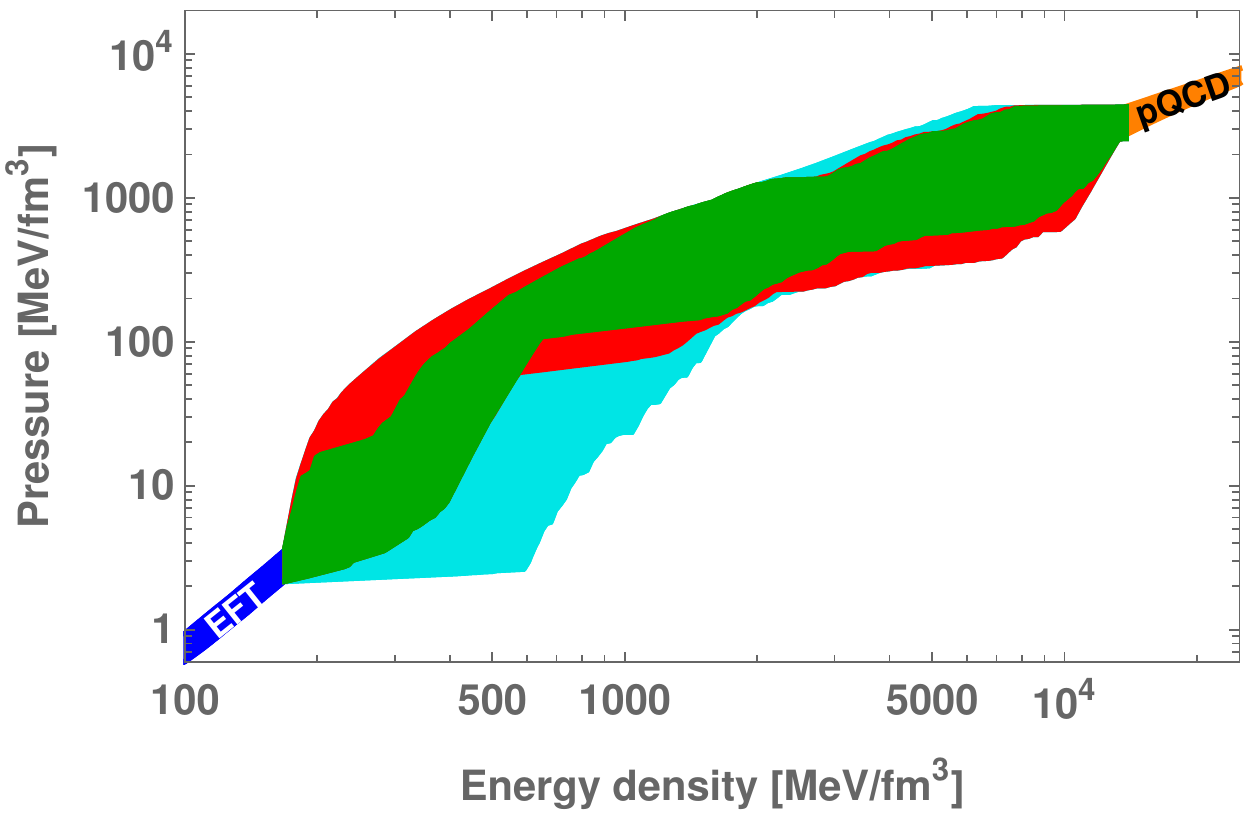}
\caption{The band spanned by polytropic interpolations of cold QCD EOS between the low density (EFT) and high density (pQCD) results. Following~\cite{Annala:2017llu}.}
\label{fig:eosinterp}       % Give a unique label
\end{figure}

In Fig.~\ref{fig:eosinterp} I show bands spanned by quadrutropic interpolations (four intervals with constant $\gamma$) between effective field theory results at low density and perturbative QCD results at high density, following the approach of~\cite{Annala:2017llu} (see also~\cite{Annala:2021gom}).\footnote{I thank the authors of~\cite{Annala:2017llu} for sharing their data which was used to generate this plot.} The full band (all colors) is spanned by all the EOSs consistent with the low and high limits (and also the causality constraint $c_s^2<1$) but without adding constraint from the measurements of neutron stars. The cyan area is then excluded by the constraint $M_\mathrm{TOV}>2 M_\odot$, and the red area is excluded by the LIGO/Virgo bound $\Lambda_{1.4}<580$.\footnote{Notice that there are various values for the bound: the early estimate of~\cite{TheLIGOScientific:2017qsa} for low-spin priors was $\Lambda_{1.4}<800$, and the analysis of~\cite{Abbott:2018wiz}, which was carried out by using a different method than in~\cite{Abbott:2018exr}, obtained $\Lambda_{1.4}<720$. All these limits are reported at 90\% confidence level.} Therefore the remaining EOSs consistent with both bounds span the green band.

Notice that the amount of polytropes used to obtain Fig.~\ref{fig:eosinterp} was relatively modest so that some corners of the bands may not be perfectly reproduced. See~\cite{Annala:2021gom} for fully up-to-date bands.
This reference also studies effects of other data, in particular the (less constraining) radius measurements and the possible upper bound of $M_\mathrm{TOV}$ that I discussed above. 

Extending the EOS to finite temperature is less well controlled, but there are several models also for the temperature dependence. See~\cite{Raduta:2021coc} for a recent overview of the temperature dependence, and~\cite{Oertel:2016bki} for a generic review of the equations of state. And apart from temperature, one can also consider dependence on charge fraction (which is important for neutron star mergers), isospin chemical potential, external magnetic field, and so on. I will not discuss such extensions in this review.

%Discuss polytropes and fig 3
%Mention HLPS also

%%%%%%%%%%%%%%%%%%%%%%%%%%%%%%%%%%%%%
%%%%%%%%%%%%%%%%%%%%%%%%%%%%%%%%%%%%%
%%%%%%%%%%%%%%%%%%%%%%%%%%%%%%%%%%%%%
%%%%%%%%%%%%%%%%%%%%%%%%%%%%%%%%%%%%%
%%%%%%%%%%%%%%%%%%%%%%%%%%%%%%%%%%%%%
%%%%%%%%%%%%%%%%%%%%%%%%%%%%%%%%%%%%%
%%%%%%%%%%%%%%%%%%%%%%%%%%%%%%%%%%%%%
%%%%%%%%%%%%%%%%%%%%%%%%%%%%%%%%%%%%%
%%%%%%%%%%%%%%%%%%%%%%%%%%%%%%%%%%%%%
%%%%%%%%%%%%%%%%%%%%%%%%%%%%%%%%%%%%%
%%%%%%%%%%%%%%%%%%%%%%%%%%%%%%%%%%%%%
%%%%%%%%%%%%%%%%%%%%%%%%%%%%%%%%%%%%%
%%%%%%%%%%%%%%%%%%%%%%%%%%%%%%%%%%%%%
%%%%%%%%%%%%%%%%%%%%%%%%%%%%%%%%%%%%%
%%%%%%%%%%%%%%%%%%%%%%%%%%%%%%%%%%%%%
%%%%%%%%%%%%%%%%%%%%%%%%%%%%%%%%%%%%%
%%%%%%%%%%%%%%%%%%%%%%%%%%%%%%%%%%%%%
%%%%%%%%%%%%%%%%%%%%%%%%%%%%%%%%%%%%%
%%%%%%%%%%%%%%%%%%%%%%%%%%%%%%%%%%%%%
%%%%%%%%%%%%%%%%%%%%%%%%%%%%%%%%%%%%%
%%%%%%%%%%%%%%%%%%%%%%%%%%%%%%%%%%%%%
%%%%%%%%%%%%%%%%%%%%%%%%%%%%%%%%%%%%%
%%%%%%%%%%%%%%%%%%%%%%%%%%%%%%%%%%%%%
%%%%%%%%%%%%%%%%%%%%%%%%%%%%%%%%%%%%%
%%%%%%%%%%%%%%%%%%%%%%%%%%%%%%%%%%%%%
%%%%%%%%%%%%%%%%%%%%%%%%%%%%%%%%%%%%%
%%%%%%%%%%%%%%%%%%%%%%%%%%%%%%%%%%%%%
%%%%%%%%%%%%%%%%%%%%%%%%%%%%%%%%%%%%%
\section{Brief review of holographic models for QCD}\label{sec:holoqcd}
%%%%%%%%%%%%%%%%%%%%%%%%%%%%%%%%%%%%%

Apart from dense QCD, the other major topic of this review is gau\-ge/gra\-vity duality. I start by giving a brief review for the basic structure of the duality in the conformal case (AdS/CFT), and go on discussing various approaches for QCD (which should include non-con\-formality, confinement, and chiral symmetry breaking). In this brief review I focus on topics relevant for QCD, and many results are stated without derivation or motivation. See~\cite{Aharony:1999ti,Natsuume:2014sfa,Ammon:2015wua} for more extensive reviews of gau\-ge/gra\-vity duality.

%%%%%%%%%%%%%%%%%%%%%%%%%%%%%%%%%%%%%
\subsection{Basics of gau\-ge/gra\-vity duality}
%%%%%%%%%%%%%%%%%%%%%%%%%%%%%%%%%%%%%

The gau\-ge/gra\-vity duality is formulated as a duality between a field theory and a higher dimensional gravitational theory. In its most commonly used form, the field theory is strongly coupled whereas the gravitational theory is a weakly coupled classical theory. Therefore the correspondence is a tool to study strongly coupled gauge theories: relatively simple computations in classical gravity can provide answers to questions in strongly coupled field theory vie the duality which would otherwise be extremely challenging.

The best known example of such a duality is the AdS/CFT correspondence between the four dimensional $\mathcal{N}=4$ Super-Yang-Mills theory with the gauge group SU($N_c$), which is a superconformal theory, and type IIB string theory on AdS$_5\times S^5$~\cite{Maldacena:1997re,Gubser:1998bc,Witten:1998qj}. In this review I will not go to the details of this example, but discuss general features of AdS/CFT. That is, quite in general one expects that CFTs in $d$-dimensions can have gravity duals with the geometry of AdS$_{d+1}$. This duality is motivated, among other things by the fact that the isometries of the AdS$_{d+1}$ space match with the $d$-dimensional conformal group SO$(2,d)$.  To be precise, if one considers a CFT in flat Minkowski space, the bulk geometry is the Poincar\'e patch of the AdS space, i.e., a coordinate patch covering a section of the full geometry.

The metric of the Poincar\'e patch 
may be written as
\be \label{eq:adsmetric}
 ds^2 = \frac{\ell^2}{r^2}\left(dr^2 + \eta_{\mu\nu} dx^\mu dx^\nu\right) \ ,
\ee
where $\ell$ is the AdS radius, $x^\mu$ are the usual space-time coordinates, the Greek indices denote the $d$-dimensional Lorentz indices, and the holographic coordinate $r$ runs from $r=0$ to $r = \infty$. I will be using mostly plus conventions for the Minkowski metric $\eta_{\mu\nu}$. It is understood that the field theory lives at the ``boundary'' of the AdS space, which is identified as the limit $r \to 0$. The (inverse of the) holographic coordinate may be interpreted as the energy scale of the field theory as suggested by the fact that the metric components of the space-time coordinates scale as $\sim 1/r^2$. For the conformal, i.e., AdS case, the metric is invariant under the mapping $x_\mu \mapsto \Lambda x_\mu$ and $r \mapsto \Lambda r$ as required by scale invariance.

The AdS/CFT correspondence is defined, among other things, by specifying the dictionary: how various bulk fields $\phi_i(r,x^\mu)$ correspond to boundary operators $\mathcal{O}_i(x^\mu)$, where $i$ indexes all the fields/operators. A special case of the dictionary is that the metric itself is dual to the energy momentum tensor $T_{\mu\nu}$ of the field theory. In order to define the correspondence concretely, one turns on source fields $J_i(x_\mu)$ for the operators $\mathcal{O}_i(x^\mu)$. The correspondence is then defined by equating the generating functional $Z_\mathrm{CFT}[\{J_i\}]$ of the field theory with the classical (on-shell) partition function of the gravity, with the boundary condition that $\phi_i$ matches with the source $J_i$ at the boundary~\cite{Gubser:1998bc}:
\be
 Z_\mathrm{CFT}[\{J_i\}] = Z_\mathrm{grav}[\{J_i\}] \ .
\ee

I will then illustrate the correspondence by considering a simple explicit example of massive bulk scalar field in AdS. Notice that the bulk theory is expected to have a dynamical (Einstein) gravity sector, to which~\eqref{eq:adsmetric} is a solution. However as the simplest illustrative example of the correspondence, it is convenient to ignore the gravity sectors and consider scalars which only probe the AdS geometry on the bulk side. That is, we may take the (probe) bulk action as
\begin{eqnarray}
 \mathcal{S}_{d+1} &=& %\frac{1}{16\pi G_{d+1}}
 - \mathcal{N}\int d^4x dr \sqrt{-\det g}\\ \nonumber
  && \times \sum_{i}\left[\frac{1}{2} g^{MN}\partial_M\phi_i\partial_N\phi_i + \frac{1}{2}m_i^2 \phi_i^2 \right]
\end{eqnarray}
where $\mathcal{N}$ is an arbitrary normalization constant, the indices $M$, $N$ run through all $d+1$ dimensions and $g_{MN}$ is the AdS metric~\eqref{eq:adsmetric}. I only consider homogeneous solutions that only depend on the holographic bulk coordinate $r$. If we parametrize 
\be
 m_i^2 = \Delta_i(\Delta_i-d) \ ,
\ee
the solutions are given by
\be \label{eq:phisol}
 \phi_i(r) = J_i r^{d-\Delta_i}+ \sigma_i  r^{\Delta_i} \ ,
\ee
where we already identified the coefficient of the dominant solution at the boundary (taking $\Delta_i>d/2$) with the source $J_i$, which is in this case $x^\mu$-independent. The coefficient of the subdominant solution is then identified by the VEV of the operator $\mathcal{O}_i(x^\mu)$, and therefore $\Delta_i$ is the dimension of the operator. 

I sketch then how the VEV of the operator arises by using the dictionary. Inserting the solution~\eqref{eq:phisol} in the action we obtain that (assuming such boundary conditions at $r\to \infty$ that no terms arise from there)
\begin{eqnarray}
 \mathcal{S}_{d+1}^{(\mathrm{o.s.})} &=& -\frac{1}{2} \mathcal{N}V_4 \ell^d \phi(r) r^{2-d} \phi'(r)\Big|_{r = \epsilon}\\ 
&=& -\frac{d}{2} \mathcal{N} V_4 \ell^d \sum_{i} J_i \sigma_i + \mathrm{divergent}
\end{eqnarray}
where $V_4$ is the volume of space-time and the divergent piece as $\epsilon \to 0$ should be removed by holographic renormalization (see~\cite{Skenderis:2002wp} for details). That is, the divergences are canceled by adding (covariant) boundary terms which do not affect the dynamics in the bulk. In this case, the necessary counterterm is
\be
 \mathcal{S}_\mathrm{ct} = \frac{1}{2}V_4 \ell^d\epsilon^{-d} \sum_{i} \phi_i(\epsilon)^2(d-\Delta_i)
\ee
and the regularized action is defined as $\widetilde{\mathcal{S}}_{d+1}^{(\mathrm{o.s.})} = \mathcal{S}_{d+1}^{(\mathrm{o.s.})}  + \mathcal{S}_\mathrm{ct}$. This UV divergence reflect a similar divergence on the field theory side.
The correspondence now states that
\be \label{eq:Zequiv}
\frac{1}{Z_\mathrm{CFT}}\! \int\! \mathcal{D}\, e^{i S_\mathrm{CFT} + i\sum_i\! J_i\! \int\! d^4x\,\mathcal{O}_i(x^\mu)} = e^{i\widetilde{\mathcal{S}}_{d+1}^{(\mathrm{o.s.})} (\{J_i\})}
\ee
where the left hand side is the generating function of the CFT and the right hand side is the on-shell bulk (gravity) partition function, or to be precise, the part of it which depends on the sources $J_i$. Moreover, $\mathcal{D}$ is the path integral measure of the CFT, $S_\mathrm{CFT}$ is the CFT action, and $Z_\mathrm{CFT} = \int \mathcal{D}\, \exp(i S_\mathrm{CFT})$. Expanding at leading nontrivial order we finally find the relation between the VEV and $\sigma_i$:
\be
 \langle\mathcal{O}_i\rangle = \left(\frac{d}{2}-\Delta_i\right) \mathcal{N} V_4 \ell^d \sigma_i \ .
\ee

Turning on finite temperature will be an important part of this review, and can be studied due to ``planar black holes''~\cite{Witten:1998zw}: The geometry~\eqref{eq:adsmetric} is a solution to the $d+1$ Einstein gravity with the cosmological constant $\Lambda = 12/\ell^2$, but there is also a more general ``black hole'' solution
\begin{eqnarray} \label{eq:adsFTmetric}
 ds^2 &=& \frac{\ell^2}{r^2}\left(\frac{dr^2}{f(r)} -f(r) dt^2 + \delta_{ij} dx^i dx^j\right) \ , \\
 f(r) &=& 1 - \left(\frac{r}{r_h}\right)^d \ , \label{eq:fCFT}
\end{eqnarray}
where the indices $i$ and $j$, run over the $d-1$ spatial coordinates, and $r=r_h$ is the location of the horizon. By the black hole being planar I mean that the horizon extends to all values of $t$ and $x^i$. %One can show that the dictionary implies that 

The thermodynamics of the field theory is obtained from the thermodynamics of the black hole. The temperature is the Hawking temperature, given by the surface gravity $\sim f'(r_h)$, and can be derived by requiring the regularity of the geometry at the horizon as follows. As in field theory, we first Wick rotate $t \mapsto -i\tau$ to obtain the Euclidean geometry and compactify on a circle. The temperature in field theory is given as the inverse of the periodicity of the Euclidean time coordinate, i.e., $\beta=1/T$. For the projection of the geometry in the time and holographic direction we then have
\begin{eqnarray}
 ds_\mathrm{2d}^2 &=& \frac{\ell^2}{r^2}\left(\frac{dr^2}{f(r)} -f(r) dt^2\right)\\\nonumber
 &\approx& \frac{\ell^2}{r_h^2}\left(\frac{dr^2}{f'(r_h)(r-r_h)}+f'(r_h)(r-r_h)d\tau^2\right) \ ,
\end{eqnarray}
where higher order corrections in $r_h-r$ were neglected. Substituting here $\rho = \sqrt{r_h-r}$, the metric becomes
\be
 ds_\mathrm{2d}^2 \approx -\frac{4\ell^2}{r_h^2f'(r_h)}\left(d\rho^2+\frac{(f'(r_h))^2}{4}\rho^2d\tau^2\right) \ ,
\ee
which we recognize as the flat space metric in radial coordinates if $-f'(r_h)\tau/2$ is identified as the angular variable. The absence of a conical singularity therefore requires that the periodicity of the angle is $2\pi$, which in turn implies that the periodicity of $\tau$ must be $-4\pi/f'(r_h)$. Since this is also the inverse of the temperature, we obtain
\be \label{eq:THaw}
 T = -\frac{1}{4\pi}f'(r_h) = \frac{d}{4\pi r_h} \ ,
\ee
where we inserted~\eqref{eq:fCFT} in the last step.
The other fundamental relation of black hole thermodynamics is the Bekenstein–Hawking formula: the entropy (density) $s$ is given by the area (element) $A$ of the black hole as
\be
 s = \frac{1}{4G_{d+1}} A = \frac{1}{4G_{d+1}}\frac{\ell^{d-1}}{r_h^{d-1}} = \frac{1}{4G_{d+1}}\frac{(4\pi T\ell)^{d-1}}{d^{d-1}} \ .
\ee
where $G_{d+1}$ is the $d+1$ dimensional Newton constant.

Finally let us recall the limitations of gau\-ge/gra\-vity duality. They are most clear in the original formulation for $\mathcal{N}=4$ SYM but apply more generally in AdS/CFT setups. 

First, the number of colors needs to be large, otherwise we would need to solve full string theory in the bulk. On the field theory side taking $N_c \to \infty$ means that we are only accounting for ``planar'' diagrams in the double line notation introduced by 't Hooft~\cite{tHooft:1973alw}. Notice that even such planar diagrams will include interactions to all orders in the gauge coupling. On the gravity or bulk side this means that we are neglecting string loops. A consequence of working in this limit is large $N_c$ factorization: all higher point correlators can be expressed in terms of one-point and two-point functions. This is already visible from the result~\eqref{eq:Zequiv} from which arbitrary-point functions can be extracted~\footnote{Notice that the result of~\eqref{eq:Zequiv} is particularly simple  because we did not include the spatial dependence of the source.}.
Second, the 't Hooft coupling $\lambda_\mathrm{'t H} = g^2 N_c$, where $g$ is the coupling of the gauge theory, needs to be large in order to validate the use of classical gravity in the bulk. This reflects the nature of the duality: computations in gauge theory are straightforward but in gravity difficult at weak coupling, and vice versa at strong coupling.

In principle one can relax these requirements of large $N_c$ and strong coupling, but then one needs to go beyond the planar classical gravity approximation, which is extremely challenging. I will not attempt to do this in this review, but only work with classical gravity.

%%%%%%%%%%%%%%%%%%%%%%%%%%%%%%%%%%%%%
\subsection{Gauge/gravity duality for QCD}
%%%%%%%%%%%%%%%%%%%%%%%%%%%%%%%%%%%%%

I then discuss applying gau\-ge/gra\-vity duality to the prime example of a strongly coupled gauge theory appearing in nature: QCD. It is however far from obvious that this duality can lead to useful results for QCD. While the standard examples with precisely established correspondence involve superconformal field theories, QCD is not supersymmetric and not conformal, but instead has discrete spectrum and confinement. Moreover regular QCD has $N_c=3$ which is not that large, and the coupling constant flows becoming small at high energies (asymptotic freedom) so that the applicability of holography in its standard form, i.e., with classical gravity, becomes questionable. 

However, research in this topic has demonstrated that despite its shortcomings, holography is extremely useful for describing the behavior of QCD. This is due to various reasons: First, since QCD is strongly coupled there is only limited information from first-principles computations, so that even rough results can be useful. For example real-time dependence of QCD plasma, which is important for heavy ion collisions, is challenging to study by using lattice QCD (or other known methods). Second, many results obtained from holography turn out to be insensitive to the precise details of the underlying strong dynamics, and therefore actually apply to a more general class of theories than just QCD. In particular, precise universal relations such as the famous result for the shear viscosity $\eta/s =1 /4\pi$~\cite{Policastro:2001yc,Kovtun:2004de} have been found.
 
The approaches for holographic QCD can be roughly divided into two categories top-down and bottom-up models. The top-down models are based on concrete (typically ten dimensional) setups in string theory, i.e., certain well-chosen supergravity backgrounds, for which also precise control of the gauge theory is possible. That is, we usually know what the gauge theory is, and it is not exactly QCD but may be similar to QCD in some well-defined sense. In some cases, the predictions of these models agree remarkably well with expectations from QCD. A well-known example in this class is the Witten-Sakai-Sugimoto model~\cite{Witten:1998zw,Sakai:2004cn,Sakai:2005yt}. The bottom-up models are, in contrast, engineered ``by hand'' with some inspiration from the top-down models. While quite a bit of guidance is provided by the (global) symmetry of QCD, which the holographic dual should respect, a precise control of the duality is lost: the gravity models are not dual to any explicitly known field theory, but instead involve parameters, which should be adjusted to obtain agreement with QCD physics. But this also means that one is free to do modifications, which may be necessary to model QCD efficiently, but are difficult to realize in the top-down framework. Moreover, it turns out that even very simple bottom-up models (such as the hard-wall model~\cite{Erlich:2005qh}) are able to describe QCD data surprisingly well.  Additional examples of models in both categories will be discussed below.

\subsubsection{Nonconformality and confinement}

I then discuss various features that need to be included in gauge/gravity duality in order to properly describe QCD, which are absent in the conformal AdS/CFT setting. First, the model should be non-conformal and confining. That is, the geometry should be deformed, such that it is no longer exactly AdS so that the conformal group is broken to the Lorentz subgroup SO$(1,d-1)$. The spectrum in CFTs is continuous which arises in the holographic model because fluctuations of the bulk fields are allowed to propagate infinitely far towards the IR endpoint $r = \infty$. One needs to introduce an IR wall which prevents this, giving rise to discrete spectrum and confinement. One often keeps the geometry close to the boundary as either exactly or asymptotically AdS$_5$ since it is expected that in the UV, the theory runs to the trivial free theory fixed point and becomes therefore asymptotically conformal. As we pointed out above, treating the weakly coupled region using gauge/gravity duality may be problematic, but using a geometry that is asymptotically AdS$_5$ is the simplest option, which also guarantees that the standard rules from AdS/CFT correspondence can be applied at the boundary. We will comment more on this below.

\begin{figure}
\centering  \includegraphics[width=0.33\textwidth]{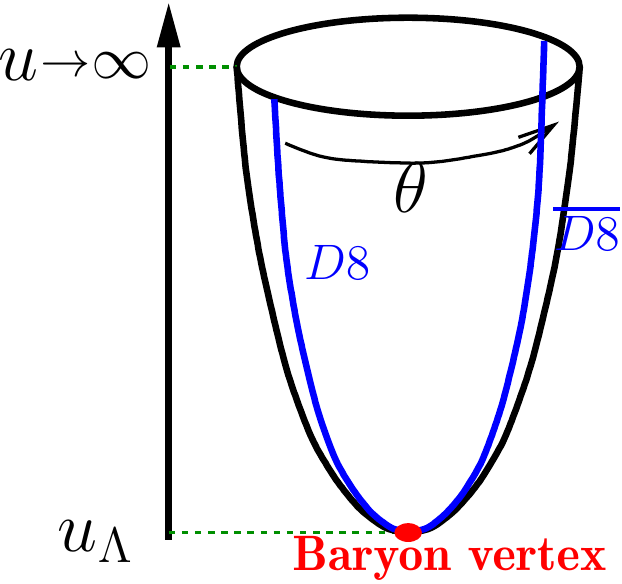}
\caption{A sketch of the confined geometry and the setup for flavors branes in the Witten-Sakai-Sugimoto model.}
\label{fig:WSSgeom}       % Give a unique label
\end{figure}

Within the top-down framework, a typical method to achieve confinement is to compactify one of the spatial directions, which gives rise to an energy scale that will be associated with the scale of confinement~\cite{Witten:1998zw}. The geometry restricted to the compactified coordinate and the holographic coordinate takes the form of a cigar (see Fig.~\ref{fig:WSSgeom}). The endpoint of the cigar creates the IR wall in this case. One can start from a 3+1 dimensional theory so that the final theory has 2+1 (uncompactified) dimensions.  In the case of $\mathcal{N}=4$ SYM this leads to the AdS$_5$ soliton geometry on the bulk side~\cite{Horowitz:1998ha}. Another possibility is to start from a 4+1 dimensional theory in which case one obtains a theory with 3+1 uncompactified dimensions. A well-known example is a geometry~\cite{Itzhaki:1998dd,Brandhuber:1998er}, which will be referred to as Witten's geometry below:
\begin{eqnarray} \label{eq:Wittengeom}
 ds^2 &=& \left(\frac{u}{R}\right)^{3/2}\left(\eta_{\mu\nu}dx^\mu dx^\nu + f(u)d\theta^2\right)\\\nonumber
 & & + \left(\frac{u}{R}\right)^{-3/2}(f(u)^{-1}du^2+u^2d\Omega_4^2)\ ,\\
 f(u)&=& 1 - \frac{u_\Lambda^3}{u^3}\ ,
\end{eqnarray}
where $\theta$ is the compactified coordinate and the holographic coordinate runs from $u=u_\Lambda$ at the tip of the cigar to $u=\infty$ at the boundary. Combined with supersymmetry breaking boundary conditions for fermions on the circle, the only light degrees of freedom are those of pure-glue Yang-Mills theory, which makes this choice particularly interesting. However at scales higher than the compactification scale $M_{KK}$ of the $\theta$ coordinate, given by the periodicity $\theta = \theta +2\pi M_{KK}^{-1}$, additional modes show up. This scale is linked to $u_\Lambda$: in analogy to the derivation of the Hawking temperature~\eqref{eq:THaw} above, the requirement of regularity of the geometry at the tip of the cigar gives 
\be
 M_{KK} = \frac{3}{2}\frac{u_\Lambda^{1/2}}{R^{3/2}}
\ee
so the additional modes cannot be eliminated simply by sending $M_{KK}$ to infinity without affecting the geometry. Actually the issue turns out to be a bit more serious: $M_{KK}$ is the only mass scale in the background so it exactly equals both the scale of the Kaluza-Klein modes and the scale of the (purely 3+1 dimensional) glueballs, as one can also check explicitly. So there is no way to separate the Kaluza-Klein modes from the glueballs. Moreover, the coupling constant does not run in this background but remains a constant parameter. Nevertheless, the phenomenology from this model has turned out to be close to that of Yang-Mills.

Apart from the method of compactifying, well studied confining backgrounds are the Klebanov-Strassler \cite{Klebanov:2000hb} and related Klebanov-Tseytlin~\cite{Klebanov:2000nc} geometries, where nonconformality and confinement is obtained by placing fractional D-branes on a conifold setup.

In bottom-up frameworks, various methods for inducing confinement are available. The simplest is to use AdS$_5$ geometry but introduce a hard cutoff in the IR. The scale of confinement is then the inverse of the coordinate value of the cutoff. Such ``hard wall'' models already produce (among other things) a good description of QCD mass spectrum. But to refine the results, one can introduce ``soft wall'' models: instead a hard cutoff one adds a dilaton field with explicit dependence on the holographic coordinate that breaks conformality and causes, in effect a softer cutoff leading to a more natural spectrum. Apart from the original hard and soft wall models, this kind of approach has been used in light front holography~\cite{deTeramond:2005su,Brodsky:2014yha}, with Einstein-dilaton actions (see, e.g.,~\cite{Li:2013oda}), and in dynamic AdS/QCD models which are inspired by the D3-D7 setup but include bottom-up elements such as IR cutoff and matching with QCD RG flow~\cite{Alvares:2012kr,Alho:2013dka,Erdmenger:2020flu}. The spectra in all these setups agree with QCD data to a good precision.

More complicated bottom-up models include a dynamical dilaton gravity sector with a nontrivial potential for the dilaton that can be adjusted to generate a nontrivial confining geometry and a dilaton profile producing effectively a soft IR wall in good agreement with QCD data. I will discuss below models in this class in more detail.

The main motivation for adding the soft IR wall is to obtain ``Regge-like'' particle spectra where squared masses are linear in excitation number and angular momentum, as observed in QCD. This behavior is reminiscent of spectrum of strings, which originally lead to to the discovery of string theory as a model of QCD at low energies (see, e.g.,~\cite{DiVecchia:2007vd}). In bottom-up models, however, the connection to string theory has been lost and linear confinement is input by adjusting the models by hand.

\subsubsection{Introducing flavors}

Let me then discuss the matter sector in QCD, i.e., quarks in the fundamental representation of the gauge group, and how chiral symmetry is broken.  Quarks are not present in SYM (which has fermions in the adjoint representation) but can be added by considering (flavor) D-branes\cite{Karch:2002sh,Sakai:2003wu,Sakai:2004cn}. Therefore it makes sense to review the brane configurations underlying the holographic models and geometries. The AdS$_5 \times S^5$ geometry arises as the near horizon limit for the (type IIB) supergravity solution of $N_c$ D3 (gauge) branes, and the geometry of~\eqref{eq:Wittengeom} arises from a setup of $N_c$ D4 gauge branes in type IIA supergravity.  

Quarks in the fundamental representation of the gauge group are identified with strings stretching between the gauge and flavor branes. In order to introduce $N_f$ flavors of light fundamental quarks (with lightness obtained when the strings are short), one therefore adds $N_f$ flavor branes that intersect with the gauge branes in the 3+1 dimensions of the field theory. Usually one assumes the probe limit $N_c \gg N_f$, in which the backreaction of the flavor branes to the geometry determined by the gauge branes can be neglected. Taking into account the backreaction is challenging, but has been considered in the literature (often resorting to approximations or special setups such as smeared flavor branes~\cite{Bigazzi:2005md,Nunez:2010sf}).

Typical brane configurations are the D3-D7 models, where one adds $N_f$ D7 flavor branes in the AdS$_5\times S^5$~\cite{Karch:2002sh}, or the Witten-Sakai-Sugimoto model based on the D4-D8-$\overline{\mathrm{D8}}$ configuration~\cite{Sakai:2004cn}, i.e., D8 flavor branes in the gravity background of~\eqref{eq:Wittengeom}. I show this latter configuration schematically in Fig.~\ref{fig:WSSgeom}. A stack of $N_f$ overlapping flavor branes implements a U$(N_f)$ flavor symmetry. One can identify the left-handed and right-handed chiral symmetries U$(N_f)_{L/R}$ in QCD with the flavor symmetries of the D8 and $\overline{\mathrm{D8}}$ branes, respectively. In the confined cigar geometry the D8 and  $\overline{\mathrm{D8}}$ join at the tip of the cigar, which breaks the chiral symmetry down to the vectorial subgroup U$(N_f)_V$. Therefore confinement triggers chiral symmetry breaking in the model.

In the probe limit $N_f \ll N_c$, the flavor branes are described through Dirac-Born-Infeld (DBI) actions, and the flavor dependent operators are dual to the gauge fields on the branes and the fluctuations of the embeddings of the branes. Apart from the D3-D7 and WSS setups, flavors have been considered in the confining Klebanov-Strassler backgrounds~\cite{Sakai:2003wu,Dymarsky:2009cm} by adding D7 branes.

In simpler models (such as the hard and soft wall models~\cite{Erlich:2005qh,Karch:2006pv}) one writes down actions for the matter sector that are polynomial in the fields, roughly corresponding to expansion of the DBI action to first nontrivial order. The typical fields then include left and right handed gauge fields, which are dual to the left and right handed currents $\bar \psi (1\pm \gamma_5)\gamma_\mu \psi$ in QCD. One also typically considers a scalar field $X$ dual to the $\bar \psi \psi$ operator. Condensation of $X$ in the bulk implies chiral symmetry breaking in the bulk, and for this to happen spontaneously at zero quark mass, a nontrivial potential for $X$ should be added. Perhaps the simplest model which achieves this is the dynamical AdS/QCD model~\cite{Alvares:2012kr,Alho:2013dka} based on the D3-D7  setup, which was already mentioned above.

\subsubsection{Asymptotic freedom} 

As I pointed out above, the geometries in holographic models for QCD are usually asymptotically AdS$_5$ near the boundary, because QCD becomes (free and) conformal at high energies. This is however a potential issue because simple formulations of AdS/CFT only work at strong coupling. So when the coupling becomes weak at high energies, applying gau\-ge/gra\-vity duality (in the classical gravity approximation) becomes questionable. Usually this is not considered as a major problem because basic observables such as decay constants and spectrum in confining backgrounds are mostly determined by the IR part of the geometry, and the results in many of the models agree well with QCD. Moreover in many top-down models such as the WSS models the coupling does not run so that this issue does not really arise. There is however also an attempt (based on semi-holography~\cite{Faulkner:2010tq}) to combine a perturbative framework of UV physics to holographic models in the IR which has been discussed in the context of quark gluon plasma~\cite{Iancu:2014ava,Mukhopadhyay:2015smb,Banerjee:2017ozx,Ecker:2018ucc}. In this review I will use a less ambitious approach (IHQCD and V-QCD) where the near-boundary behavior of the geometry of the holographic model is tailored to agree with basic results on perturbative QCD. A somewhat similar approach is taken in the dynamic AdS/QCD model where one inputs the perturbative running of the quark mass.

%%%%%%%%%%%%%%%%%%%%%%%%%%%%%%%%%%%%%
\subsection{Phases of holographic QCD}
%%%%%%%%%%%%%%%%%%%%%%%%%%%%%%%%%%%%%

I then discuss the holographic realization of the phases of QCD at finite temperature and density. Starting from the structure at zero density, the basic idea is (as already pointed out above) that for nontrivial finite temperature configurations, one needs to consider (planar) black hole configurations. For confining backgrounds one typically obtains a Hawking-Page transition~\cite{Hawking:1982dh}, where at low temperature one has a geometry similar to that of the zero temperature vacuum, and at high temperatures in the quark-gluon plasma phase, the geometry is the black hole geometry. The critical temperature of the deconfinement transition is set by the confinement scale.

A nice geometric picture arises in the WSS model, where at low temperatures the geometry is that of~\eqref{eq:Wittengeom}, but at high temperature the roles of the compactified coordinate and time have been changed~\cite{Aharony:2006da}:
\begin{eqnarray} \label{eq:WSSFT}
 ds^2 &=& \left(\frac{u}{R}\right)^{3/2}\left(-f(u)dt^2+\delta_{ij}dx^i dx^j + d\theta^2\right)\\\nonumber
 & & + \left(\frac{u}{R}\right)^{-3/2}(f(u)^{-1}du^2+u^2d\Omega_4^2)\ ,\\
 f(u)&=& 1 - \frac{u_T^3}{u^3}\ ,
\end{eqnarray}
with the Hawking temperature $T = \frac{3}{4\pi}\frac{u_T^{1/2}}{R^{3/2}}$. The phase transition is found when $u_T=u_\Lambda$, so $T_c=M_{KK}/2\pi$. In the high temperature phase, the geometry for the compactified $\theta$ and holographic $u$ coordinates takes the form of a cylinder, so that the D8 and $\overline{\mathrm{D8}}$ branes are no longer connected and chiral symmetry is restored at the transition.\footnote{Notice however that the work of~\cite{Mandal:2011ws} suggests that the phase transition should proceed through the Gregory-Laflamme instability~\cite{Gregory:1993vy}.}

Going to finite baryon number chemical potential is in principle straightforward following the dictionary. 
That is, the temporal component of the (vectorial) gau\-ge field is dual to $\bar \psi \gamma_0 \psi = \psi^\dagger \psi$ (summed over flavors), to which the baryon number chemical potential couples in field theory, so it is enough to turn on a boundary value for the temporal component on the holographic side. At high densities one finds charged black hole solutions: the baryon number density arises from behind the horizon of the black hole. The field theory interpretation of such configurations is that the nonzero baryon number emerges from free quarks, i.e., these solutions are dual to the quark gluon plasma (or quark matter) phase. Baryon number can also arise from baryons, which are the only possible source in the confined phases, but their realization in holography is more involved. I will discuss it in Sec.~\ref{sec:baryons}. Notice however that in probe brane setups there is no backreaction of the baryon number to the geometry so that the charged geometry actually does not differ from the neutral geometry. For nontrivial charged black hole solutions, backreaction is required.

A simple explicit example of a (backreacted, deconfined) charged solution is obtained by coupling Einstein gravity to a quadratic action for the gauge field, i.e., Einstein-Maxwell gravity
\begin{eqnarray}
\mathcal{S}_\mathrm{EM} &=& \frac{1}{16\pi G_5}\int d^5x\sqrt{-\det{g}}\\\nonumber
&&\times \left[R +\frac{12}{\ell^2} - \frac{1}{4} F_{MN}F^{MN}\right] \ .
\end{eqnarray}
The charged (Reissner-Nordstr\"om) solution is of the form~\eqref{eq:adsFTmetric} but with the blackening factor modified to
\be
 f(r) = 1-\frac{r^4}{r_h^4} -2 \hat Q^2 \left(\frac{r^6}{r_h^6}-\frac{r^4}{r_h^4}\right)
\ee
where we set $d=4$ and the charge was normalized such that $f'(r_h) \to 0$ as $\hat Q \to 1$ so that the blackening function exhibits a double root and the black hole becomes extremal. Therefore we should take $0 \le \hat Q \le 1$. The geometry is supported by the gauge field
\be 
A_t(r) = \mu \left(1-\frac{r^2}{r_h^2}\right) \ , \qquad \mu = \frac{\sqrt{6}\ell}{r_h} \hat Q \ .
\ee
The rest of the thermodynamics is determined by the relations
\be
 T =\frac{1-\hat Q^2}{\pi r_h} \ , \quad s = \frac{1}{4G_5}\frac{\ell^3}{r_h^3} \ .
\ee
Notice that all positive values of $T$ and $\mu$ are covered for positive $r_h$ and $0<\hat Q<1$. 
Recall however that the Reissner-Nordstr\"om black hole is unlikely to be a realistic model for QCD as it corresponds to adding an ad-hoc five dimensional gauge field term to the background for an exactly conformal ($\mathcal{N}=4$ SYM) theory. Nevertheless, it illustrates the general properties of charged solutions.

Finite density phase diagrams have been studied in the literature in D3-D7 setups, both in the probe limit~\cite{Kobayashi:2006sb,Nakamura:2007nx,Karch:2007br,Ammon:2011hz}, and taking into account the backreaction~\cite{Bigazzi:2011it,Bigazzi:2013jqa} by using a method where flavor branes are smeared~\cite{Bigazzi:2005md}. The probe analysis shows a second order phase transition at zero temperature when the chemical potential equals the quark mass (given by the UV separation of the D3 and D7 branes) which may be used as a rough model for the deconfinement transition in QCD. Moreover, turning on background fields leads to interesting effects (see, e.g.,~\cite{Erdmenger:2007bn,Evans:2010iy,Evans:2011mu} and the review~\cite{Bergman:2012na}). Another interesting possibility is the spontaneous creation of inhomogeneous phases in the region of low temperatures and high densities, in the D3-D7 setup with backreacted smeared branes~\cite{Faedo:2017aoe}.

Also the WSS model at finite density has been studied extensively. For WSS the backreaction is even more tricky~\cite{Burrington:2007qd,Bigazzi:2014qsa,Li:2016gtz} because the background breaks all supersymmetries. However interesting phenomenology arises by considering probe brane setups where the D8 branes of Fig.~\ref{fig:WSSgeom} are not antipodal on the compactified circle but separated by a distance $L$ at the boundary which is taken as a free parameter~\cite{Aharony:2006da}. In this generalization the chiral and deconfinement transition need not take place at the same value, but a more complicated phase diagram arises. In the limit of small $L$~\cite{Horigome:2006xu,Li:2015uea} chirally broken phase is found only in the region where both $T$ and $\mu$ are small even at zero quark mass. Effects of finite quark mass can be analyzed~\cite{Kovensky:2019bih} by considering effects from the strings between the D8 and $\overline{\mathrm{D8}}$ branes, either through a ``tachyonic'' bifundamental scalar field~\cite{Bergman:2007pm,Dhar:2007bz,Dhar:2008um,Jokela:2009tk} or through an open Wilson line between the branes~\cite{Aharony:2008an,Hashimoto:2008sr}. The model is also known to contain instabilities towards inhomogeneous phases which have been studied in the WSS model~\cite{Ooguri:2010xs} (see also~\cite{Domokos:2007kt}).
 
Another interesting direction is turning on nonzero isospin chemical. This is relevant for neutron star applications as the matter inside neutron stars is isospin asymmetric. Isospin chemical potential, and consequent condensation of charged pions and $\rho$ mesons, has been considered both in the WSS model~\cite{Parnachev:2007bc,Aharony:2007uu,Kovensky:2021ddl}, in the D3-D7 model~\cite{Erdmenger:2007ap,Erdmenger:2008yj}, as well as in bottom-up models~\cite{Kim:2009yf,Albrecht:2010eg,Nishihara:2014nva,Cao:2020ryx}.

At low temperatures and high densities one expects quark pairing to take place in quark matter. Model computations suggest that the phase diagram in this regime contains various different paired phases, including color superconducting and/or color-flavor locked phases~\cite{Alford:2007xm}. Such ``exotic'' phases are challenging to describe with gau\-ge/gra\-vity duality, because they involve spontaneous breaking of the gauge group SU$(N_c)$. For standard holographic geometries, conservation of SU$(N_c)$ is manifest and the dictionary only contains operators that are singlets under the gauge group. In the language of the brane setups, breaking the gauge symmetry would mean pulling a significant fraction of the $N_c$ gauge branes apart from the overlapping stack of branes, which understandably leads to a complex geometry. Despite this fact there is a wide literature working toward the holographic realization of these phases.  Color superconductors have been analyzed in the probe  D3-D7 model at finite baryon~\cite{Chen:2009kx} and isospin densities~\cite{Faedo:2018fjw,Faedo:2019jlp}. Higgsing of the SU$(N_c)$ in both pure gauge (probe) top-down and bottom-up setups was analyzed in~\cite{Rozali:2012ry}. The recent work~\cite{Henriksson:2019zph} found color superconducting solutions with probe color D3 branes in AdS$_5\times S^5$ that rotate in the internal directions (see also~\cite{Henriksson:2021zei}).

Another approach is to use the holographic superconductor model of~\cite{Hartnoll:2008vx} where the breaking of the gauge group is modeled through breaking of a global symmetry. This has been studied both in the D3-D7 motivated setup~\cite{BitaghsirFadafan:2018iqr}, and by backreacting the condensing scalar field action of~\cite{Hartnoll:2008vx} in five dimensional Einstein gravity~\cite{Basu:2011yg}, in six dimensional gravity with the AdS soliton background~\cite{Ghoroku:2019trx}, as well as in Gauss-Bonnet gravity~\cite{Nam:2021qwv,Fadafan:2021xcw}.

There is a priori no reason why the chiral and deconfinement transitions should take place simultaneously, and it is known that in many strongly interacting theories they are separate (see, e.g.,~\cite{Evans:2020ztq}).
A possible scenario in QCD is that the deconfinement and chiral transitions are separate in the regime of high density.  Such a behavior may be related to the quarkyonic phase, which is confined but chirally symmetric and shares features of both nuclear and quark matter~\cite{McLerran:2007qj}. It is expected to be present at least in the limit of large $N_c$ in QCD. Separate chiral and deconfinement transitions are also found in NJL models (see~\cite{McLerran:2008ua}). It is also possible to generate a phase which is chirally symmetric and confined by using holography: this has been demonstrated by carefully chosen bottom-up model in~\cite{Chen:2019rez}.
But in holographic models it is actually easier to generate deconfined chirally broken phases. An example is the non-antipodal WSS model mentioned above. In a class of phenomenologically adjusted D3-D7 models there is a chirally broken finite density phase which appears at intermediate chemical potentials~\cite{Evans:2011eu,BitaghsirFadafan:2019ofb}. The V-QCD models, which will be discussed below, can also support a chirally broken deconfined phase, but this phase will be absent for the potentials we will be using in this review~\cite{Alho:2013hsa}.

%%%%%%%%%%%%%%%%%%%%%%%%%%%%%%%%%%%%%
%%%%%%%%%%%%%%%%%%%%%%%%%%%%%%%%%%%%%
%%%%%%%%%%%%%%%%%%%%%%%%%%%%%%%%%%%%%
%%%%%%%%%%%%%%%%%%%%%%%%%%%%%%%%%%%%%
%%%%%%%%%%%%%%%%%%%%%%%%%%%%%%%%%%%%%
%%%%%%%%%%%%%%%%%%%%%%%%%%%%%%%%%%%%%
%%%%%%%%%%%%%%%%%%%%%%%%%%%%%%%%%%%%%
%%%%%%%%%%%%%%%%%%%%%%%%%%%%%%%%%%%%%
%%%%%%%%%%%%%%%%%%%%%%%%%%%%%%%%%%%%%
%%%%%%%%%%%%%%%%%%%%%%%%%%%%%%%%%%%%%
%%%%%%%%%%%%%%%%%%%%%%%%%%%%%%%%%%%%%
%%%%%%%%%%%%%%%%%%%%%%%%%%%%%%%%%%%%%
%%%%%%%%%%%%%%%%%%%%%%%%%%%%%%%%%%%%%
%%%%%%%%%%%%%%%%%%%%%%%%%%%%%%%%%%%%%
%%%%%%%%%%%%%%%%%%%%%%%%%%%%%%%%%%%%%
%%%%%%%%%%%%%%%%%%%%%%%%%%%%%%%%%%%%%
%%%%%%%%%%%%%%%%%%%%%%%%%%%%%%%%%%%%%
%%%%%%%%%%%%%%%%%%%%%%%%%%%%%%%%%%%%%
%%%%%%%%%%%%%%%%%%%%%%%%%%%%%%%%%%%%%
%%%%%%%%%%%%%%%%%%%%%%%%%%%%%%%%%%%%%
%%%%%%%%%%%%%%%%%%%%%%%%%%%%%%%%%%%%%
%%%%%%%%%%%%%%%%%%%%%%%%%%%%%%%%%%%%%
%%%%%%%%%%%%%%%%%%%%%%%%%%%%%%%%%%%%%
\section{Nuclear matter in gau\-ge/gra\-vity duality} \label{sec:baryons}
%%%%%%%%%%%%%%%%%%%%%%%%%%%%%%%%%%%%%

Understanding the description of nuclear matter in gau\-ge/gra\-vity duality requires first understanding the description of its constituents, baryons. They are special in particular because their mass grows linearly with $N_c$ and becomes infinite in the limit of large $N_c$, where gau\-ge/gra\-vity duality works. Therefore the behavior of baryons and nuclear matter depends more strongly on $N_c$, which leads to complications when applying the holographic results to real-world matter, as I shall discuss below. But in order to motivate the holographic baryons, I will start by discussing the descriptions of baryons in large $N_c$ QCD by using effective field theory.

\subsection{QCD and baryons at large $N_c$}

As is well known, the low energy physics of QCD is well described by the effective theory of the states having the lowest masses, i.e., chiral perturbation theory of pions. Pions are the Goldstone bosons of the spontaneously broken chiral symmetry so they map to the generators of the (broken) axial SU$(N_f)$ symmetry. That is, they are in the adjoint of SU$(N_f)$. 
If one turns on light quark masses which break the axial SU$(N_f)$ explicitly, the pions become ``pseudo'' Goldstone bosons, i.e., their masses are nonzero but they are still anomalously light compared to the other mesons.
The axial U$(1)$ symmetry is broken by the axial ``triangle'' anomaly and the state $\eta'$ mapping to its generator is in general not a Goldstone boson. In the large $N_c$ limit however the axial anomaly is suppressed, the flavor symmetry is enhanced from SU to U, and there is an additional Goldstone boson~\cite{Veneziano:1979ec}. 

The leading order chiral action can be written as (assuming flavor independent quark masses)
\begin{eqnarray} \label{eq:chiralaction}
 \mathcal{L}_\pi &=& \frac{f_\pi^2}{4}\mathrm{tr}\left(L_\mu L^\mu\right) +\frac{f_\pi^2m_\pi^2}{2}\mathrm{tr}\left(U+U^\dagger\right) \ , \\\nonumber
 U &=& e^{i\pi_a\tau_a/f_\pi}\ , \qquad L_\mu = U^{\dagger}\partial_\mu U \ .
\end{eqnarray}
where $m_\pi$ is the pion mass, $f_\pi$ is the pion decay constant, and $\tau_a$ are the generators of U$(N_f)$ so that the $\eta'$ is included in the pions. The Lagrangian can be systematically extended to include higher order terms in momenta and pion masses.

The fact that the baryon becomes infinitely massive in the large $N_c$ limit suggests that it can be described as a soliton of the pion fields. I will here discuss the standard picture which requires $N_f>1$. See~\cite{Komargodski:2018odf} for the realization at $N_f=1$. The baryon number should be conserved, and indeed one can find a conserved current
\be \label{eq:JBSk}
 J_\mu^B = \frac{1}{24\pi^2}\epsilon_{\mu\nu\rho\sigma}\mathrm{tr}\left(L^\nu L^\rho L^\sigma\right) \ .
\ee
The solitons are topologically protected: they carry a nontrivial winding number under the third homotopy group $\pi_3(\mathrm{SU}(N_f))=\mathbb{Z}$, which is identified with the baryon number defined through the temporal component of the current~\eqref{eq:JBSk}.

However it is immediate that~\eqref{eq:chiralaction} does not have such solitonic solutions. Applying a simple scaling argument shows that the energy is minimized when the size of the soliton goes to zero. The situation is however changed if one adds derivative corrections to the Lagrangian including the last term in
\begin{eqnarray} \label{eq:Lskyrme}
 \mathcal{L}_\mathrm{Skyrme} &=& \frac{f_\pi^2}{4}\mathrm{tr}\left(L_\mu L^\mu\right) +\frac{f_\pi^2m_\pi^2}{2}\mathrm{tr}\left(U+U^\dagger\right) \\ \nonumber
 &&+\frac{1}{32e^2}\mathrm{tr}[L_\mu,L_\nu][L^\mu,L^\nu] \ .
\end{eqnarray}
Now the action has a nontrivial (charge one) solution, the skyrmion~\cite{Skyrme:1961vq}, which is identified with the baryons. At large $N_c$, we have $f_\pi^2 \sim e^{-2} \sim N_c$, and consequently the size of the skyrmion is independent of $N_c$ and determined by the QCD scale as $1/\Lambda_\mathrm{QCD}$, and the energy is $\sim N_c$, matching with the expected behavior for baryons. Notice that because the size is $N_c$ independent, one should in principle consider derivative corrections to all orders. However, including only the leading nontrivial correction gives a phenomenologically successful description~\cite{Adkins:1983ya,Zahed:1986qz}. 

\subsection{Holographic description of a single baryon}

It was understood early on how baryons can be included in gau\-ge/gra\-vity duality, and the description turns out to be closely related to the Skyrmion picture. First recall that a baryon in QCD with the SU$(N_c)$ gauge group is a color singlet state composed of $N_c$ quarks, and that quarks (in the fundamental representation) are obtained from strings stretching between the flavor and gauge branes. The holographic dual to a single baryon is then obtained through a baryon vertex in the bulk, which is an object where $N_c$ fundamental strings can end. In the dual geometry AdS$_5 \times S^5$ of $\mathcal{N}=4$ SYM, the baryon vertex can be identified with a D5 brane wrapping the $S^5$ part of the geometry~\cite{Witten:1998xy,Gross:1998gk}. It is therefore localized in the spatial and holographic directions (for a baryon at rest), and extended along the time direction as well as along the angular coordinates of the $S^5$. Notice that inclusion of flavors is necessary in order to have physical dynamical baryons: without flavors branes the strings emerging from the baryon vertex can only end at the boundary, and are long, so this configuration is dual to a baryon made of external heavy quarks. When flavor branes are added, it is possible to create dynamical baryon configurations with short strings ending on the brane.

The realization of baryons has been studied extensively in the WSS model, in which case the baryon vertex is obtained by wrapping a D4 brane over the $S^4$ of the geometry~\cite{Sakai:2004cn,Kim:2006gp,Hata:2007mb,Kim:2007zm}. The configuration for a single baryon then consist of the vertex and $N_c$ strings starting from the vertex and ending on the flavor branes. The tension of the strings pulls the vertex on the flavor branes so that the D4 brane is dissolved in the flavor branes. Such dissolved branes are, in this case, described through solitonic configurations of the non-Abelian gauge-fields living on the D8 branes. The energy of the soliton is minimized when the baryon vertex lies at the tip of the cigar, as sketched in Fig.~\ref{fig:WSSgeom}. 

As it turns out, in the limit of large 't Hooft coupling $\lambda$ the WSS solitonic baryons are small. Assuming that the soliton is located exactly at the tip, it is then simply described in terms of a five dimensional Yang-Mills action in flat space plus a Chern-Simons action
\begin{eqnarray} \label{eq:5DYM}
S &=& -\frac{\lambda M_{KK}N_c}{432\pi^3} \int d^5 x\, \mathrm{tr}\left( F_{MN}F^{MN}\right) \\\nonumber
&&+\frac{iN_c}{24\pi^2} \int \mathrm{tr}\left(AF^2-\frac{i}{2}A^3F-\frac{1}{10}A^5\right) \ .
\end{eqnarray}
To arrive at this five dimensional action we chose a holographic coordinate for the D8 embeddings which is smooth at the tip and rescaled so that the metric at the tip is flat. The field strengths $F_{MN}$ are small for the soliton which allowed as to replace the DBI action by the first nontrivial term in the expansion, i.e., $F^2$, giving the first term in~\eqref{eq:5DYM}.

It is the Chern-Simons action which prevents the solitons from collapsing in this case, in analogy to the Skyrme term in~\eqref{eq:Lskyrme}, and balance between the terms sets the size to be $\sim 1/\sqrt{\lambda}$. More precisely, the contribution from the non-Abelian Maxwell term balances with the ``Coulomb interaction'' contribution arising from the Chern-Simons term\footnote{Actually it is the curvature corrections to the DBI term, which we have omitted in~\eqref{eq:5DYM}, that balance against the Coulomb contribution. These corrections are irrelevant for functional form of the leading order soliton solution and only affect its size. The size is determined by contributions suppressed by $1/\l$ to the energy arising both from the CS term and from the curvature corrections to the DBI term. See~\cite{Hata:2007mb} for details.}
\be
 \int \widehat A_t\, dt \wedge \mathrm{tr}\left(F\wedge F\right) \ ,
\ee
where $\widehat A$ is the Abelian part of the gauge field. Notice that the baryons therefore become point-like in the strong coupling limit.
The soliton configurations are again topologically protected and carry the same winding number of $\pi_3(\mathrm{SU}(N_f))$ as the Skyrmions. The baryon number is
\be \label{eq:NB}
 N_B = \frac{1}{32\pi^2}\int d^3 xdz \,\epsilon_{mnpq}\mathrm{tr}\left(F^{mn}F^{pq}\right) \ ,
\ee
where the integration is over the spatial coordinates and the holographic coordinate $z$ which is smooth over the tip, and the indices are summed over the same four directions, i.e., excluding the time direction.
One can indeed show that this is the same winding number as the baryon number arising from~\eqref{eq:JBSk} in the Skyrme picture~\cite{Son:2003et}. Moreover, the pion effective Lagrangian derived from the WSS model matches with the Skyrme Lagrangian~\eqref{eq:Lskyrme} and gives a prediction for the value of the coupling $e$.

In the special case of $N_f=2$ the classical soliton solution can be found explicitly. It matches exactly with the  Belavin-Polyakov-Schwarz-Tyupkinin (BPST) instanton of four dimensional Yang Mills theory~\cite{Belavin:1975fg} except that the time coordinate is replaced by $z$ so that the solution is a soliton (localized in the holographic coordinate) rather than an instanton (localized in time). The solution can be written as
\begin{eqnarray} \label{eq:BPST}
 A_m(x) &=& -i \frac{\xi(x)^2}{\xi(x)^2+\rho^2}\, g(x) \partial_m g(x)^{-1} \ , \\
 g(x) &=& \frac{(z-Z) -i (\vec x -\vec X)\cdot \vec \sigma}{\xi(x)}  \ , \\
 \xi(x) &=& \sqrt{(\vec x -\vec X)^2+(z-Z)^2}
\end{eqnarray}
where $\sigma$ denotes the Pauli matrices, the constant $\rho \propto 1/\sqrt{\lambda}$ gives the size of the soliton, small coordinates give the coordinate dependence of the soliton, and the capital coordinates denote the location of the center of the soliton. 

Quantization of the soliton can be carried out using the moduli space approximation method (see~\cite{Hata:2007mb}), where one considers slow variation of time of the parameters of the moduli space and obtains the Hamiltonian from the variation of the soliton energy. The moduli space is a product of the Minkowski space, parameterized by the location of the soliton, and the orientation space which corresponds to the spin and isospin degrees of freedom as well as the size of the soliton. For $N_f=2$, these latter degrees of freedom, i.e., the variation of $\rho$ and the SU$(2)$ gauge transformations of the soliton, form the space $\mathbb{R}^4/\mathbb{Z}_2$. The rotations in the four dimensional space include the spin and isospin rotations: SO$(4) \simeq (\mathrm{SU}(2)_I\times \mathrm{SU}(2)_J)/\mathbb{Z}_2$.
The Hamiltonian is then quantized by using standard rules of quantum mechanics. In the case of the BPST soliton, the Schr\"odinger equation for the wave function may be solved analytically by using separation of variables.

The properties of the Sakai-Sugimoto soliton have been studied extensively in the literature. An effective 5D theory for the solitons was derived in~\cite{Hong:2007kx}, shown to lead to vector meson dominance~\cite{Hong:2007ay}, and used to analyze form factors~\cite{Hong:2007dq}. Form factors, among other things, were also analyzed directly by using the soliton solution~\cite{Hashimoto:2008zw,Kim:2008pw}. The long-distance properties were analyzed in~\cite{Cherman:2009gb,Cherman:2011ve} and the solution in all regimes, including the complete numerical solution, was analyzed in~\cite{Bolognesi:2013nja}.  Deformed generalizations were studied in~\cite{Rozali:2013fna}.

While the above discussion was specific to the WSS model, the construction works in a similar way in other models. In particular, baryonic solitons have been studied in bottom-up models, where one introduces explicitly five dimensional Maxwell actions separately for left and right handed gauge fields, corresponding roughly to the gauge fields on the D8 and $\overline{\mathrm{D8}}$ branes of the WSS model, respectively. In a hard wall setup~\cite{Pomarol:2007kr,Pomarol:2008aa}, the action is
\begin{eqnarray} \label{eq:PomarolYM}
 S_\mathrm{hw} &=& -\frac{M}{2}\! \int\! d^5x \sqrt{-\det g}\\\nonumber
 &&\times \mathrm{tr}\left[F_{L\,MN}F_L^{MN}+F_{R\,MN}F_R^{MN}\right] \\
  \label{eq:PomarolCS}
 &&+\frac{iN_c}{24\pi^2}\int \mathrm{tr}\bigg[\left(A_LF_L^2-\frac{i}{2}A_L^3F_L-\frac{1}{10}A_L^5\right) \\\nonumber
&&\ \ \  - (L\leftrightarrow R)\bigg] \ ,
\end{eqnarray}
where the metric is AdS$_5$ and one also introduces UV and IR cutoffs for the holographic coordinate which we have not included explicitly. The soliton is found through an Ansatz which respects parity, so that the left and right handed gauge fields are simply related. In bottom-up frameworks the size of the soliton is not parametrically small. Consequently one needs to take into account the variation of the metric over the soliton, and the solution can only be found numerically. The properties of the soliton, such as the electric and magnetic radii (defined through the coupling of electric and magnetic currents to the nucleon, respectively), agree well with experimental data~\cite{Pomarol:2008aa,Pomarol:2009hp}. Properties of the soliton at long distance were analyzed in~\cite{Cherman:2009gb} and compared to other solutions.  Notice that the action of~\eqref{eq:PomarolYM} and~\eqref{eq:PomarolCS} does not contain a scalar degree of freedom (dual to $\bar \psi \psi$) even though chiral symmetry is broken in the nuclear matter phase (at least at not too high densities). Coupling of the soliton to such a scalar field was studied in~\cite{Domenech:2010aq,Gorsky:2012eg,Gorsky:2013dda,Gorsky:2015pra}.

Apart from soliton solutions, fermionic fluctuations at brane intersections~\cite{Kirsch:2006he} may effectively show some properties of baryons. A class of fermionic meson states was identified in~\cite{Abt:2019tas} that composed of three elementary fermionic fields, and it was shown in~\cite{Nakas:2020hyo} that the masses of these state have the same scaling with number of colors, $M \sim N_c$, as baryons in QCD. Another approach describes baryons as light objects in an alternative large $N_c$ limit, so that baryons and mesons are treated in the same footing, which is made possible in the bulk by considering orientifolds in addition to brane intersections~\cite{Hoyos-Badajoz:2009zmh,Hoyos:2016ahj}.

\subsection{Nuclear matter from holographic baryons}

Constructing holographic nuclear matter properly requires considering configurations of several solitons, including all their interactions. Rather obviously, this is technically extremely challenging. Much of the physical picture can however be figured out without solving the configurations explicitly.

Holographic baryons are heavy with their masses being $\mathcal{O}(N_c)$, so they behave non-relativistically. Their average momentum can be estimated from the (inverse of the) diameter of the potential well where the baryon lies in a dense configuration, and it is independent of $N_c$ which is also plausible as the size of the baryons is independent of $N_c$. Therefore their kinetic energy is $\sim 1/N_c$ while the potential energy (e.g. from meson exchange) is known to be $\mathcal{O}(N_c)$. The suppression of the kinetic energy means, together with the fact that the interactions between the solitons are repulsive, that nuclear matter at large $N_c$ is a crystal, i.e., different from the $N_c=3$ phase which is a liquid. The location of the liquid to solid phase transition can be estimated to lie at around $N_c = 8$ by using analogue to condensed matter~\cite{Kaplunovsky:2010eh}.

At low densities the crystal consist of a layer of instantons in the holographic direction: the location of each soliton is minimized independently so for the (antipodal) WSS model, for example, they are found at the tip of the cigar as shown in Fig.~\ref{fig:WSSgeom}. However, when the density becomes larger, the solitons are expected start to populate the holographic direction. Details are model dependent, but estimates suggest that there are ``popcorn'' transitions with increasing density where the configuration changes from a single layer to a multi layer crystal or more complicated four dimensional crystal~\cite{Kaplunovsky:2012gb,Kaplunovsky:2013iza}. Crystals in a setup motivated by the single layer configuration in the WSS model were constructed recently in~\cite{Jarvinen:2020xjh}. It is possible that the solitons break into dyonic half-solitons (or half-instantons) as the density increases~\cite{Rho:2009ym}.

A basic property characterizing the system is the two-body potential between the solitons, i.e., the holographic nuclear force. In the WSS model, the behavior of the potential reflects the presence of two scales, the confinement scale $M_{KK}$ and the size of the soliton $1/(\sqrt{\lambda}M_{KK})$. For short distances, $x\ll 1/(\sqrt{\lambda}M_{KK})$ the two solitons overlap which creates a repulsive interaction. For intermediate distances, $1/(\sqrt{\lambda}M_{KK}) \ll x \ll 1/M_{KK}$, the solitons can be treated as pointlike but curvature effects of the geometry can still be neglected so that the solitons are effectively in flat space. In this range the interaction potential can be solved analytically by analyzing the asymptotics of the Atiyah-Drinfeld-Hitchin-Manin (ADHM) construction~\cite{Atiyah:1978ri} for two solitons~\cite{Kim:2008iy,Kaplunovsky:2013iza}. The solution is a repulsive $1/r^2$ potential, with the strength depending on the relative orientation difference between the solitons.
For long range interactions, $x \gg 1/M_{KK}$, curvature effects are important. In this range the potential is obtained as a sum of Yukawa potentials from the exchange of mesons and is found to be repulsive~\cite{Kaplunovsky:2010eh}.

The analysis of the potential therefore gives rise to another difficulty in view of applications to real nuclear matter: the WSS nucleon-nucleon potential is always repulsive, unlike the nucleon nucleon potential of regular $N_c=3$ QCD. This is perhaps unsurprising because the nuclear binding energy is much smaller than the nucleon masses, which suggests that the potential involves a delicate cancellation between attractive and repulsive forces that is easily disturbed by approximations.  Consequently, the transition from vacuum to nuclear matter in the WSS model is apparently of second order which means that there is no nuclear saturation density (or it is zero) and no nuclei. An attractive component of the potential can be obtained by allowing for non-antipodal embeddings of the D8 branes~\cite{Kaplunovsky:2010eh}. For long range interactions, such an attractive component is due to isoscalar scalar exchange while the repulsive force arises from isoscalar vector exchange. However, the vector always dominates even in the non-antipodal WSS case, and the total force is repulsive. But we learn that a possible key to attractive forces are light scalars, which could help the scalar exchange to become the dominant channel. And indeed attractive potential was found by considering a probe D7 brane setup in the Klebanov-Strassler geometry which contains a parametrically light scalar meson~\cite{Dymarsky:2010ci}. The masses of the mesons can be tuned such that an almost cancellation between the attractive and repulsive forces takes place, leading to small nuclear binding energy, similarly to regular QCD. However, while the light scalar of this setup therefore works as a model for the light $\sigma$ meson of QCD, other properties are perhaps not as close to QCD as in WSS.

The difficulties described above indicate that it is challenging to learn much from real world  nuclear matter at low densities by using holography. But this is not really an issue since, as I pointed out in the introduction, the region with densities up to around the nuclear saturation density can be analyzed by using traditional methods such as effective theory, which are also supported by experimental data from heavy elements and scattering of nuclei. The situation is different for densities well above the saturation density, where reliable predictions cannot be obtained by using such methods. This motivates us to consider the phase diagram and basic observables such as the EOS by using different approximations in the WSS model. And actually, as we shall discuss below, quite a bit of progress has been made in the WSS model even at low densities.

Early attempts to study the phase diagram with nuclear matter ignored interactions and used simply pointlike solitons, obtained by taking the strict $\lambda \to \infty$ limit, in the antipodal and non-antipodal versions of the WSS model~\cite{Bergman:2007wp,Rozali:2007rx}. The crystal of nuclear matter was considered in the Wigner-Seitz approximation in~\cite{Kim:2007vd}. 
Effects of finite widths for the antipodal embeddings were considered in~\cite{Ghoroku:2012am}.

As I pointed out above, the non-antipodal embedding at small separation $L$ leads to a phase diagram where chiral symmetry is restored at high densities (and low temperatures)~\cite{Horigome:2006xu,Li:2015uea}. This configuration is interpreted as adding a four-fermion operator in the field theory, similarly to the NJL model~\cite{Antonyan:2006vw}. Finite width effects in this limit were considered in~\cite{Li:2015uea,Preis:2016fsp}. Point-like solitons lead to a phase diagram where nuclear matter is present up to arbitrary high densities, but after including finite width effects, the nuclear to quark matter transition takes place at finite density. Two-body interactions from exact instanton solution were considered in this setup in~\cite{BitaghsirFadafan:2018uzs}. Desired low-density features, such as correct nuclear saturation density, could be obtained, and the construction was seen to be suggestive of quark-hadron continuity, i.e., absence of (first order) phase transition between the nuclear and quark matter phases~\cite{Schafer:1998ef}. The effect of turning on a quark mass was analyzed in~\cite{Kovensky:2019bih}.

Another direction is the the analysis of the quarkyonic phase~\cite{McLerran:2007qj} in holography, which has been studied in the WSS model in~\cite{deBoer:2012ij,Kovensky:2020xif}.

Apart from the WSS model, nuclear matter in terms of noninteracting soliton liquid has been considered in the D3-D7 model~\cite{Gwak:2012ht,Evans:2012cx}, with somewhat similar results as in the WSS model.

\subsection{Nuclear matter from homogeneous bulk fields}\label{ssec:homog}

Another approach to dense nuclear matter would be to abandon the picture of individual solitons, and try to model the phase as a homogeneous configuration. 

Restricting to $N_f =2$, and to the WSS model, we can consider the Ansatz~\cite{Rozali:2007rx}
\be \label{eq:homogAn}
 A_i(z) = h(z) \sigma_i
\ee
where $i$ is the (spatial) Lorentz index, $z$ is the holographic coordinate, and the Pauli matrices $\sigma_i$ give rise to nontrivial flavor structure. The non-Abelian terms (i.e., terms spanned by the Pauli matrices) of the temporal components as well as $A_z$ are set to zero. It is clear that~\eqref{eq:homogAn} is the only simple consistent non-Abelian Ansatz: for a homogeneous configuration, there is no other vector that $\sigma_i$ could be contracted with. Notice that also that the behavior under rotations is trivial, as the rotations of the $\sigma_i$ can be removed by SU$(2)$ transformations of the gauge field. These considerations suggest that~\eqref{eq:homogAn} is the correct expression for nuclear matter in the limit of infinite density (if such a limit exists)~\cite{Rozali:2007rx}. 

Another motivation for the Ansatz can be obtained by smearing the BPST soliton, given in~\eqref{eq:BPST}. That is, we want to integrate the solution over all $\vec X$, and study the resulting configuration. However it is immediate that the integral does not converge, because the soliton decays quite slowly at large $|\vec X|$. The tail of the soliton is however pure gauge. The expression of~\eqref{eq:BPST} is given in regular Landau gauge, where the solution is free of singularities, but the slowly decaying tail is present. The tail is actually important since it carries information on the nontrivial winding number of the solution. It can be gauged away only with the cost of creating a singularity somewhere in the solution. Another commonly used expression for the soliton is that in singular Landau gauge:
\be
 A_m^\mathrm{sing}(x) = -i \frac{\rho^2}{\xi(x)^2+\rho^2}\, g(x)^{-1} \partial_m g(x) \ .
\ee
Notice that here the numerator of the first fraction we now have $\rho^2$ instead of $\xi^2$ so that the singularity at $\xi \to 0$, arising form the derivative, is no longer canceled. The long distance tail is gone, but at the price of generating this singularity. In this form of the solution, the winding number arises from it.

For the singular gauge expression, it is straightforward to compute the smeared soliton:
\begin{eqnarray} \label{eq:smearedA}
\int d^3X\, A_i^\mathrm{sing} &=& -\frac{2\pi^2\rho^2(z-Z)}{|z-Z|+\sqrt{(z-Z)^2+\rho^2}}\sigma_i \ ,\\  
\int d^3X\, A_z^\mathrm{sing} &=& 0  \ ,
\end{eqnarray}
which is, as expected, of the form~\eqref{eq:homogAn}. This is of course a handwaiving argument, since the soliton is a solution to nonlinear equations, so smearing it, which amounts to adding (linearly) a large amount of solitons, does not strictly make sense. 

The Ansatz~\eqref{eq:homogAn} leads to an issue when we try to compute the baryon number density. Namely, using~\eqref{eq:NB} we find that
\be
 N_B \propto \int dz\, \partial_z\left[ h(z)^3\right] \ .
\ee
The path of integration goes along the D8 brane embedding in Fig.~\ref{fig:WSSgeom}, and only gives terms at the UV boundary. But $h(z)$ must vanish at the boundaries, otherwise the dictionary tells us that we are turning on a peculiar source for the non-Abelian currents $\bar\psi\sigma_i(1\pm\gamma_5)\gamma_i\psi$, where $\sigma_i$ acts on the SU$(2)$ flavor indices. Therefore it seems that the baryon number must vanish. 

There is however a way to include a nonzero winding number: $h(z)$ may have a discontinuity in the bulk, which gives rise to the winding number. This may sound like a completely ad-hoc approach, but it is actually well motivated. The Ansatz should arise as a high density limit of the nuclear matter configuration, where the density of the solitons is so high that they largely overlap. The winding number of these solitons could arise from the asymptotic pure gauge behavior of the field but such pure gauge asymptotics is not allowed by the Ansatz~\eqref{eq:homogAn}, so it might be better to consider the dense configuration in a gauge (analogous to the singular gauge of a single instanton) where each soliton center has a singularity giving rise to the winding number. Indeed, I demonstrated above explicitly that the homogeneous Ansatz is similar to the smeared BPST instanton in the singular gauge. The idea is then that the discontinuity of $h(z)$ effectively arises from smearing the singularities of the dense soliton configuration. We also note that the expression in~\eqref{eq:smearedA} behaves at small $\rho$ as 
\be
 \int d^3X\, A_i^\mathrm{sing} \sim - \pi^2\rho^2 \mathrm{sgn}(z-Z)
\ee
so there is indeed a discontinuity, even though it is suppressed by $\rho^2$.
These motivations also tell us where we should place the discontinuity: at the tip of the cigar, where the solitons are expected to lie (at least at smaller densities). This is natural also because due to parity the value of $h(z)$ should be opposite on the two ``branches'' of the D8 brane.

The homogeneous approach has been studied in the WSS model, originally for the antipodal case~\cite{Rozali:2007rx} and later for the non-antipodal configuration~\cite{Li:2015uea} as well as in the presence of an isospin chemical potential~\cite{Kovensky:2021ddl}. The non-antipodal case shows an interesting phase diagram, where baryonic matter is created through a first order phase transition with increasing chemical potential, for large enough values of the 't Hooft coupling. For small values of the coupling, baryons do not appear but the phase at high chemical potentials is the quark matter phase.  Notice also that the Wigner-Seitz approximation for the instanton crystal considered in~\cite{Kim:2007vd} can be used to analyze nuclear matter at high density. This approximation captures a nontrivial interplay between the chiral condensate and the crystal. Interestingly, it produces the nonrelativistic scaling for the energy density as a function of the baryon number density, $\epsilon \sim n_B^{5/3}$, at large densities.

Homogeneous nuclear matter has also been considered by using a different approach where one first allows a spatial dependence in the Ansatz, so that the solution has a pure gauge tail at the boundary, and later (after imposing the zero curvature condition) averaging the action over the spatial directions~\cite{Elliot-Ripley:2016uwb}. This way one avoids the need for the discontinuity, but there is no clear action principle for the final solution.

I will discuss the homogeneous approach for the V-QCD model in Sec.~\ref{sec:VQCDNM}.

%%%%%%%%%%%%%%%%%%%%%%%%%%%%%%%%%%%%%
%%%%%%%%%%%%%%%%%%%%%%%%%%%%%%%%%%%%%
%%%%%%%%%%%%%%%%%%%%%%%%%%%%%%%%%%%%%
%%%%%%%%%%%%%%%%%%%%%%%%%%%%%%%%%%%%%
%%%%%%%%%%%%%%%%%%%%%%%%%%%%%%%%%%%%%
%%%%%%%%%%%%%%%%%%%%%%%%%%%%%%%%%%%%%
%%%%%%%%%%%%%%%%%%%%%%%%%%%%%%%%%%%%%
%%%%%%%%%%%%%%%%%%%%%%%%%%%%%%%%%%%%%
%%%%%%%%%%%%%%%%%%%%%%%%%%%%%%%%%%%%%
%%%%%%%%%%%%%%%%%%%%%%%%%%%%%%%%%%%%%
%%%%%%%%%%%%%%%%%%%%%%%%%%%%%%%%%%%%%
%%%%%%%%%%%%%%%%%%%%%%%%%%%%%%%%%%%%%
%%%%%%%%%%%%%%%%%%%%%%%%%%%%%%%%%%%%%
%%%%%%%%%%%%%%%%%%%%%%%%%%%%%%%%%%%%%
%%%%%%%%%%%%%%%%%%%%%%%%%%%%%%%%%%%%%
%%%%%%%%%%%%%%%%%%%%%%%%%%%%%%%%%%%%%
%%%%%%%%%%%%%%%%%%%%%%%%%%%%%%%%%%%%%
%%%%%%%%%%%%%%%%%%%%%%%%%%%%%%%%%%%%%
%%%%%%%%%%%%%%%%%%%%%%%%%%%%%%%%%%%%%
%%%%%%%%%%%%%%%%%%%%%%%%%%%%%%%%%%%%%
%%%%%%%%%%%%%%%%%%%%%%%%%%%%%%%%%%%%%
%%%%%%%%%%%%%%%%%%%%%%%%%%%%%%%%%%%%%
%%%%%%%%%%%%%%%%%%%%%%%%%%%%%%%%%%%%%
%%%%%%%%%%%%%%%%%%%%%%%%%%%%%%%%%%%%%
%%%%%%%%%%%%%%%%%%%%%%%%%%%%%%%%%%%%%
%%%%%%%%%%%%%%%%%%%%%%%%%%%%%%%%%%%%%
%%%%%%%%%%%%%%%%%%%%%%%%%%%%%%%%%%%%%
%%%%%%%%%%%%%%%%%%%%%%%%%%%%%%%%%%%%%
%%%%%%%%%%%%%%%%%%%%%%%%%%%%%%%%%%%%%
%%%%%%%%%%%%%%%%%%%%%%%%%%%%%%%%%%%%%
%%%%%%%%%%%%%%%%%%%%%%%%%%%%%%%%%%%%%
%%%%%%%%%%%%%%%%%%%%%%%%%%%%%%%%%%%%%
\section{Holographic models for QCD in the Veneziano limit}\label{sec:vqcd}
%%%%%%%%%%%%%%%%%%%%%%%%%%%%%%%%%%%%%

In this section, I will review the V-QCD model~\cite{Jarvinen:2011qe}. It is a relatively complex bottom-up holographic model for QCD which contains both gluon and flavor sectors, with full backreaction between the two sectors. The backreaction arises naturally in the Veneziano limit~\cite{Veneziano:1976wm,Veneziano:1979ec}:
\be
 N_c \to \infty \ , \quad N_f \to \infty\ , \quad \mathrm{with}\ x_f \equiv \frac{N_f}{N_c}\ \mathrm{fixed.}
\ee
Here $N_c$ is the number of colors and $N_f$ is the number of flavors. Notice that the ``V'' in the name of the model refers to the use of this limit.

The approach follows that of improved holographic QCD (IHQCD)~\cite{Gursoy:2007cb,Gursoy:2007er,Gursoy:2010fj}, but extends it to full flavored QCD by including backreacted branes.
The action of the model is inspired by string theory, but perhaps more importantly it contains various potentials which are determined by comparing to QCD data, in rough analogy to effective field theory. Indeed, the aim of the V-QCD model is to provide a general framework which allows modeling the physics of QCD as closely as possible with holography. Apart from qualitative features such as confinement and chiral symmetry breaking, the current version of the model already describes to a good precision, among other things, vacuum properties of QCD such as hadron spectra, as well as physics at finite temperature and density, in particular the equation of state at finite $T$ and $\mu$. Good description of the thermodynamics means a major improvement over simpler bottom-up models~\cite{Gursoy:2009zza}. I will focus on results at finite density in this review.

I start analyzing the holographic V-QCD model by discussing separately the different sectors of the model. The main building blocks of the V-QCD model are
\begin{enumerate}
 \item The model for the glue sector: Improved holographic QCD~\cite{Gursoy:2007cb,Gursoy:2007er}.
 \item The model for the flavor sector: tachyonic Dirac-Born-Infeld (DBI) action~\cite{Bigazzi:2005md,Casero:2007ae}.
\end{enumerate}
Let me then go through the structure of both these building blocks in detail. 
I present all terms relevant for the computations in the rest of the review for completeness. A reader not interested in the details may wish to jump to section~\ref{sec:VQCDNM}.

\begin{table}
\caption{The dictionary of V-QCD. Here $\eta_{\mu\nu}$ is the metric of field theory, $\theta$ is the Yang-Mills $\theta$-angle, $M_q$ is the (possible complex) quark mass matrix, and the external gauge fields contain chemical potentials and generalizations of electric and magnetic fields. See Appendix~A for the precise coupling between QCD and V-QCD.} \label{tab:dictionary}
\begin{center}
\begin{tabular}{|c|c|c|}
\hline
  field & operator  & source \\
\hline
\hline
 $\l=e^\phi$ & $\mathbb{T}r\,G^2$ & $\l_\mathrm{'tH}=g^2N_c$ \\
\hline
 $g_{MN}$ & $T_{\mu\nu}$ &  $\eta_{\mu\nu}$ \\
\hline 
 $a$ & $\mathbb{T}r\,G \wedge G$ &  $\theta$ \\
\hline 
\hline 
  $T^{ij}$ & $\bar \psi^i \psi^j$ & $M_q^{ij}$ \\
\hline
  $A_L^{ij}$ & $\bar \psi^i(1+\gamma_5)\gamma_\mu \psi^j$ & $A^{(\mathrm{ext})ij}_{L\,\mu}$ \\
\hline
  $A_R^{ij}$ & $\bar \psi^i(1-\gamma_5)\gamma_\mu \psi^j$ & $A^{(\mathrm{ext})ij}_{R\,\mu}$ \\
\hline

\end{tabular}
\end{center}
\end{table}

%%%%%%%%%%%%%%%%%%%%%%%%%%%%%%%%%%%%%
\subsection{Improved holographic QCD} 
%%%%%%%%%%%%%%%%%%%%%%%%%%%%%%%%%%%%%

Improved holographic QCD (IHQCD) is a model for Yang-Mills theory, inspired by five-dimensional noncritical string theory~\cite{Gursoy:2007cb,Gursoy:2007er} (see also~\cite{Gubser:2008ny}).  The action of the model is given in terms of dilaton-axion gravity in the Einstein frame:
\be
 S_\mathrm{IHCQD} = S_g +S_\mathrm{GH}+ S_a
\ee
where 
\be
 S_g = M_\mathrm{p}^3 N_c^2 \!\int\!\! d^5x\sqrt{\!-\!\det g} \left[R - \frac{4}{3}\left(\partial_\mu \phi\right)^2 + V_g(\phi) \right]
\ee
is the piece governing the physics of the gluon sector,
\be
 S_\mathrm{GH} = 2 M_\mathrm{p}^3 N_c^2 \int d^4x\,\sqrt{-\det h}\, K 
\ee
is the corresponding Gibbons-Hawking term defined on the four dimensional UV boundary,
and 
\be \label{eq:Saglue}
 S_a = -\frac{M_\mathrm{p}^3}{2} \int d^5x\, \sqrt{-\det g}\, Z(\phi)\, (\partial_\mu a)^2
\ee
is the CP-odd piece governing the physics of the $\theta$-angle. Here $M_p$ is the five-dimensional Planck mass, $N_c$ is the number of colors, $R$ is the scalar curvature, and $K$ is the extrinsic curvature. The (five-dimensional) Lorentz indices are contracted with the full metric for which we use the Ansatz
\be \label{eq:metric}
 ds^2 = e^{2A(r)} \left(\frac{dr^2}{f(r)}-f(r)dt^2 + d\mathbf{x}^2 \right) \ .
\ee
The five dimensional metric is denoted by $g$ and its reduction on the boundary is denoted by $h$.
Here the factor $A(r)$ is understood as the dual of the logarithm of energy in field theory, $A \sim \log \mu$, which defines the mapping between the holographic RG flow and the RG flow in field theory. For this Ansatz, the boundary is at a finite minimum value of the holographic coordinate $r$ which I choose to be at $r=0$. Near the boundary, where the geometry will be asymptotically AdS$_5$, the coordinate is therefore the inverse of the energy scale. The blackening factor $f(r)$ is equal to one for simplest (zero temperature vacuum) backgrounds. For finite temperature backgrounds one finds quite in general a zero of $f(r)$ in the bulk, which is interpreted as the horizon of a planar bulk hole. The thermodynamics of field theory are then obtained from the thermodynamics of the black hole~\cite{Gursoy:2008bu,Gursoy:2008za}. I will elaborate on these points below. 

The dilaton field $\phi$ is dual to the $\mathrm{Tr} G^2$ operator in Yang-Mills theory. Here the trace is over the color degrees of freedom. By using the holographic dictionary, homogeneous solutions $\phi(r)$ contain information on the corresponding VEV and the source, which in this case are $\langle \mathrm{Tr} G^2\rangle$ and the 't Hooft coupling $\lambda_\mathrm{'t H} = g^2 N_c$, with $g$ being the Yang-Mills coupling constant. One can show that it is actually the exponential of the dilaton $\lambda = e^{\phi}$ which equals the 't Hooft coupling near the UV boundary (where the 't Hooft coupling is well defined by Yang-Mills perturbation theory). The field-operator correspondence is also summarized in Table~\ref{tab:dictionary}. Notice that it includes all (relevant and) marginal operators of the field theory, which are expected to capture the most important features. 

As usual in gau\-ge/gra\-vity duality, the metric $g_{\mu\nu}$ is dual to the energy-momentum tensor $T_{\mu\nu}$ of Yang-Mills theory. The source is the metric of the field theory.
The holographic axion field $a$ is dual to the CP-odd operator $\mathrm{Tr}\, G \wedge G$. In this case therefore the source is the $\theta$-angle of Yang-Mills theory. 
Our normalization is such that for homogeneous solutions, $a(r) \approx \theta$ near the boundary.

Finally, the functions $V_g(\phi)$ and $Z(\phi)$ need to be determined to pin down the model. They can be in principle derived from five dimensional string theory which gives
$ 
 V_g(\phi) \propto \exp(4\phi/3) + \cdots
$
where the neglected terms are suppressed at small $\lambda = e^{\phi}$, and $Z(\phi) = \mathrm{const}$. However, this choice for $V_g$ does not agree nicely with known phenomenology of Yang-Mills: it would not lead to asymptotically AdS$_5$ geometries at the boundary. 
Therefore we will at this point switch to bottom-up approach and treat the potentials as free functions, which need to be determined through comparison with data.  This can be done such that the model agrees well with known weak and strong coupling properties of QCD such asymptotic freedom and confinement~\cite{Gursoy:2008bu,Gursoy:2008za}. Moreover, the potentials can be tuned so that both the zero temperature vacuum properties (glueball spectra) and finite temperature thermodynamics both agree remarkably well with results from lattice simulations~\cite{Gursoy:2009jd} (see also~\cite{Ballon-Bayona:2017sxa,Ballon-Bayona:2021tzw}). I will discuss this in more detail in connection to the determination of the full V-QCD model below. For a more detailed account on the IHQCD part, see the reviews~\cite{Gursoy:2010fj,Gursoy:2016ebw}. But before determining the model through comparison to data I review the general structure of the flavor sector.

%%%%%%%%%%%%%%%%%%%%%%%%%%%%%%%%%%%%%
\subsection{Flavor sector: tachyonic brane action}
%%%%%%%%%%%%%%%%%%%%%%%%%%%%%%%%%%%%%

The flavor sector of V-QCD is based on a setup a pair of space filling $D4 - \overline{D4}$ branes~\cite{Bigazzi:2005md,Casero:2007ae}. The brane action includes two terms,
\be \label{eq:Sfdef}
 S_f = S_\mathrm{TDBI} + S_\mathrm{CS} \ .
\ee
The former term is the tachyon DBI action~\cite{Casero:2007ae,Arean:2013tja},
\begin{eqnarray} \label{eq:STDBI}
 S_\mathrm{TDBI} &=& - \frac{1}{2} M^3 N_c \, {\mathrm{tr}}\! \int d^5x\,
\Big(V_f(\l,T^\dagger T)\sqrt{-\det {\bf A}_L} \nonumber \\
&&\quad+ V_f(\l, TT^\dagger)\sqrt{-\det {\bf A}_R}\Big)\,,
\label{generalact}
\end{eqnarray}
where
\begin{eqnarray}
{\mathbf{A}}_{L\,\mu\nu} &=&g_{\mu\nu} + w(\l,T^\dagger T) F^{(L)}_{\mu\nu}
\\\nonumber
&+& {\kappa(\l,T^\dagger T) \over 2 } \left[(D_\mu T)^\dagger (D_\nu T)+
(D_\nu T)^\dagger (D_\mu T)\right] \,,\\
{\mathbf{A}}_{R\,\mu\nu} &=&g_{\mu\nu} + w(\l,TT^\dagger) F^{(R)}_{\mu\nu}\\\nonumber
&+& {\kappa(\l,TT^\dagger) \over 2 } \left[(D_\mu T) (D_\nu T)^\dagger+
(D_\nu T) (D_\mu T)^\dagger\right] \,,
\label{Senaction}
\end{eqnarray}
the determinants are taken over the five dimensional Lorentz indices, and the covariant derivative is
\be
D_\mu T = \partial_\mu T + i  T A_\mu^L- i A_\mu^R T\,.
\ee
The fields  $A_{L}$, $A_{R}$ and $T$ are $N_f \times N_f$ matrices in the flavor space and $\mathrm{tr}$ denotes trace over flavors. The structure of the action is therefore largely dictated by the $U(N_f)_L \times U(N_f)_R$ flavor symmetry, but a precise definition (in the generic case where the fields are nontrivial matrices) requires a prescription for the trace in particular due to the presence of the square root factors. That is, we need to decide how the various matrices are order when taking the trace in~\eqref{eq:STDBI}. The standard is to use the symmetrized trace prescription~\cite{Tseytlin:1997csa}. However, as it turns out, the prescription does not play a role for the results discussed in this review: the matrices in all calculations will be so simple that the trace unambiguously reduces to the standard trace. 
 
The gauge fields $A_{L}^\mu$, $A_{R}^\mu$ are dual to the left and right handed flavor currents $\bar \psi (1\pm \gamma_5)\gamma^\mu \psi$ of the theory while $T$ is dual to the quark mass operator $\bar \psi \psi$. Therefore the source for $T$ is the (possibly complex) quark mass matrix (see Table~\ref{tab:dictionary}).

The remaining task is to determine the potentials $V_f(\lambda,TT^\dagger)$, $w(\lambda,TT^\dagger)$ and $\kappa(\lambda,TT^\dagger)$. As for the dependence for the tachyon, the idea is to use a Sen-like exponential potential in the squared tachyon, i.e., $V_f \propto \exp(-TT^\dagger)$, while the tachyon dependence of the other functions will be dropped~\cite{Sen:2002in,Sen:2004nf}. I will comment on this choice below. The dilaton dependence is known for probe branes, but we follow the philosophy of bottom-up holography and leave the dependence free for the moment. It will be determined by comparing to known features of QCD and to QCD data, as will be discussed below. We will mostly be considering the simple configuration where the tachyon is real, homogeneous, and proportional to the unit matrix: $T = \tau(r) \mathbb{I}$. Because the source corresponding to the tachyon field is the quark mass, this means that the quark mass matrix is likewise real and proportional to the unit matrix, i.e., the quark mass is independent of the flavors. Actually, we will be setting the mass mostly to zero. Using this Ansatz, we denote 
\be
 V_f(\l,TT^\dagger) = V_f(\l,\tau) = V_{f0}(\l)e^{-\tau^2}
\ee
and assume that $\kappa = \kappa(\l)$ and $w = w(\l)$. I will discuss how these functions are determined in section~\ref{ssec:datacomp}.

The latter term in~\eqref{eq:Sfdef} is the Chern-Simons action which has been constructed explicitly (assuming that the tachyon is proportional to a unitary matrix, $TT^\dagger = T^\dagger T \propto \mathbb{I}$) starting from a general expression derived in boundary string field theory~\cite{Casero:2007ae}. It can be divided into separate contributions, $S_\mathrm{CS} = S_\mathrm{CS\,1}+S_\mathrm{CS\,3}+S_\mathrm{CS\,5}$ which are responsible, among other things, for the implementation of the axial anomaly and the flavor anomalies of QCD in the model. Also the chiral magnetic effect and the baryon (instanton) number, which will be analyzed below, arise from these terms. I do not show all the lengthy expressions here (see~\cite{Casero:2007ae}), but the terms relevant at finite density will be discussed below. 
The first term transforms nontrivially under the axial $U(1)_A$. It can be written as
\begin{eqnarray}
 S_\mathrm{CS\,1} &=& T_4 \int C_3 \wedge d\Omega_1 \,,
\\
\Omega_1& = & \mathrm{tr}\Big[V_a(TT^\dagger)i(A_L-A_R)\\\nonumber
&&-\frac{1}{2}(\log T-\log T^\dagger) \,dV_a(TT^\dagger)\Big]  
\end{eqnarray}
where $T_4$ is the four brane tension, $V_a$ is a tachyon potential, and the Ramond-Ramond three form $C_3$ can be related to the axion field through
\be
 \frac{dC_3}{Z(\lambda)} = *\,(da+i\Omega_1)
\ee
where $*$ denotes Hodge dual. If we assume that the tachyon is proportional to the unit matrix, $T = \tau e^{i\xi }\mathbb{I}$, one can combine this term with the glue term~\eqref{eq:Saglue} as~\cite{Arean:2013tja}
\begin{eqnarray} \label{eq:CPoddS}
 &&S_a +  S_\mathrm{CS\,1} = -\frac{M_\mathrm{p}^3}{2}\int d^5x\,\sqrt{-\det g}\, Z(\lambda) \\\nonumber
 &&\quad \times\left[\partial_\mu a -V_a(\tau)\,\mathrm{tr}(A_{L\mu}-A_{R\mu})+N_f\xi\partial_\mu V_a(\tau) \right]^2 \ .
\end{eqnarray}
This term implements the axial $U(1)_A$ anomaly and its relation to the QCD $\theta$-angle and the phase of the quark mass matrix. In the rest of the review I will not discuss CP-odd physics and the $\theta$-angle is set to zero. That is, one can choose a gauge such that $a$ and $\xi$ vanish, and the CP-odd terms of~\eqref{eq:CPoddS} do not play any role.

Finally, let me comment on the term $S_\mathrm{CS\,5}$, which has the form
\be \label{eq:CS5}
 S_\mathrm{CS\,5} = \frac{iN_c}{4\pi^2} \int \Omega_5 \ ,
\ee
where the five-form $\Omega_5$ is single trace in flavor space, and can be expressed (see~\cite{Casero:2007ae}) in terms of $T$, $DT$, $A_{L/R}$, and $F_{L/R}$, but the expression is lengthy. Actually, following the spirit of bottom-up holography, one might consider even more general choices than the form derived in this reference, but for the purposes of this review the explicit choice of~\cite{Casero:2007ae} will be enough. 

%%%%%%%%%%%%%%%%%%%%%%%%%%%%%%%%%%%%%
\subsection{Definition of V-QCD} \label{ssec:vqcddef}
%%%%%%%%%%%%%%%%%%%%%%%%%%%%%%%%%%%%%

V-QCD is obtained by putting together the two building blocks (IHQCD and tachyon brane actions) discussed in detail above. That is, the full model action is~\cite{Jarvinen:2011qe}
\be
S_\mathrm{V-QCD} = S_\mathrm{IHQCD} + S_f
\ee
where the two terms describe the two sectors (gluon and flavor) of the model.
The sectors are fully backreacted in the Veneziano limit: 
\be
 N_c \to \infty \ , \quad N_f \to \infty\ , \quad \mathrm{with}\ x_f \equiv \frac{N_f}{N_c}\ \mathrm{fixed}
\ee
and also keeping the 't Hooft coupling $g^2 N_c$ fixed.
Notice that the presence of backreaction is clear as the number of gluonic degrees of freedom is $\mathcal{O}(N_c^2)$ whereas the number of quark degrees of freedom is $\mathcal{O}(N_f N_c)$. For the full dictionary of the model, see Appendix~A.

But why do I consider Veneziano limit? Reliable use of gau\-ge/gra\-vity duality requires that $N_c$ is large, but I could also consider the standard 't Hooft limit, i.e.,
\be
 N_c \to \infty\ , \quad \mathrm{with}\ N_f\ \mathrm{and}\ g^2N_c\ \mathrm{fixed}
\ee
so that $N_f/N_c \to 0$. This limit can also be called to probe limit because the leading order physics is dominated by gluons, and it is enough to treat the quarks in the probe approximation in this limit. However, in regular QCD we have $N_c=3$ with $N_f = 2$ or 3 light quark depending on whether the strange quark is taken into account or not, so $N_f/N_c$ is not small and backreaction is expected to be significant. Moreover (even though it is not the topic of this review), QCD has an interesting phase diagram in the Veneziano limit, which is in part determined by strongly coupled physics and can be studied using gau\-ge/gra\-vity duality. In particular, the phase diagram as a function of $x_f$ has a quantum phase transition (or potentially several transitions) from the phase similar to regular QCD to the ``conformal window'', where the theory flows to a conformal fixed point in the IR~\cite{Jarvinen:2011qe,Jarvinen:2009fe,Arean:2012mq,Arean:2013tja}. As for this review, however, the motivation of backreacting the quarks is simply to mimic real QCD as closely as possible. See~\cite{Gursoy:2007er} for a brief analysis of the flavor setup in the probe limit in IHQCD, and~\cite{Iatrakis:2010zf,Iatrakis:2010jb} for a more extensive analysis of the flavor setup using the background geometry of~\cite{Kuperstein:2004yf}.

Considering the standard variation of V-QCD already introduced above we can write down the full action for the typical homogeneous backgrounds that we need. That is, we assume a flavor independent tachyon $T =\tau(r) \mathbb{I}$, only turn on vectorial Abelian gauge fields, and set the axion field to zero. The action is then
\begin{eqnarray} \label{eq:vqcdgl}
 S_\mathrm{V-QCD}^\mathrm{bg} &=& M_\mathrm{p}^3 N_c^2 \!\int\! d^5x\,\sqrt{-\det g} \\\nonumber
 &&\times \left[R - \frac{4\left(\partial \l\right)^2}{3\l^2} + V_g(\l) \right]\\
 \label{eq:vqcdfl}
 &-& x_f M_\mathrm{p}^3 N_c^2 \!\int\! d^5x\,V_{f0}(\l)e^{-\tau^2}\\\nonumber 
 &&\times \sqrt{-\det(g_{\mu\nu}+\kappa(\l)\partial_\mu\tau \partial_\nu\tau + w(\l)\hat F_{\mu\nu})}
\end{eqnarray}
where $\hat F$ denotes 
the Abelian component of the field strength tensor and the trace over flavors has already been taken. 
Notice that the gluon~\eqref{eq:vqcdgl} and flavor~\eqref{eq:vqcdfl} terms are indeed of the same order in the Veneziano limit. 

The main features of V-QCD, which are required when it is used to describe real QCD, are confinement and chiral symmetry breaking. Before going to the details, I will sketch how this works in the model. With a natural choice of $V_g$, which has large $\l$ asymptotics similar to what is found from noncritical string theory, 
the glue sector, IHQCD, confines and produces a mass gap for glueballs. The resulting geometry caps off at a singularity in the bulk. This singularity is of the ``good'' kind in the classification of Gubser, i.e., it can be cloaked by horizons to construct regular black hole solutions~\cite{Gubser:2000nd}. The geometry provides, in effect, a soft wall for the fluctuations of the action in the IR.

With the confining geometry, and for a large class of the functions $V_{f0}$, $\kappa$, and $w$, the solution for the tachyon field in the flavor sector (with smallest action) diverges in the IR. That is, the tachyon condenses in the bulk, and this actually happens for any value of the quark mass, including zero. The nonzero value of the tachyon implies that the chiral symmetry is broken down to the vectorial subgroup, $U(N_f)_L\times U(N_f)_R \to U(N_f)_V$, and (via the dictionary) that the chiral condensate $\langle \bar \psi \psi\rangle$ is nonzero. Since this happens even at zero quark mass, chiral symmetry is spontaneously broken. That is, confinement triggers chiral symmetry breaking via tachyon condensation in the model.

In the Sen-like picture, the growth of the tachyon in the bulk corresponds to the annihilation of the $D4-\overline{D4}$ brane pair in the IR. Indeed, when the tachyon becomes large, the flavor action in~\eqref{eq:vqcdfl}
is driven to zero due to the exponential dependence on the squared tachyon. The behavior of the model is somewhat similar to, for example, the Witten-Sakai-Sugimoto model~\cite{Sakai:2004cn,Sakai:2005yt} where for a confining ``cigar'' geometry the $D8 -\overline{D8}$ flavor branes join at the tip of the cigar, which produces chiral symmetry breaking as I explained in Sec.~\ref{sec:holoqcd}, see Fig.~\ref{fig:WSSgeom}. In V-QCD this joining of the branes in the bulk is replaced by the brane ``annihilation'' driven by the tachyon.

%%%%%%%%%%%%%%%%%%%%%%%%%%%%%%%%%%%%%
\subsection{Comparing IHQCD and V-QCD with data} \label{ssec:datacomp}
%%%%%%%%%%%%%%%%%%%%%%%%%%%%%%%%%%%%%
 
Let me then review in more detail how the parameters of the model are chosen to ensure that it agrees with QCD data. Recall that V-QCD is an effective framework, and depending on the choice of potentials, it should be able to describe also other behavior of the field theory than what is found in regular QCD (in particular different IR flows, for example flows ending on fixed points). Therefore, to pin down the model that is actually close to QCD, it is essential to compare all free parameters of the model carefully to QCD data, in rough analogy to effective field theory approaches for QCD.

First, we will set $N_c=3$ (effectively neglecting $1/N_c$ corrections) and $x_f=1$, roughly corresponding to $N_f=3$ light flavors in QCD. Notice that we might want to work in the Veneziano limit with large $N_c$, but most available data has $N_c=3$ and small $N_f$, so we will need to do some compromises. The free parameters of the action are $M_\mathrm{p}$ and the four functions $V_g$, $V_{f0}$, $\kappa$, and $w$. In addition, there are the usual ``knobs'' given by the sources of the various fields as well as the temperature which is parametrized by the size of the bulk black hole. Their values need to be chosen appropriately when comparing to the data.

Among these parameters, there is one special parameter, 
namely the overall energy scale. Notice that the action is invariant under
\begin{eqnarray} \label{eq:Lambdatr}
 &&g_{\mu\nu} \to \Lambda^2 g_{\mu\nu} \ , \quad x_\mu \to \Lambda^{-1}  x_\mu \ , \nonumber\\ 
 &&\partial_\mu \to \Lambda  \partial_\mu \ , \quad  \hat A_\mu \to \Lambda \hat A_\mu 
\end{eqnarray}
where $\hat A$ is the Abelian vectorial gauge field. We can actually identify the overall energy scale with this invariance. To make this more concrete, consider backgrounds depending only on the holographic coordinate $r$, and assume the Ansatz~\eqref{eq:metric}. Then the following transformations
\be \label{eq:scaleinv}
 A \to A + \log\Lambda \ , \quad r \to \frac{r}{\Lambda} \ , \quad \hat A \to \Lambda \hat A  
\ee
leave the action invariant. The consequences of this are most obvious when considering the case where no scale is introduced through the sources, i.e., setting $\hat A$ and the quark mass to zero, or simply by considering the IHQCD sector only. In this case any nontrivial background solution $A(r)$, $\l(r)$ is automatically extended to a family of solutions by the transformations~\eqref{eq:scaleinv}, which only differ by the value of the overall energy scale. This is similar to the dimensional transmutation in QCD: The action of QCD (at zero quark mass) does not fix any energy scale, but the scale of the RG flow ($\Lambda_\mathrm{QCD}$) 
arises from quantum corrections. 
Different RG flows form a family of solutions, which are naturally indexed by the values of $\Lambda_\mathrm{QCD}$.

I will make the analogy even more explicit below by demonstrating that the RG flow of the coupling in QCD is mapped to the flow of the dilaton field near the boundary, therefore implementing asymptotic freedom. 
After this, the source parameter of the dilaton field can be identified as an energy scale $\Lambda_\mathrm{UV}$ (rather than the value of the coupling) in direct analogy to $\Lambda_\mathrm{QCD}$ in field theory side.

\subsubsection{Weak coupling behavior}\label{ssec:potsUV}

Let me start the detailed comparison to QCD by studying the weak coupling behavior. Considering this in detail may be surprising given that gau\-ge/gra\-vity duality is expected not to work at weak coupling. It is, however, important that the weak coupling behavior of the various fields agrees with QCD in order to have the best possible ``boundary conditions'' for the potentially more interesting strong coupling behavior of the model. Indeed usually in holographic bottom-up models of QCD one makes sure that the leading behavior of the bulk fields near the boundary agrees with the leading (free field) UV dimensions of the dual operators. Here I will go one step further by requiring that also the first few quantum corrections, and the RG flow imposed by them, agrees with the near-boundary holographic RG flow of the bulk fields. 

The boundary conditions for the tachyon are such that it vanishes near the boundary (see~\eqref{eq:tauUV} below). It also turns out that the gauge field is irrelevant for the boundary behavior of the metric. Setting $\tau=0$ in the action~\eqref{eq:Sfdef} we see that the geometry is determined by the effective potential
\be \label{eq:Veff}
 V_\mathrm{eff}(\l) = V_g(\l) - x_f V_{f0}(\l)\ .
\ee
For the geometry to be asymptotically AdS$_5$ at the boundary, $V_\mathrm{eff}$ needs to go to a constant at small coupling:
\be
 V_\mathrm{eff}(\l) \to \frac{12}{\ell^2} \qquad \mathrm{as} \quad  \l \to 0 \ ,
\ee
where $\ell$ is the AdS$_5$ radius for the UV geometry. It is natural to assume a Taylor series around $\l=0$, i.e.,
\be
 V_\mathrm{eff}(\l) = \frac{12}{\ell^2}\left[1 + v_1 \frac{\l}{\l_0} +v_2 \left(\frac{\l}{\l_0}\right)^2 + \mathcal{O}\left(\l^3\right) \right] \ ,
\ee
where the constant $\l_0 = 8 \pi^2$ was introduced for later convenience.
Then the near boundary asymptotics of the geometry is AdS$_5$ with logarithmic corrections~\cite{Gursoy:2007cb}: 
\begin{eqnarray} \label{eq:UVA}
 &&A(r) = -\log\frac{r}{\ell_0} + \frac{4}{9\log(r\Lambda)} \\\nonumber 
 &&\quad + \frac{\left(\frac{95}{162}\!-\!\frac{32v_2}{81v_1^2}\right)+\left(-\frac{23}{81}\!+\!\frac{64v_2}{81v_1^2}\right)\log(-\log(r\Lambda))}{(\log(r\Lambda))^2}\\\nonumber
 &&\quad +\mathcal{O}\left(\frac{1}{(\log(r\Lambda))^3}\right) \\
 &&\frac{v_1\l(r)}{\l_0} = -\frac{8}{9\log(r\Lambda)} \\\nonumber
 &&\quad+\frac{\left(\frac{46}{81}\!-\!\frac{128v_2}{81v_1^2}\right)\log(-\log(r\Lambda))}{(\log(r\Lambda))^2}
 +\mathcal{O}\left(\frac{1}{(\log(r\Lambda))^3}\right) \ . \label{eq:UVla}
\end{eqnarray}
Notice that the blackening function of~\eqref{eq:metric} is set to one in the vacuum,  $f(r)=1$.
The flow behaves as expected: the source term of the dilaton has become logarithmically flowing instead of a constant, and the value of the source is now identified with the scale $\Lambda = \Lambda_\mathrm{UV}$. This scale defines the units for all dimensionful quantities in our analysis.
There are also $\mathcal{O}(r^4)$ VEV terms that we did not write down because they involve mixing of the geometry with the asymptotic tachyon solution. 

We then require~\cite{Gursoy:2007cb,Jarvinen:2011qe} that the holographic RG flow agrees with the perturbative QCD flow, i.e., that the holographic $\beta$-function
\be
 \frac{d\l}{dA} = \frac{\l'(r)}{A'(r)}
\ee
matches when evaluated on the solution with the two-loop perturbative $\beta$ function of QCD  in the Veneziano limit:
\be
 \beta(\l_\mathrm{'t H})\equiv \frac{d\l_\mathrm{'t H}}{d\log\mu} = -b_0 \l_\mathrm{'t H} + b_1 \l_\mathrm{'t H}^2 + \mathcal{O}(\l_\mathrm{'t H}^3)
\ee
where
\be
 b_0 = \frac{88-16 x_f}{192\pi^2} \ , \quad b_1 = \frac{34-13 x_f}{384 \pi^4} \ .
\ee
Agreement is found for
\be 
 v_1 = \frac{88-16 x_f}{27} \ , \quad v_2 = \frac{4619-1714 x_f+92 x_f^2}{729} \ .
\ee
Since $V_g$ must be independent of $x_f$, setting $x_f=0$ implies that
\be
 V_g(\l) = \frac{12}{\ell_0^2}\left[1 + \frac{11 \l}{27 \pi^2} + \frac{4619}{729}\left(\frac{\l}{8\pi^2}\right)^2 + \mathcal{O}\left(\l\right)^3 \right]
\ee
where $\ell_0 =\ell(x_f=0)$. Finally solving for $V_{f0}$ gives
\begin{eqnarray}
 &&V_{f0}(\l) = \frac{1}{\ell_0^2}\Bigg[W_0 +\frac{((11 -2x_f)W_0 +24)\l}{27 \pi ^2}\\\nonumber
 && + \frac{W_0 \left(4619-1714 x_f+92 x_f^2\right)+24 (857-46 x_f)}{729} \\\nonumber
&& \times \left(\frac{\l}{8\pi^2}\right)^2 +  \mathcal{O}\left(\l\right)^3
 \Bigg] \ ,
\end{eqnarray}
and the AdS radius is given by
\be
 \ell^2 = \frac{\ell_0^2}{1-x_f W_0/12} \ .
\ee

Notice that the constant $W_0$ remains a free parameter. It should satisfy $0<W_0<12/x_f$ for the geometry to be asymptotically AdS. 

Apart from $W_0$, also the AdS radius $\ell_0$ appears in the expressions but we can set it to one without loss of generality. This point may need a careful explanation because the AdS radius is often used to set the scale of dimensionful quantities in gau\-ge/gra\-vity duality which we have chosen not to do here. This is already visible from~\eqref{eq:Lambdatr} where we rescaled the metric by a dimensionful quantity even though the metric is dimensionless. This works because there are other dimensionful quantities that we chose not to rescale, i.e., the potentials. In fact, there is another transformation 
\begin{eqnarray} \label{eq:elltr}
 &&V_g \to \Lambda^2 V_g \ , \quad V_{f0} \to \Lambda^2 V_{f0} \ , \quad g_{\mu\nu} \to \Lambda^{-2} g_{\mu\nu} \ , \\\nonumber
 &&\kappa \to \Lambda^{-2} \kappa \ , \quad w \to \Lambda^{-2} w \ , \quad  M_\mathrm{p} \to \Lambda M_\mathrm{p}\ , 
\end{eqnarray}
which also leaves the action invariant. This latter transformation can be used to absorb changes of $\ell$ in to the potentials, so we are free to set its value to any number. Alternatively we could combine the transformations~\eqref{eq:Lambdatr} and~\eqref{eq:elltr} so that dimensionful quantities would be measured in units of $\ell$ rather than $\Lambda_\mathrm{UV}$. We however choose to use $\Lambda_\mathrm{UV}$ because we find it simpler: unlike~\eqref{eq:Lambdatr}, the transformation~\eqref{eq:elltr} affects the potentials and therefore modifies the action rather than just the fields.

Apart from the metric, the tachyon field also has nontrivial asymptotics near the boundary, which depends on the function $\kappa$ in addition to the potentials $V_g$ and $V_{f0}$. Remarkably, the choice for $\kappa$ which gives the desired boundary behavior is also simply a Taylor series:
\be
 \kappa(\l) = \frac{2\ell^2}{3}\left[1+\frac{\kappa_1 \l}{\l_0} + \mathcal{O}(\l^2)\right] \ ,
\ee
where the leading term was already chosen such that the leading dimension of the quark mass and the $\bar\psi\psi$ operator is reproduced. 
This leads to the following asymptotics of the tachyon~\cite{Jarvinen:2011qe}:
\begin{eqnarray} \label{eq:tauUV}
 \frac{\tau(r)}{\ell} &=& m_q r (-\log(r\Lambda))^{\gamma_\tau} \left[1+\mathcal{O}\left(\frac{1}{\log(r\Lambda)}\right)\right] \\\nonumber
 &+& \sigma r^3 (-\log(r\Lambda))^{-\gamma_\tau} \left[1+\mathcal{O}\left(\frac{1}{\log(r\Lambda)}\right)\right] \ ,
\end{eqnarray}
where 
\be
 \gamma_\tau = \frac{4}{3}\left(1+\frac{\kappa_1}{v_1}\right) \ .
\ee
Near the boundary we therefore find that the solution satisfies
\begin{eqnarray}
 \frac{d\log \tau}{dA} = \frac{\tau'(r)}{\tau(r) A'(r)} &=& -1 +\frac{9 v_1 \gamma_\tau \l(r)}{8 \l_0} + \mathcal{O}(\l^2) \\\nonumber
 &=& -1 + \frac{3}{2}(v_1+\kappa_1) \frac{\l(r)}{\l_0} + \mathcal{O}(\l^2)
\end{eqnarray}
where the first term is the leading dimension of the quark mass and the second term is interpreted as the anomalous dimension~\cite{Jarvinen:2011qe}. Comparing to the expression for the perturbative one-loop anomalous dimension for QCD in the Veneziano limit
\be
\gamma(\l) = \gamma_0 \l + \mathcal{O}(\l^2) = \frac{3\l}{16\pi^2}+ \mathcal{O}(\l^2) \ ,
\ee
one finds agreement if $v_1 + \kappa_1 = -1$. Therefore we choose
\be
 \kappa_1 = -\frac{115-16x_f}{27} \ .
\ee
Notice also that the anomalous dimension for $\bar\psi\psi$ has the opposite sign, as it should, which can be seen by comparing the logarithmic flows of the source and VEV terms in~\eqref{eq:tauUV}.

Finally, the boundary values of the potentials and the value of $M_\mathrm{p}$ can also be estimated by comparing two-point functions (at high momentum)~\cite{Kajantie:2011nx,Arean:2013tja} and the thermodynamics of the model (at high temperature and density) to predictions of perturbative QCD~\cite{Gursoy:2008bu,Gursoy:2008za,Alho:2012mh,Alho:2013hsa}. For example, analysis of the thermodynamics and its $x_f$-dependence suggests that $W_0 \approx 6$ for $x_f \approx 1$. Agreement also requires that the function $w(\l)$ must go to a constant as $\l \to 0$. Some of the results obtained by comparison to perturbation theory are in mutual agreement while others are not: For example, the values of $M_\mathrm{p}$ obtained from correlators of the energy momentum tensor and from high temperature thermodynamics differ by a constant factor. However, the result for $w_0$ from the vector-vector correlator at high momentum agrees with the results obtained from thermodynamics at high temperature and density. 

So far we did not consider the function $Z(\l)$ and $V_a(\tau)$ appearing in the CP-odd terms~\eqref{eq:Saglue} and~\eqref{eq:CPoddS}. As it runs out, $Z(\l)$ should approach a constant fixed by the Yang-Mills topological susceptibility~\cite{Gursoy:2007er} as $\l\to0$, and we must have $V_a(\tau=0)=1$ for consistency with the axial anomaly in QCD~\cite{Arean:2013tja,Arean:2016hcs}.

The general conclusion from the analysis is therefore that the UV behavior works nicely if the potential functions go to (well chosen) constants as $\l \to 0$, and have regular Taylor series around $\l=0$. The functions can be analytic at $\l=0$ but this is not required: there can be nonanalyticity if it vanishes fast enough as $\l \to 0$ so that the above analysis is not affected.

One should however recall that all results obtained by comparing the model to QCD perturbation theory should be taken with a grain of salt due to the fact that gau\-ge/gra\-vity duality is not expected to work at weak coupling. Therefore we will only use the results for the asymptotics of the potentials at small $\l$ which were obtained by comparing to the RG flow of the coupling and the quark mass as detailed above. The other parameters (including those that could in principle be obtained by studying the results at weak coupling) will be determined by comparing the model to QCD data at strong coupling, in particular to lattice QCD data.

\subsubsection{Strong coupling behavior}

Let me then discuss the more interesting region from the viewpoint of holography, i.e., the region of strong coupling. The various potentials at large $\l$ can be determined quite precisely by requiring agreement with known features of QCD. In most cases the asymptotics of the form $\sim \l^{c_1}(\log\l)^{c_2}$ is required to meet all constraints. I will only give an overview of the results without explicit derivation.

I will start with the constraints in the gluon sector, i.e., IHQCD, assuming the asymptotics
\be
 V_g(\l) \sim \l^{g_p}(\log\l)^{g_\ell} \ , \quad (\l \to \infty) \ .
\ee
The main constraints affecting the potential are the following~\cite{Gursoy:2007er}:
\begin{itemize}
 \item \emph{Confinement.} The quark-antiquark potential for a pair of probe quarks in the pure Yang-Mills background can be found by the Wilson loop test, i.e., by computing the action of a fundamental string embedded in the string frame background (see~\cite{Kinar:1998vq} for details). Confinement is found for $g_p>4/3$ or when $g_p=4/3$ and $g_\ell\ge 0$. All confining geometries end in a ``good'' kind of IR singularity according to the classification by Gubser~\cite{Gubser:2000nd}.
 \item \emph{Mass gap.} The glueball spectrum is found to be gapped, remarkably, for the same values of parameter as imposed by confinement, i.e., when  $g_p>4/3$ or when $g_p=4/3$ and $g_\ell\ge 0$. 
 \item \emph{Magnetic screening.} Magnetic screening between a pair of point sources can be probed by considering the embedding a D-string instead of a fundamental string. Screening is the expected behavior in QCD, and it is found for $g_p<8/3$.
 \item \emph{Linear glueball trajectories.} For the confining geometries (which all have gapped glueball spectrum) one can further study the asymptotic spectrum for radial excitations. One finds that when $g_p=4/3$ and $0 \le g_\ell < 1$, the masses behave as $m_n^2 \sim n^{2g_\ell}$ as the excitation number $n \to \infty$. For other confining cases, one finds the ``hard wall'' or harmonic oscillator spectrum, $m_n^2 \sim n^2$.
\end{itemize}
For QCD, we want to have confinement, mass gap, magnetic screening, and linear ``Regge-like'' asymptotic radial trajectories. The choice which produces all of these is~\cite{Gursoy:2007er}
\be \label{eq:gchoices}
 g_p = \frac{4}{3} \ , \quad g_\ell = \frac{1}{2}\ .
\ee
In this case the IR geometry at the good kind of IR singularity obeys
\begin{eqnarray} \label{eq:IRA}
 A(r) &=& -(r\Lambda_\mathrm{IR})^2 + \frac{1}{2} \log (r\Lambda_\mathrm{IR}) + A_c +\mathcal{O}\left(r^{-2}\right) \\  \label{eq:IRla}
 \log \l(r) &=& \frac{3}{2}(r\Lambda_\mathrm{IR})^2 + \l_c +\mathcal{O}\left(r^{-2}\right)
\end{eqnarray}
as $r \to \infty$, where $A_c$ and $\l_c$ are calculable constants that depend on the higher order terms in the asymptotic expansion for $V_g(\l)$, and $\Lambda_\mathrm{IR}$ is a constant of integration.

One might be worried that the curvature singularity at $r \to \infty$, or the divergence of the dilaton in the IR, leads to the model not being self-consistent and to the breakdown of the classical analysis in the IR. However, a detailed analysis of the IR asymptotics, which we do not attempt to cover here, suggests that there are no major issues with the IR behavior (see~\cite{Gursoy:2007er,Kiritsis:2009hu}). A key observation is that, while the curvature invariants diverge for the Einstein frame metric considered above, in the string frame, which is obtained by multiplying the metric by a factor of $\l^{4/3}$, they vanish asymptotically. For example, the string frame scalar curvature goes to zero as $R_s \sim 1/r^3$~\cite{Gursoy:2007er}. This leads to the suppression of higher derivative terms of the gravity action in the IR.

The choice of the function $Z(\l)$ affects (apart from the spectrum of pseudoscalar glueball states) the topological susceptibility and the Chern-Simons diffusion rate~\cite{Gursoy:2012bt}. In order to obtain correct behavior at finite $\theta$-angle and universal glueball trajectories also in the pseudoscalar sector, we need that $Z(\l)\sim \l^4$ in the IR, $\l \to \infty$. Notice that the choice of $Z(\l)$ will not play a major role in the analysis of this review, because we keep the $\theta$-angle at zero.
 
Let me then go on discussing the flavor sector. I parametrize the asymptotics as
\begin{eqnarray}
 V_{f0}(\l) &\sim& \l^{v_p}(\log \l)^{v_\ell} \ ,\quad \kappa(\l) \sim \l^{-\kappa_p}(\log\l)^{-\kappa_\ell} \ , \\\nonumber
 w(\l) &\sim& \l^{-w_p}(\log \l)^{-w_\ell} \ .
\end{eqnarray}
I will first discuss the constraints from the analysis of the chirally broken phase with bulk tachyon condensate. This will constrain the flavor potential $V_{f0}$ and the coupling function of the tachyon $\kappa$ (but not the coupling of the gauge field $w$ as the gauge fields vanish for the background).  We need the following basic properties~\cite{Jarvinen:2011qe,Arean:2013tja}:
\begin{itemize}
 \item \emph{``Good'' kind of IR behavior.} The choice of $V_g$ following~\eqref{eq:gchoices} guarantees that the geometry ends in a good kind of singularity, so long as the tachyon diverges fast enough near it, which causes the flavor sector to decouple from the flow of the metric. However the tachyon behavior itself should also be ``regular'': requiring the flavor action to remain finite in the IR should remove a nonnormalizable mode, leaving a one-parameter family of solutions. Then this single parameter is understood as the quark mass, and the chiral condensate is fixed in term of the quark mass for regular solutions. This kind of behavior is not guaranteed for all actions, but we require it: finiteness of the action should set the boundary condition of the tachyon uniquely without the need of imposing extra conditions by hand. 
 \item \emph{Annihilation of the brane action in the IR.} As I discussed above, following the Sen-like picture requires that the $D4-\overline{D4}$ brane pair annihilates in the IR, and the brane action vanishes. This also makes it sure that the holographic dictionary works and there are no undesired surface term contributions to correlators arising from the IR endpoint. 
 \item \emph{Discrete meson spectrum.} The growing tachyon in the IR should reduce the IR fluctuations of the flavor fields such that the spectrum is discrete. 
\end{itemize}
As it turns out, the second of these conditions is the most restrictive one so that the other conditions usually follow from it. We find that all these conditions hold for the following parameter values: 
\begin{eqnarray} \label{eq:fcons1}
 v_p&\le &\frac{10}{3} \ , \quad \kappa_p>4/3 \quad \mathrm{or} \\
 v_p&<&\frac{10}{3} \ , \quad \kappa_p=4/3 \ , \quad \kappa_\ell \ge -\frac{1}{2}  \quad \mathrm{or} \\
\label{eq:fcons3}
v_p&=&\frac{10}{3} \ , \quad \kappa_p=4/3 \ , \quad \kappa_\ell \ge -\frac{3}{2} \ ,
\end{eqnarray}
(when $v_p=10/3$ there is also an additional constraint for $v_\ell$, see~\cite{Arean:2016hcs}). 

Having identified the range of potentially reasonable asymptotics, we have a look at more detailed features of the model. These include the following: 
\begin{itemize}
 \item \emph{Linear meson trajectories.} The masses of the mesons are regulated by a soft IR wall created by the solution for the tachyon field. Interestingly, all of the choices listed above give roughly linear radial trajectories but typically with logarithmic corrections (most usually $m_n^2 \sim n  \log n$). The logarithmic corrections are absent only for specific choices: $v_p<10/3$, $\kappa_p=4/3$, and $\kappa_\ell=-1/2$, or alternatively for $v_p=10/3$, $\kappa_p=4/3$, and $\kappa_\ell=-3/2$~\cite{Arean:2013tja,Iatrakis:2015rga}. The latter choice has however an issue: it is challenging to construct complete backgrounds with $v_p=10/3$ as I shall discuss below. 
 \item \emph{Universal slopes of asymptotic trajectories.} All the slopes of the (linear) meson trajectories are the same, if a single additional constraint is satisfied, namely that
 \be
  \frac{\kappa(\l)}{w(\l)} \to 0 \ , \quad (\l \to \infty) \ .
 \ee
 Otherwise the slopes of the axial vector and pseudoscalar mesons will be larger than those of the vector and scalar mesons. If we require this and choose exactly linear trajectories, so that $\kappa_p =4/3$, This means that $w_p \le 4/3$ (with the understanding that for $w_p=4/3$ we choose $w_\ell<\kappa_\ell$)~\cite{Arean:2013tja}.  
 \item \emph{Phase diagram at finite chemical potential.} It turns out that for some choices of $w(\l)$ the phase diagram at low temperatures has undesired features as any small chemical potential is turned on. The background changes from a horizonless geometry to a black hole geometry implying an immediate transition to a quark matter phase at unphysically low values of the chemical potential. In order to avoid this, $w(\l)$ should vanish fast enough as $\l \to \infty$. If we choose $\kappa_p=4/3$ following the constraints discussed above, we need $w_p \ge 4/3$~\cite{Ishii:2019gta}.  
\end{itemize}

Apart from these precisely defined constraints, there are other constraints that are somewhat less precise. First, the phase diagram of QCD in the Veneziano limit as a function of $x_f$ (at zero quark mass and temperature) includes the so called ``conformal window'' for $3 \lesssim x_f <11/2$  where the theory flows to a fixed point  in the IR. In order to describe this, the effective potential of~\eqref{eq:Veff} should have a maximum at some $\l=\l_*$, implementing the fixed point. The fixed point is perturbative near $x_f=11/2$, and it is ``automatically'' implemented in the model if we choose the potentials which agree with the RG flow of QCD as discussed in Sec.~\ref{ssec:potsUV}. As $x_f$ decreases, the fixed point should move to higher values of the coupling which eventually triggers a ``conformal transition'' to a theory with IR physics similar to regular QCD. It turns out that it is difficult to make this picture to work unless $V_{f0}$ grows faster than $V_g$ with increasing $\l$, i.e., we need that $v_p>g_p=4/3$~\cite{Jarvinen:2011qe}. This is however not a strict requirement because the fixed point only flows to values $\l_* =\mathcal{O}(1)$ before the transition, so the asymptotic region of $V_{f0}$ is never probed. However working potentials with $v_p<4/3$ may need to have a rather peculiar structure at intermediate coupling, which may lead to other issues with the spectrum or with the phase diagram.
Second, it is difficult to find complete solutions that interpolate between the IR singularity and AdS$_5$ in the UV if $v_p$ is close to the upper bound $10/3$. This is particularly challenging at finite $\theta$-angle~\cite{Arean:2016hcs}. Therefore in practice the upper bound for $v_p$ is lower than what is suggested by the analytic IR analysis in~\eqref{eq:fcons1}--\eqref{eq:fcons3}. We have found numerically that regular, complete solutions can be constructed when $v_p \approx 2$. 

Notice that the Ansatz for the flavor action~\eqref{eq:vqcdfl} involved several assumptions, for example the factorized $\l$ and $\tau$ dependence of the tachyon potential, and the independence of $\kappa$ and $w$ on $\tau$. One might wonder if such assumptions constrain the action too much. This is an important question and it is difficult to answer it decisively, but we have carried out quite extensive tests. First, we considered a more general nonfactorizable Ansatz of the form
\be
 V_f(\l,\tau) = V_{f0}(\l)e^{-a(\l)\tau^2}
\ee
for the tachyon potential. Analysis of the behavior of the background and meson spectrum (similar to what was outlined above but more general) suggests however that $a(\l)$ should be constant~\cite{Arean:2013tja}. A specific observation is that for the mass gap of mesons to grow linearly (or even as a power law) with quark mass, $a(\l)$ should be a constant \emph{and} the asymptotic behavior of $\log V_f$ at large $\tau$ needs to be $\sim \tau^2$~\cite{Jarvinen:2015ofa}. Second, we have considered more general power laws than the standard square root for the DBI action. Interestingly, it turns out that the requirements discussed above can be satisfied simultaneously only for the square root action~\cite{Arean:2013tja}. 

I am now ready to summarize the results of the IR analysis. An optimal choice for the asymptotics of the functions was found to be
\begin{eqnarray} \label{eq:finalpowers}
 V_g &\sim& \l^{4/3}(\log \l)^{1/2} \ , \quad V_{f0}(\l)\sim \l^{v_p} \\\nonumber
 \kappa(\l) &\sim& \l^{-4/3}(\log \l)^{1/2} \ , \quad w(\l) \sim \l^{-4/3} (\log \l)^{-w_\ell} \ ,
\end{eqnarray}
with
\be 
 v_p \approx 2 \ , \quad w_\ell < -\frac{1}{2} \ .
\ee
Interestingly, the final results for the powers $g_p$, $\kappa_p$ and $w_p$ in~\eqref{eq:finalpowers} exactly agree with the powers expected from string theory, even if they were derived purely based on phenomenology~\cite{Arean:2013tja}. Even the result for $v_p$ is close to the number expected for five dimensional DBI in the Einstein frame, i.e., $v_p=7/3$. That is, taking the potentials predicted from string theory at large coupling, and modifying them such that they go to constants at small coupling, already gives a good first guess for all the potentials. One just needs to add subleading logarithmic corrections in the IR in order to improve the agreement with QCD.

One remark is in order: the logarithmic corrections in~\eqref{eq:finalpowers} mostly arose from the requirement of linear confinement and Regge-like behavior of the spectra. That is, while linear confinement arises from stringy behavior of QCD in the IR, in the V-QCD model it is obtained by tuning the asymptotic potentials without any obvious connection to string dynamics. Therefore the motivation to choose exactly the logarithmic behavior  of~\eqref{eq:finalpowers} is perhaps not that strong. Nevertheless, I choose to use exactly these asymptotics in order to have phenomenology which is as close to QCD as possible. Notice also that choosing the logarithmic corrections differently would mean only a minor modification in the IR asymptotics in the potentials, which is unlikely to cause significant changes in the next steps of the analysis which I will discuss below. 

The tachyon asymptotics for the preferred choice of potentials is a power law in $r$~\cite{Arean:2013tja}:
\be \label{eq:tauIR}
 \tau(r) \sim \tau_0 (r\Lambda_\mathrm{IR})^{C_\tau}\left[1+\mathcal{O}\left(\frac{1}{r^2}\right)\right]
\ee
Here the coefficient $C_\tau$ is fixed: it can be determined in terms of the subleading terms in the IR asymptotics of the background and the potentials. The single free parameter $\tau_0$ indexes the regular solutions, and can be mapped to the quark mass when the whole solution is know. 

The summary for the geometry is then the following: the UV asymptotics of the chirally broken vacuum solution is given by~\eqref{eq:UVA},~\eqref{eq:UVla}, and~\eqref{eq:tauUV}. The divergence of the tachyon in the IR decouples the flavor so that the geometry is independent of the tachyon and given asymptotically in~\eqref{eq:IRA} and~\eqref{eq:IRla}, while the tachyon obeys~\eqref{eq:tauIR}. The background is parametrized in terms of the sources $\Lambda_\mathrm{UV}$ for the geometry and $m_q$ for the tachyon, but thanks to the scale invariance, the only nontrivial parameter is the dimensionless ratio $m_q/\Lambda_\mathrm{UV}$ as in QCD. Equivalently, the solution may be parametrized by the IR parameters $\Lambda_\mathrm{IR}$ and $\tau_0$, where only varying $\tau_0$ affects the solution nontrivially due to the scale invariance. This latter parametrization is what one often uses when solving the background numerically due to practical reasons. When the whole solution is known, the IR parameters can be mapped to the UV parameters (and vice versa).

\subsubsection{Comparison with lattice} \label{ssec:latticefit}

Having determined the asymptotics of the potentials both at small and large $\l$, one needs to tune the potentials in the middle so that the model agrees quantitatively with QCD. The main available data are lattice data for the thermodynamics of QCD (at small density) and experimental data for QCD spectrum. The thermodynamics and the spectrum have been fitted separately in~\cite{Jokela:2018ers} and in~\cite{Amorim:2021gat}, respectively, and were shown to lead to remarkably similar fit results for the potentials. The spectrum fit was carried out in connection to the analysis of Regge trajectories in IHQCD and V-QCD~\cite{Iatrakis:2015rga,Ballon-Bayona:2015wra,Ballon-Bayona:2017vlm,Amorim:2018yod,Amorim:2021ffr,Amorim:2021qlu}. An overall fit which would include all available finite temperature (thermodynamical) data and zero temperature data for the spectrum and decay constants, is work in progress. Because the main topic of this review is thermodynamics at finite temperature and density, I will here concentrate on the fit to the finite temperature lattice data. 

For the fit we will use Ans\"atze having separate UV and IR terms:
\begin{eqnarray}
 &&F(\l) = \sum_{k=0}^{N_\mathrm{UV}}F_k \left(\frac{\l}{\hat \l_0}\right)^k\\\nonumber
 && \ \ + e^{-\hat \l_0/\l}\left(\frac{\l}{\hat \l_0}\right)^{f_p}\left(\log\left(1+\frac{\l}{\hat \l_0}\right)\right)^{f_\ell} \sum_{k=0}^{N_\mathrm{IR}}f_k \left(\frac{\hat \l_0}{\l}\right)^k
\end{eqnarray}
Here $F(\l)$ stands for any of the functions $V_g(\l)$, $V_{f0}(\l)$, $1/\kappa(\l)$, and $1/w(\l)$, and $N_\mathrm{UV}$ and $N_\mathrm{IR}$ are small integers (we will use slightly different values for different potentials). For $\kappa$ and $w$ we use the reciprocal because these functions vanish in the IR rather than blowing up. If $N_\mathrm{UV}>f_p$ one should modify the UV term for example by adding extra suppression factors (e.g. $(1+\l/\hat \l_0)^{-1}$) to make sure that it is subleading with respect to the IR term at large $\l$. Notice that the Ansatz satisfies the requirements for both UV and IR asymptotics for an appropriate choice of the coefficients. We choose $v_p =2$, $v_\ell =0$, and $w_\ell = -1$. Those coefficients $F_k$, $f_k$ which are not fixed by the analysis of the asymptotics will be treated as fit parameters. Apart from these values, we also determine the Planck mass $M_\mathrm{p}$ and the overall energy scale $\Lambda_\mathrm{UV}$.  

The fitting procedure is sequential: we first fit $V_g(\l)$ to lattice data for pure Yang-Mills, then $V_{f0}(\l)$ for lattice data for full QCD at zero density, and finally $w(\l)$ to data at small density. The remaining function $\kappa(\l)$ is not fitted directly, but it needs to satisfy a requirement arising from the deconfinement transition temperature at zero density. The explicit Ans\"atze and fit parameters are given in Appendix~B. I will give more details on how the thermodynamics is computed in the holographic model in Sec.~\ref{ssec:QM}. I will only discuss the fit results here.

\begin{figure}
\centering  \includegraphics[width=0.48\textwidth]{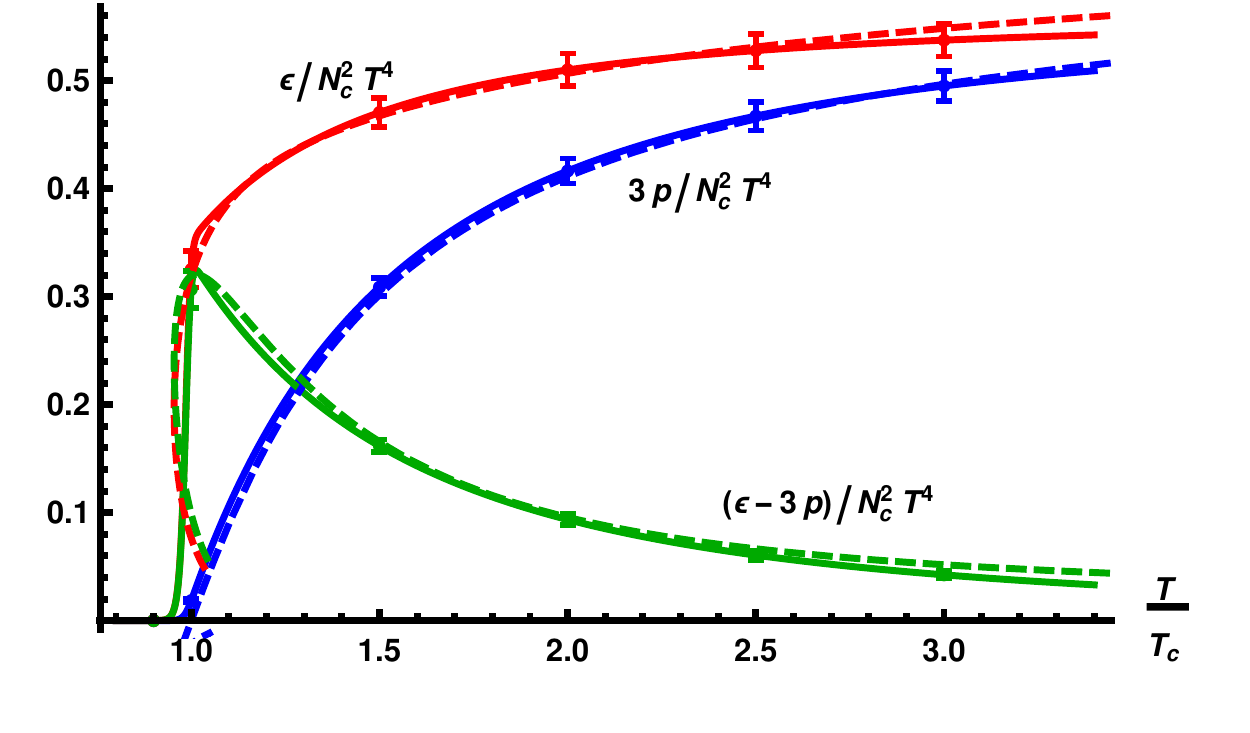}
\caption{Fitting $V_g(\l)$ to the large $N_c$ lattice data for pure Yang Mills. Red, blue, and green curves are for (normalized) energy density, pressure, and interaction measure $(\epsilon-3p)/T^4$, respectively. Solid curves and error bars show the lattice data~\protect\cite{Panero:2009tv}, and the dashed curves are our fit. Figure taken from~\protect\cite{Jokela:2018ers}.}
\label{fig:Vgfit}       % Give a unique label
\end{figure}

The function to be fitted first is $V_g$. It is determined~\cite{Alho:2015zua,Jokela:2018ers} by comparing to pure Yang-Mills lattice data~\cite{Panero:2009tv} in the limit of large $N_c$, see Fig.~\ref{fig:Vgfit}. This is a three-parameter fit, with one parameter being the overall normalization, i.e., $M_\mathrm{p}$, and two parameters affect the shape of the curve. The value of $T_c$ determines $\Lambda_\mathrm{UV}$ -- this is not a real fit but merely a choice of units. Notice that the data is well reproduced within the error bars of the lattice data.

Then I move to the flavor sector where the most important potential to be fitted is $V_{f0}$. I will determine it by comparing to lattice data for thermodynamics of full (quarks+gluons) QCD at zero density. There are however a few complications. First, there is useful lattice data only for $N_c=3$ so the fitting cannot be done directly in the Veneziano limit, so one may suspect that some nontrivial $1/N_c$ effects are lost. But for pure Yang-Mills it was shown in~\cite{Panero:2009tv} that $1/N_c$ effects are small: fitting the data at $N_c=\infty$ and at $N_c=3$ would give almost identical results in that case. This suggests that the same is true in the Veneziano limit. Second, the lattice data uses finite ``physical'' quark masses. We will use data with 2+1 flavors, i.e., results from simulations with two light quarks and the strange quark~\cite{Borsanyi:2013bia}. But in the holographic model we are working with the simplifying assumption that all quarks have the same mass, and actually we will set the mass to zero when fitting the data. We do this because the light quark mass is much smaller than $\Lambda_\mathrm{QCD}$ so adding it would be such a small correction that it is not necessary within the precision of our approach (this is not true for specific observables such as the pion mass, but should hold for all data we are using here). We also expect that the effects of finite  strange quark mass can be largely absorbed into our fit parameters.   
Third, the QCD deconfinement phase transition at physical quark masses and zero density is a crossover, whereas the holographic model has a first order phase transition. The phase transition in the model arises from change in the geometry (as I shall discuss below) from a horizonless geometry ending in the good singularity of~\eqref{eq:IRA} to a planar black hole, and it is difficult (albeit possible~\cite{Gursoy:2010jh}) to construct the theory such that this is not a first order transition. However there are physics reasons that suggest that one should not even try to make the transition a crossover. Namely, the pressure in the low temperature confined phase is dominated by the pressure of (essentially free) mesons, with the pions giving the most important contribution as they are the lightest states in the spectrum. The pressure from the mesons is $O((N_c)^0)$, so it is suppressed with respect to the $O(N_c^2)$ pressure of the quark-gluon plasma phase in the large $N_c$ limit. Therefore it is not captured by holographic models. This holds even for V-QCD: even though the meson pressure $\sim N_f^2$ is comparable to the gluon pressure $\sim N_c^2$ in the Veneziano limit, it arises from string loop corrections that are not included in the model. One can include these corrections by hand and show that this can turn the first order transition into a higher order transition~\cite{Alho:2015zua}. We therefore anticipate that these loop corrections will modify the result in the confined phase, and only fit the thermodynamics in the deconfined phase, i.e., for $T>T_c$ where $T_c \approx 155$~MeV. Due to the anticipated loop corrections, the crossover temperature of the lattice data will also be different from the transition temperature of the holographic model without the loops (which will be around 120~MeV).

\begin{figure}
\centering  \includegraphics[width=0.48\textwidth]{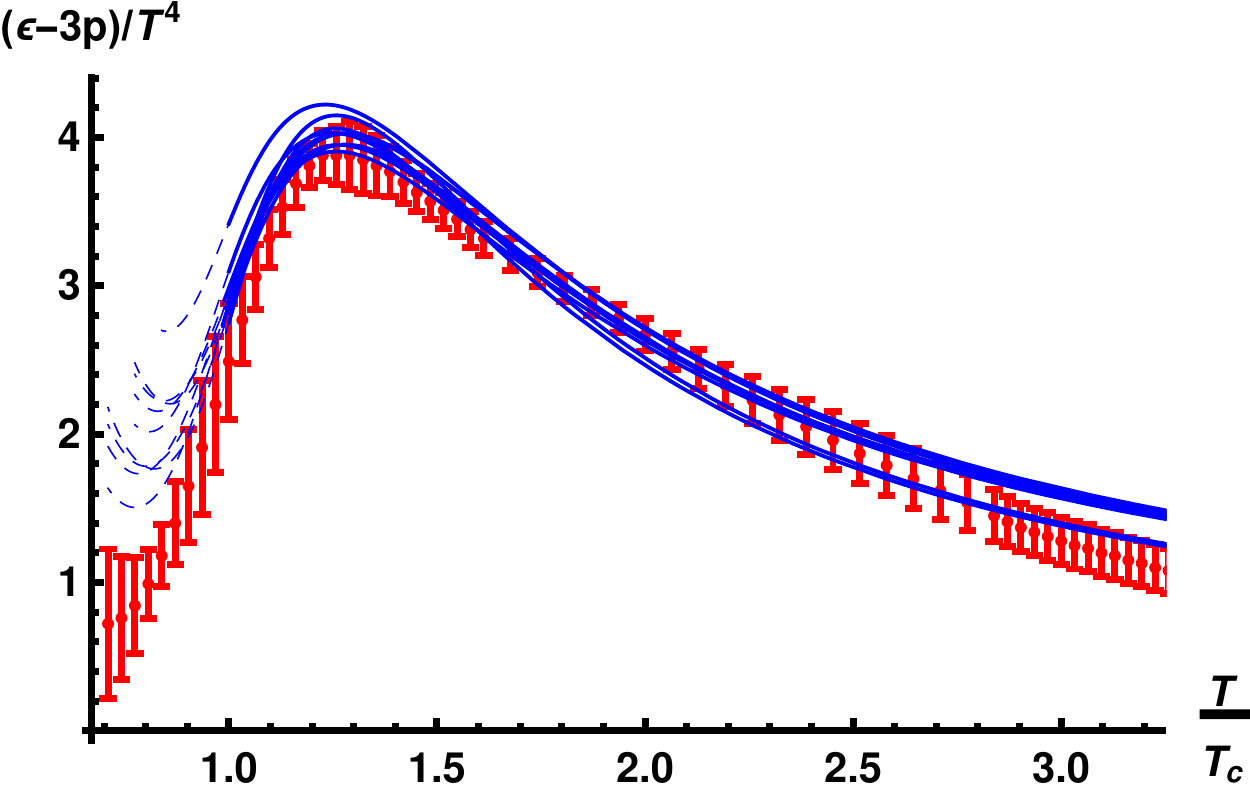}
\caption{Fitting $V_{f0}(\l)$ to the QCD lattice data for the interaction measure. The red dots and error bars show the lattice data~\protect\cite{Borsanyi:2013bia}, and blue curves are our fits. Figure taken from~\protect\cite{Jokela:2018ers}.}
\label{fig:Vf0fit}       % Give a unique label
\end{figure}

The results of fitting $V_{f0}$ to the $N_f=2+1$ lattice data for the interaction measure $(\epsilon-3p)/T^4$ are shown in Fig.~\ref{fig:Vf0fit}. There are several curves because, as it turns out, the fit has a flat direction: the data can be fitted well for a one-parameter curve in the parameter space, and this one-parameter freedom is represented by a family of different fits. 
The solid parts of the curves are for $T>T_c$ and are fitted to lattice data, whereas the dashed curves are not fitted. The dashed curves are expected to be replaced by thermodynamics of meson gas after their contribution is included in the model. As we set the quark mass to zero, the curves arise from a chirally symmetric configuration with $\tau=0$, so that the curves indeed only depend on $V_{f0}(\l)$ (and $V_g(\l)$). This is however not completely true: the pressure in the model is actually the difference of the pressures between the deconfined and confined phases, and the (constant) pressure of the chirally broken confined phase also depends on $\kappa(\l)$. That is, there is dependence on $\kappa(\l)$ through one additional parameter, which is determined so that the transition temperature is consistent with the fit. Notice that it is also possible to fit~\cite{Misra:2019thm} the 2+1 flavor QCD lattice data for the interaction measure in top-down frameworks based on modified Klebanov-Strassler constructions~\cite{Mia:2009wj,Dhuria:2013tca}.

\begin{figure}
\centering  \includegraphics[width=0.48\textwidth]{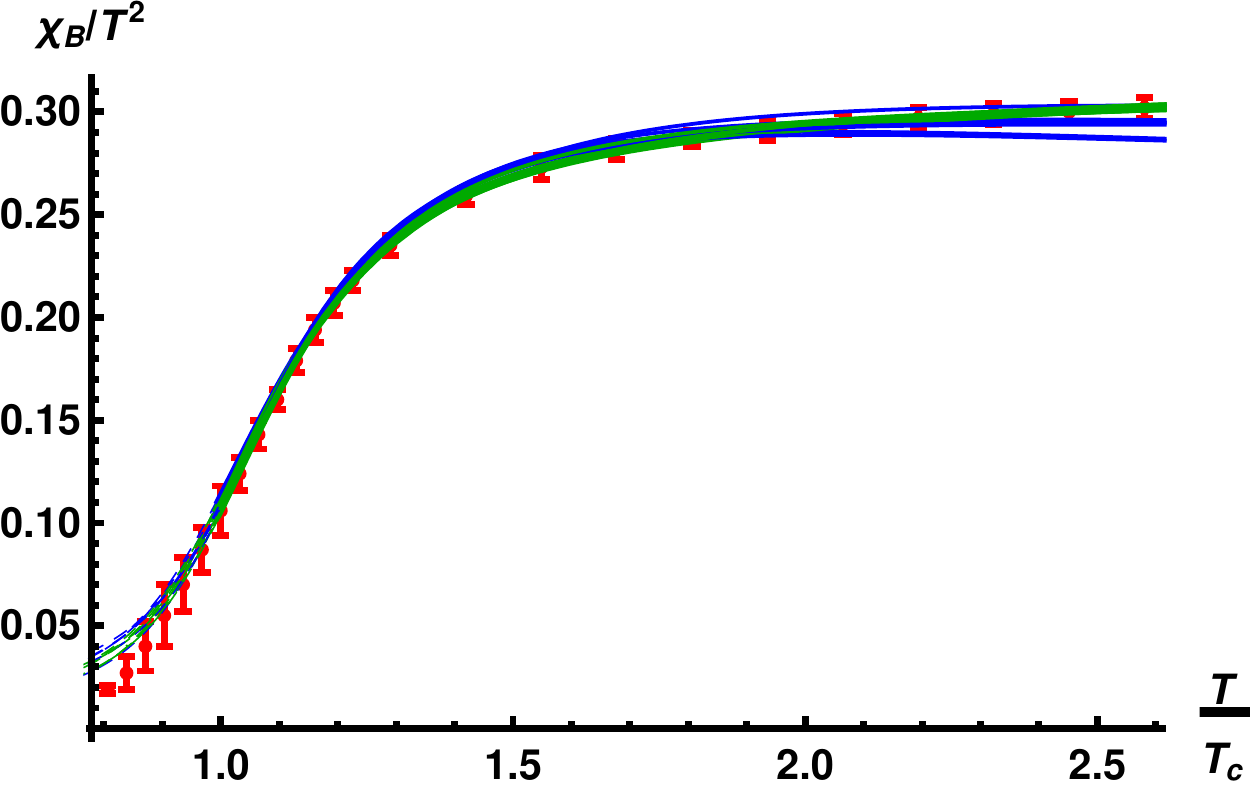}
\caption{Fitting $w(\l)$ to the QCD lattice data~\protect\cite{Borsanyi:2011sw} for the cumulant $\chi_B$. The red dots and error bars show the lattice data, and blue and green curves are our fits. Figure taken from~\protect\cite{Jokela:2018ers}.}
\label{fig:wfit}       % Give a unique label
\end{figure}

Finally, we fit the function $w(\l)$. This function is the coupling of the gauge fields, and therefore controls the thermodynamics at finite density. It is also important for the direct photon production in heavy-ion collisions~\cite{Iatrakis:2016ugz}. We fit it to the leading nontrivial cumulant of the pressure, i.e., the baryon number susceptibility:
\be
 \chi_2(T) = \chi_B(T) = \frac{d^2p}{d\mu^2}\bigg|_{\mu=0} \ .
\ee
The fits are shown in Fig.~\ref{fig:wfit} with lattice data from~\cite{Borsanyi:2011sw}. The blue (green) curves are fits with three (four) parameters. 

For the various fits shown in Figs.~\ref{fig:Vf0fit} and~\ref{fig:wfit} we have chosen a sample of three potential sets having different value of the most important parameter $W_0$. See Appendix~B. These potentials will be referred to as 5b, 7a, and 8b in the following. 

\begin{figure}
\centering  \includegraphics[width=0.48\textwidth]{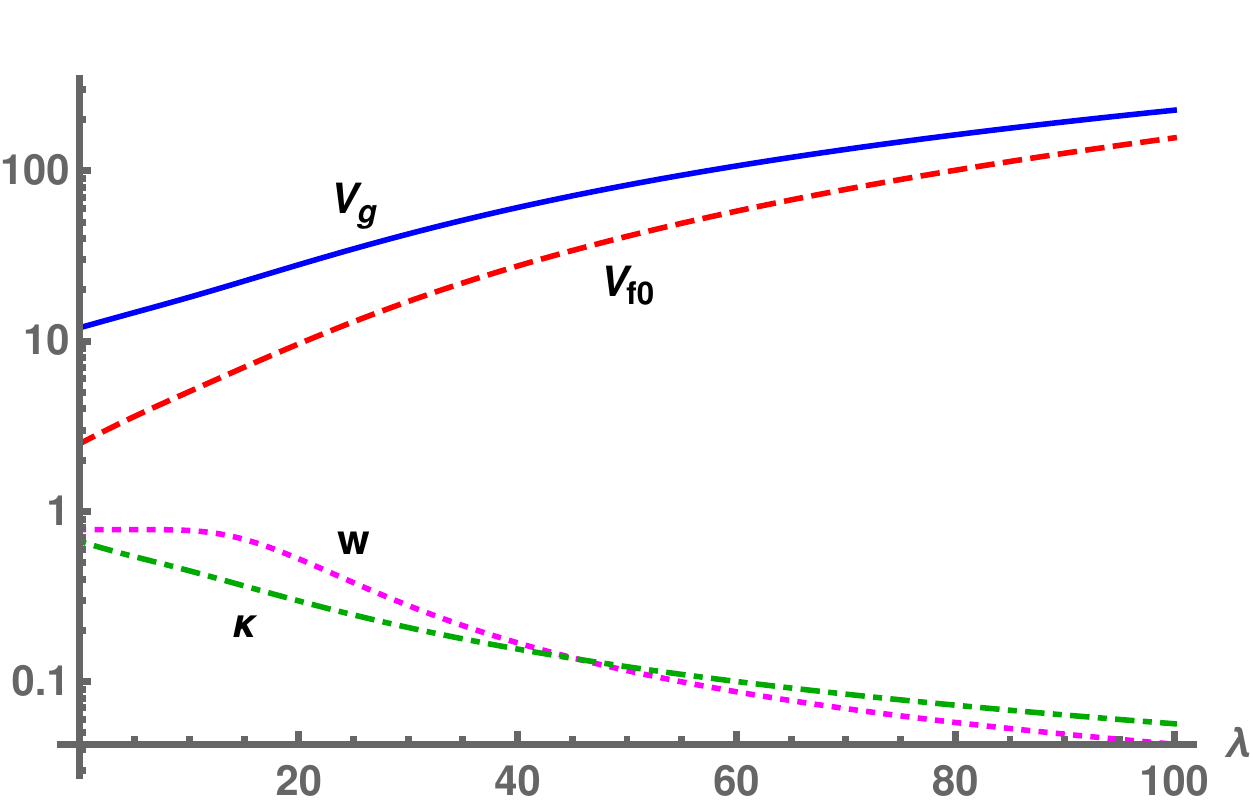}
\caption{Final potentials after the fit (potentials 7a).}
\label{fig:potentials}       % Give a unique label
\end{figure}

To conclude this section, let me comment on the quality of the fits. A priori, one might think that since the number of fitted parameters is large, we would have been able to fit any data, and the seemingly good quality of the fits is rather unremarkable. However this is not the case: our fits are very stiff, meaning that the dependence on all parameters is weak. The shape of the curves for all parameters is first and foremost a prediction of gau\-ge/gra\-vity duality, and precise match with QCD data only requires only a small tuning of the fits. The family of curves that the model is able to produce is strictly limited, and it is therefore remarkable that the lattice data happens to be among the curves that can be fitted. Moreover, we note that all the potentials resulting from the fit are simple monotonic functions, see an example in Fig.~\ref{fig:potentials}, rather than complicated ``tailored'' functions required to reproduce the data. In particular, notice that the bump of the interaction measure slightly above the critical temperature in Fig.~\ref{fig:Vf0fit} is nicely reproduced without any need of special tuning of $V_{g}$ and $V_{f0}$.

%%%%%%%%%%%%%%%%%%%%%%%%%%%%%%%%%%%%%
%%%%%%%%%%%%%%%%%%%%%%%%%%%%%%%%%%%%%
%%%%%%%%%%%%%%%%%%%%%%%%%%%%%%%%%%%%%
%%%%%%%%%%%%%%%%%%%%%%%%%%%%%%%%%%%%%
%%%%%%%%%%%%%%%%%%%%%%%%%%%%%%%%%%%%%
%%%%%%%%%%%%%%%%%%%%%%%%%%%%%%%%%%%%%
%%%%%%%%%%%%%%%%%%%%%%%%%%%%%%%%%%%%%
%%%%%%%%%%%%%%%%%%%%%%%%%%%%%%%%%%%%%
%%%%%%%%%%%%%%%%%%%%%%%%%%%%%%%%%%%%%
%%%%%%%%%%%%%%%%%%%%%%%%%%%%%%%%%%%%%
%%%%%%%%%%%%%%%%%%%%%%%%%%%%%%%%%%%%%
%%%%%%%%%%%%%%%%%%%%%%%%%%%%%%%%%%%%%
%%%%%%%%%%%%%%%%%%%%%%%%%%%%%%%%%%%%%
%%%%%%%%%%%%%%%%%%%%%%%%%%%%%%%%%%%%%
%%%%%%%%%%%%%%%%%%%%%%%%%%%%%%%%%%%%%
%%%%%%%%%%%%%%%%%%%%%%%%%%%%%%%%%%%%%
%%%%%%%%%%%%%%%%%%%%%%%%%%%%%%%%%%%%%
%%%%%%%%%%%%%%%%%%%%%%%%%%%%%%%%%%%%%
%%%%%%%%%%%%%%%%%%%%%%%%%%%%%%%%%%%%%
%%%%%%%%%%%%%%%%%%%%%%%%%%%%%%%%%%%%%
\section{Thermodynamics and dense nuclear matter in V-QCD} \label{sec:VQCDNM}
%%%%%%%%%%%%%%%%%%%%%%%%%%%%%%%%%%%%%

In this section, I will discuss the applications of the V-QCD model to the thermodynamics of hot and dense QCD. The stress will be in the implementation of nuclear matter and applications to neutron stars. But before moving to nuclear matter, it is necessary to review the standard formulation of the model at finite temperature and chemical potential, as well as the description of quark matter in the model.

%%%%%%%%%%%%%%%%%%%%%%%%%%%%%%%%%%%%%
\subsection{Dense quark matter in V-QCD} \label{ssec:QM}
%%%%%%%%%%%%%%%%%%%%%%%%%%%%%%%%%%%%%

I will now discuss the ``standard'' V-QCD setup at finite temperature and density~\cite{Alho:2012mh,Alho:2013hsa}:  
\begin{eqnarray} \label{eq:vqcdgls5}
 S_\mathrm{V-QCD}^\mathrm{bg} &=& M_\mathrm{p}^3 N_c^2 \!\int\! d^5x\,\sqrt{-\det g} \\\nonumber
 &&\times \left[R - \frac{4\left(\partial \l\right)^2}{3\l^2} + V_g(\l) \right]\\
\label{eq:vqcdfls5}
 &-& x_f M_\mathrm{p}^3 N_c^2 \!\int\! d^5x\,V_{f0}(\l)e^{-\tau^2}\\\nonumber 
 &&\times \sqrt{-\det(g_{\mu\nu}+\kappa(\l)\partial_\mu\tau \partial_\nu\tau + w(\l)\hat F_{\mu\nu})}
\end{eqnarray}
where going to finite charge and chemical potential requires turning on the temporal component of the gauge field:
\be
 A_t(r) = \Phi(r) \ , \quad F_{rt} 
 = \Phi'(r) \ , \quad \mu = \Phi(r=0)
\ee
and for the metric we use, as above,
\be
 ds^2 = e^{2A(r)}\left(\frac{dr^2}{f(r)}-f(r)dt^2 +d\mathbf{x}^2 \right) \ .
\ee
Our convention is that the UV boundary lies at $r=0$.

The first task is to identify all possible geometries appearing at various values of $T$ and $\mu$ (restricting to homogeneous and time-independent configurations). As it turns out, there are two possibilities for the geometry:
\begin{enumerate}
 \item Horizonless geometry ending at a ``good'' kind of IR singularity at $r=\infty$. We call such geometries ``thermal gas'' solutions because they are dual to the confined (hadron gas) phase. This geometry is independent of the temperature and chemical potential (the solution for the gauge field is $\Phi(r) = \mathrm{const.} = \mu$). 
 The entropy is zero (as there is no black hole).
 \item Black hole geometry with a ``planar'' horizon, i.e., a horizon at constant $r$ for all values of $t$ and $x^i$. That is, $f(r_h) = 0$ for some $r=r_h$. The temperature and entropy density are determined through black hole thermodynamics:
 \be
  T = \frac{1}{4\pi}|f'(r_h)| \ , \quad s = \frac{1}{4G_5} e^{3A(r_h)} \ ,
 \ee 
 with $M_\mathrm{p}^3N_c^2 = 1/(16\pi G_5)$. Recall that black holes are dual to deconfined phases in QCD.
\end{enumerate}
At zero quark mass both geometries come as two variants: hairless $\tau=0$ and hairy $\tau\ne 0$ solutions, corresponding to chirally symmetric backgrounds and to backgrounds with spontaneous chiral symmetry breaking, respectively. At finite quark mass there is, naturally, only chirally broken solutions with $\tau \ne 0$ because the tachyon has a nonzero source. Depending on potentials and parameter values, there may be several solutions with $\tau \ne 0$ at the same quark mass, but the solution without tachyon nodes is unique and has the lowest action~\cite{Alho:2012mh,Jarvinen:2015ofa}.

I will be mostly discussing the solutions at zero quark mass in this review. There are therefore four different phases chirally symmetric and broken thermal gas phases, and hairy and hairless black holes phases. However, for the potentials obtained in Sec.~\ref{sec:vqcd}, only two of these phases appear: the chirally broken (tachyonic) thermal gas, which is dual to the chirally broken confined phase, and the hairless, chirally symmetric black hole solution, which is dual to the chirally symmetric quark-gluon plasma phase. 

Let me then discuss the solutions at finite chemical potential more closely. The equation of motion for $\Phi$ can be integrated to give
\be
 -\frac{e^{A}V_fw^2\Phi'}{\sqrt{1\!+\!e^{-2A}f\kappa (\tau')^2\! -\! e^{-4A}w^2 (\Phi')^2}} = \mathrm{const.} \equiv \hat n
\ee
so that 
\be
 \Phi'(r) = -\frac{\hat n}{e^A w^2V_f}\sqrt{\frac{1+e^{-2A}f\kappa(\tau')^2}{1+\frac{\hat n^2}{e^{6A}w^2V_f^2}}}
\ee
Recall that here $V_f(\l,\tau) =V_{f0}(\l)e^{-\tau^2}$. The constant $\hat n$ is related to the charge density $n$ (or to be precise, the quark number density) by
\be 
 n = \frac{\hat n}{16 \pi G_5} \ .
\ee

For the thermal gas backgrounds, the only regular solution has $\hat n=0$ so that $\Phi = \mathrm{const}$. This is natural as there is no black hole to source the charge. One can show that the IR behavior with constant $\Phi$ is regular in the sense that it can be obtained from charged small black holes in the limit where their size goes to zero. The thermodynamics in the thermal gas phase is therefore trivial: the pressure is constant (which will be set to zero) and there is no matter in this phase. This is a large $N_c$ effect that I already discussed in Sec.~\ref{sec:vqcd}: the pressure in this phase comes from hadron gas, is suppressed by $1/N_c^2$ with respect to the pressure in the high temperature phase, and would arises from stringy loop corrections that are not included in the model (as they are subleading in $1/N_c$).

The charged black hole backgrounds, we require that $\Phi(r=r_h)=0$, so that 
\begin{eqnarray}
 \mu &=& -\int_0^{r_h} dr\, \Phi' \\\nonumber
 &=& \int_0^{r_h}dr\, \frac{\hat n}{e^A w^2V_f}\sqrt{\frac{1+e^{-2A}f\kappa(\tau')^2}{1+\frac{\hat n^2}{e^{6A}w^2V_f^2}}} \ .
\end{eqnarray}
That is, in this case nonzero chemical potential requires nonzero charge and vice versa. 

The phase diagram can be obtained by numerically solving the equations of motion. The easiest way is to use a shooting method starting from the IR: either from the IR singularity (in the case of thermal gas backgrounds) or the horizon (in the case of black holes). The grand potential density $\Omega = -p$ can be in principle computed from the (regularized) on-shell action. However, in practice it turns out to be easier to integrate the first law of thermodynamics:
\be
 d\Omega  = -s dT - nd\mu \ .
\ee
Since, as I pointed out above, the thermal gas solutions can be obtained as a limit from the black hole solutions, the first law is enough to determine the differences between the grand potentials in various phases, which is all that is needed for the phase diagram. As the grand potential (or equivalently pressure) in the thermal gas phase is constant, it is convenient to normalize the results such that this constant is zero. This also agrees with the usual conventions on QCD side.

\begin{figure}
\centering  \includegraphics[width=0.48\textwidth]{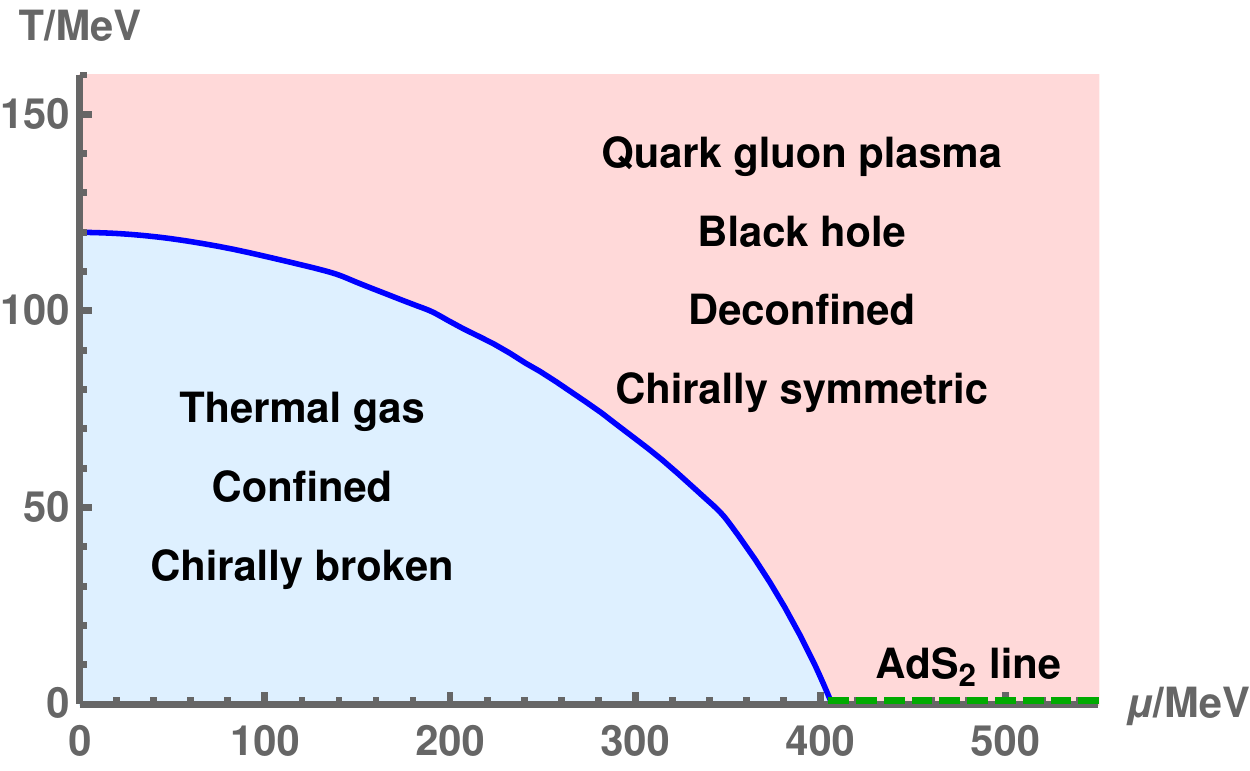}
\caption{The phase diagram for a set of potentials fitted to QCD data (potentials 7a).}
\label{fig:7aphase}       % Give a unique label
\end{figure}

The resulting phase diagram for a choice of potentials, constructed as outlined in Sec.~\ref{sec:vqcd}, is shown in Fig.~\ref{fig:7aphase}.\footnote{This diagram differs from that of~\protect\cite{Alho:2013hsa} because different potentials were used.} I remind that the transition temperature $T \approx 120$~MeV at $\mu=0$ differs from the crossover temperature $\approx$ 155~MeV of QCD due to the missing pressure contribution from meson loops in the confined phase. 
Moreover, there is not yet nuclear matter in this diagram, it will be considered below.

There is one more interesting feature which deserves attention. Namely, the zero temperature limit of the deconfined phase is a ``quantum critical'' line realized through an AdS$_2$ geometry~\cite{Alho:2013hsa}. The AdS$_2$ geometry is given by 
\be \label{eq:ads2geom}
  ds^2 =\frac{dr^2}{c_f(r_0-r)^2} - c_f \hat \Lambda^4 (r_0-r)^2 dt^2 + \hat\Lambda^2 d\mathbf{x}^2
\ee
where $\hat \Lambda$ is an energy scale and $c_f$ is a constant which can be computed from the action. At $T=0$ exactly, the flows ending in the AdS$_2$ geometry are realized through corrections to~\eqref{eq:ads2geom} in a transseries form, i.e., in powers of $r_0-r$ and $(r_0-r)^\alpha$ where $\alpha$ is a positive coefficients which can be expressed in term of the potentials and their derivatives (see~\cite{Hoyos:2021njg} for details).

An important point about the AdS$_2$ geometry is that it has nonvanishing entropy. Consequently, entropy remains nonzero as $T \to 0$ in the quark matter phase. This may signal that this phase is not the true vacuum in this region, but should be replaced by a different background that could be dual to a color superconducting or color-flavor locked phase. Indeed it is know that the AdS$_2$ geometry can lead to instabilities. Studying these is a topic of ongoing research.

%%%%%%%%%%%%%%%%%%%%%%%%%%%%%%%%%%%%%
\subsubsection{Equation of state in quark matter phase}
%%%%%%%%%%%%%%%%%%%%%%%%%%%%%%%%%%%%%

I then turn my attention to the equation of state. I described in Sec.~\ref{ssec:latticefit} how the model is fitted to precisely agree with the thermodynamics at small chemical potentials obtained from lattice simulations of QCD with $N_f=2+1$ flavors. Therefore agreement with the QCD equation of state is guaranteed in the region where lattice data is available, but it is interesting to see how well extrapolations of the equation of state based on holography work at higher values of chemical potentials, up to the region relevant for neutron stars.

Extrapolations of lattice data to finite chemical potential by using holography have been carried out in the literature by using the Einstein-Maxwell-dilaton models~\cite{DeWolfe:2010he,DeWolfe:2011ts,Knaute:2017opk,Critelli:2017oub,Grefa:2021qvt}. In these studies, main topics have been locating the QCD critical point and studying the properties of the plasma near the critical point. In this review, I am however mainly interested in another region, namely the region of low temperatures and high densities, which is relevant for neutron star cores.

\begin{figure}
\centering  \includegraphics[width=0.48\textwidth]{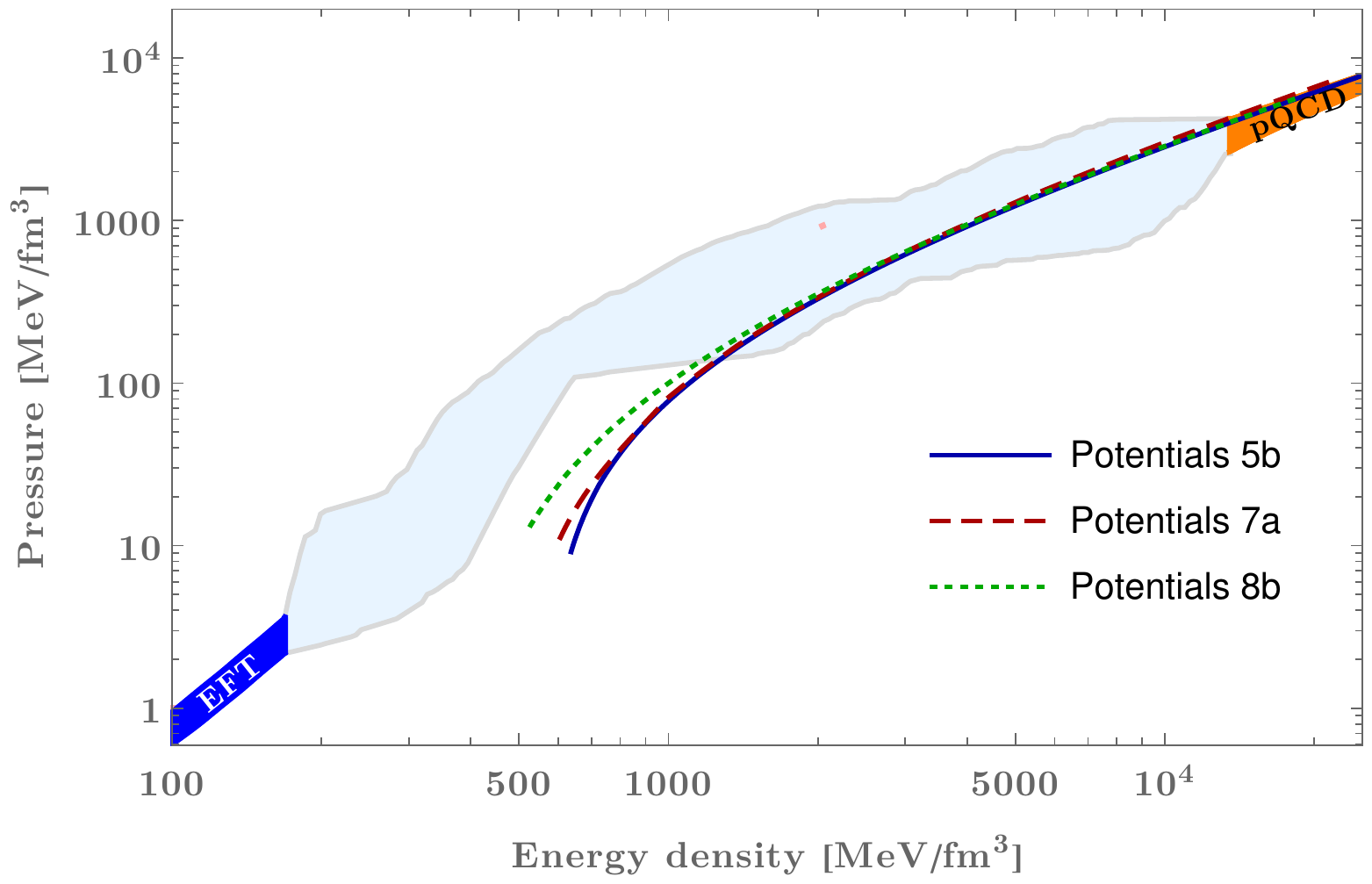}
\caption{The quark matter EOS from the V-QCD model for a selection of potentials compared to the band of allowed EOSs from polytropic interpolations.}
\label{fig:QMEOS}       % Give a unique label
\end{figure}

The equation of state for dense and cold V-QCD quark matter can be extracted by using the numerical techniques which were sketched above. I show the results for the three potentials of Appendix~B in Fig.~\ref{fig:QMEOS}. The EOSs from all these potentials agree with the polytropic interpolations (see Sec.~\ref{sec:qcd}) at high density, i.e., they stay within the blue band spanned by phenomenologically viable polytropic EOSs. At low densities the curves exit the band: this is not a problem but expected since at low density the thermodynamically dominant phase should have nuclear matter, which we have not considered yet. Therefore the quark matter EOS is feasible for essentially all values of temperature and chemical potential (apart from subleading disagreement with perturbation theory at asymptotically high $T$ and $\mu$). At this point it can be already used to make predictions of the whole phase diagram when combined with other EOS for the nuclear matter side. Such analysis was carried out at zero temperature in~\cite{Jokela:2018ers} and at finite temperature in~\cite{Chesler:2019osn}.

%%%%%%%%%%%%%%%%%%%%%%%%%%%%%%%%%%%%%
\subsubsection{Magnetic field and anisotropy}
%%%%%%%%%%%%%%%%%%%%%%%%%%%%%%%%%%%%%

Before going to the discussion of nuclear matter, let me briefly comment on two other applications of the model, which may also be also be extended for nuclear matter in the future. The first is the effect of adding external (homogeneous) magnetic field and anisotropy in the quark gluon plasma~\cite{Drwenski:2015sha,Demircik:2016nhr,Gursoy:2016ofp,Gursoy:2017wzz,Gursoy:2018ydr,Gursoy:2020kjd}. See also the review~\cite{Gursoy:2021efc}. The study is motivated in part by non-central heavy ion collisions, where the collision creates a strong magnetic field perpendicular to the beam and impact parameter, and the plasma is anisotropic since it expands faster in the beam direction, and there is also anisotropy in the transverse plane due to the collision being off-central or if Uranium ions are used. Another motivation is to understand the so-called ``inverse magnetic catalysis'': Normally, the chiral condensate in QCD is expected to be enhanced with increasing magnetic field in QCD, which is a well-understood an model-independent result. Lattice studies however show that near the critical deconfinement temperature of QCD, increasing magnetic field suppresses the condensate~\cite{Bali:2011qj,Bali:2012zg,DElia:2012ems}. This unexpected phenomenon is the inverse magnetic catalysis, and it remains poorly understood.

Choosing the magnetic field to lie in the $x_3$ direction, it can be added in the model simply by turning on the corresponding bulk gauge field
\be
 F_{12}= - F_{21} = B
\ee
with no dependence on the holographic coordinate. An anisotropy may be imposed by adding an ``external'' axion field $a$ with action as in~\eqref{eq:Saglue}, and considering a linear background solution $a(x) = a_\perp x_2$ or $a(x) = a_\parallel x_3$ depending on the orientation of the anisotropy. Such linear solutions create an anisotropy without introducing $x$-dependence in the EOMs~\cite{Mateos:2011ix,Mateos:2011tv}.

Some of the main results from analyzing such configurations in V-QCD were the following:
\begin{itemize}
 \item We showed~\cite{Gursoy:2016ofp} that the V-QCD models are able to produce (magnetic and) inverse magnetic catalysis in good agreement with lattice data, and demonstrated in the holographic model that backreaction of the flavors to the gluon dynamics is important for this phenomenon, again in agreement with lattice results~\cite{Bruckmann:2013oba}. 
 \item We predicted that the inverse catalysis is weakened with increasing chemical potential and absent for high chemical potentials~\cite{Gursoy:2017wzz}.
 \item We showed that turning on an anisotropy (while setting the magnetic field to zero) creates a similar effect as the magnetic field, i.e., we confirmed the ``inverse anisotropic catalysis''~\cite{Gursoy:2018ydr} that was conjectured in~\cite{Giataganas:2017koz}. This suggests that the inverse magnetic catalysis is due to the anisotropy created by the magnetic field rather than direct interaction with the field. We demonstrated that turning on the magnetic field and the anisotropy at the same time supports this idea, leading to expected strong interference (only) when the anisotropies created by the magnetic field and the axion are in the same direction~\cite{Gursoy:2020kjd}.
 \end{itemize}
 
The geometry of the model encodes the parameter dependence in an interesting way. Turning on any nonzero $a_\perp$ or $a_\parallel$ causes the vacuum IR geometry to change from that of~\eqref{eq:IRA} to AdS$_4 \times \mathbb{R}$. Turning on a strong magnetic field gives rise to an approximate intermediate geometry AdS$_3 \times \mathbb{R}^2$ along the holographic RG flow. The changes in the geometry can be probed by analyzing the quark-antiquark potential and the entanglement entropy for a stripe for various values of the parameters~\cite{Gursoy:2020kjd}.

Our results complement those obtained in the literature in other models, see, e.g.,~\cite{Mamo:2015dea,Evans:2016jzo,Dudal:2015wfn,Rougemont:2015oea,Bohra:2019ebj,Arefeva:2020vae}. 
 
\begin{figure*}
\centering  \includegraphics[width=0.49\textwidth]{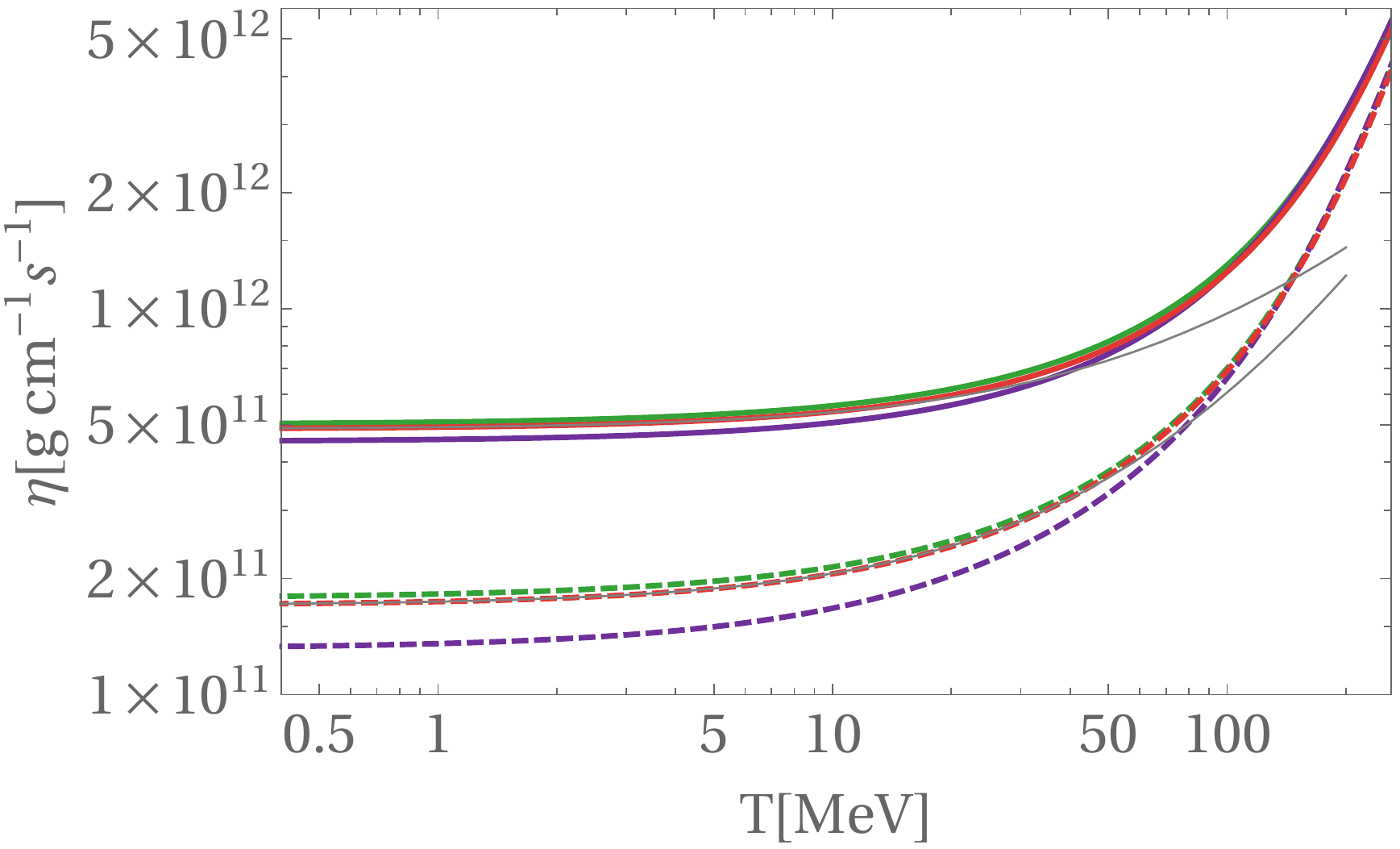}%
\hspace{3mm}  \includegraphics[width=0.47\textwidth]{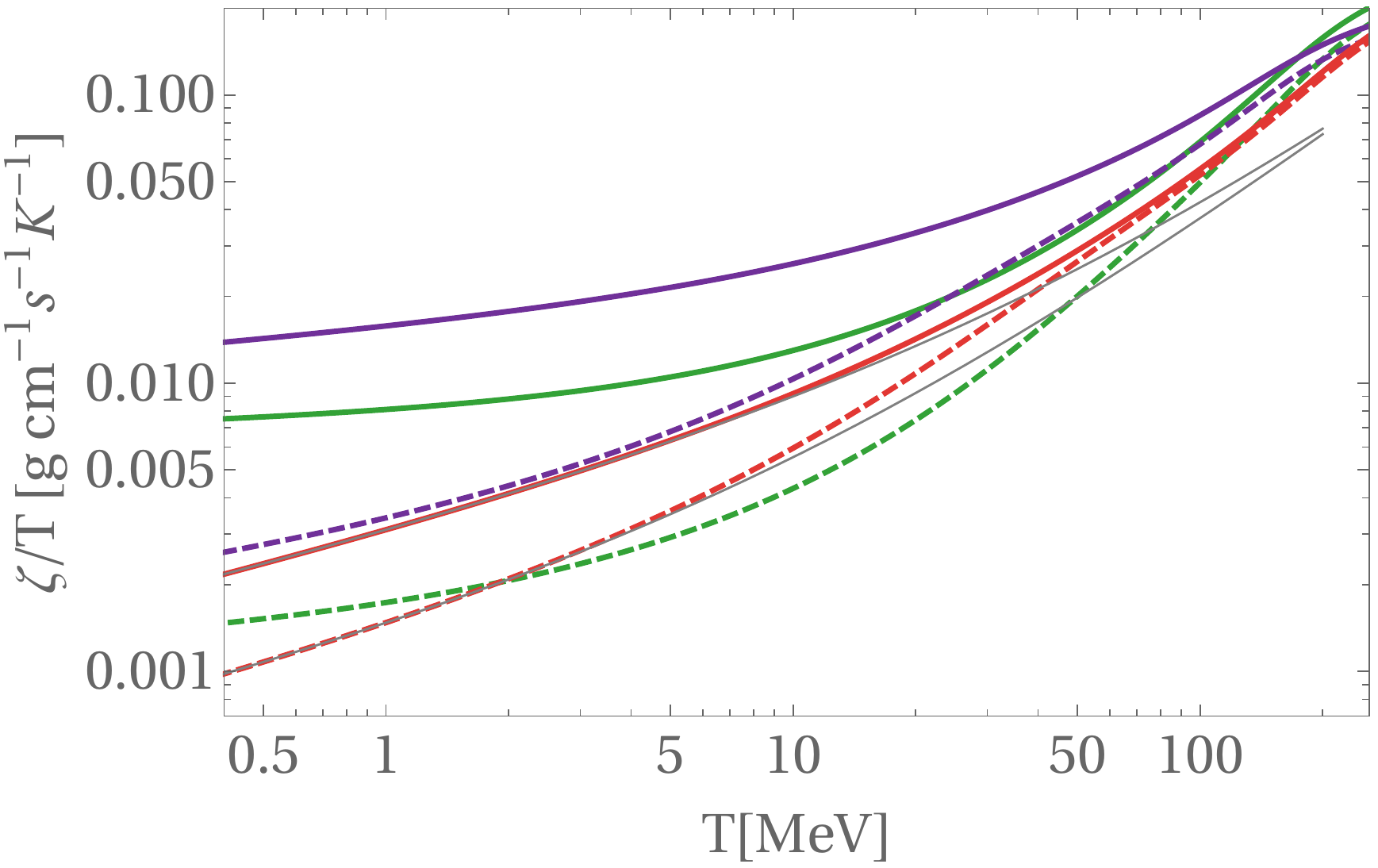}
\caption{The shear viscosity (left) and bulk viscosity divided by temperature (right) as a function of temperature. Dashed (solid) curves are for $\mu_B=450$~MeV ($\mu_B=600$~MeV). Green, red, and violet curves use potentials 5b, 7a, and 8b, respectively. The thin gray curves are given by transseries fits to the 7a data. }
\label{fig:viscosities}       % Give a unique label
\end{figure*}

%%%%%%%%%%%%%%%%%%%%%%%%%%%%%%%%%%%%%
\subsubsection{Transport in quark matter phase}
%%%%%%%%%%%%%%%%%%%%%%%%%%%%%%%%%%%%%

The second application is transport in quark matter. By using the holographic dictionary, it is straightforward to analyze (among other things) the viscosities and conductivities in the high density phase, see~\cite{Hoyos:2020hmq,Hoyos:2021njg} where the results for the D3-D7 model were also analyzed. (For analysis of transport and thermalization at zero density, see~\cite{Gursoy:2009kk,Gursoy:2010aa,Iatrakis:2014txa,Ishii:2015gia,Alho:2020gwl}). For example, the shear viscosity satisfies the standard relation $\eta = s/4\pi$, and the bulk viscosity may be found by employing the method of Eling and Oz~\cite{Eling:2011ms}. 

I present the result for two values of chemical potential and low values of the temperature in the quark matter phase (with zero quark mass so that chiral symmetry is intact) in Fig.~\ref{fig:viscosities}. Different colors present the three different choices of potentials given in Appendix~B. The finite value of $\eta$ as $T \to 0$ reflects the fact that the entropy is finite as I pointed out above. That is, these results are for the unpaired quark matter phase, and for very low values of $T$ one expects pairing which is not (yet) included in the model. The plots also include the thin gray curves which are four parameter fits to the potential 7a data by transseries, i.e., (integer) powers of $T$ and $T^\alpha$, with $\alpha$ being the parameter defined through the asymptotic flow around the AdS$_2$ point I discussed above. The convergence of the fit confirms the expectation from the asymptotic analysis around the AdS$_2$ point that the low temperature asymptotics is indeed a transseries.
 
The behavior of the transport coefficients in V-QCD is totally different from the predictions of perturbative QCD. For the shear viscosity, for example, the leading order perturbative QCD result is known~\cite{Heiselberg:1993cr} and it is larger than the V-QCD curves by several orders of magnitude at small temperatures, decreases rapidly with temperature, and is smaller then the V-QCD prediction by orders of magnitude at temperatures around the QCD scale. This night not be particularly surprising, since the plotted region is far from the region where perturbative analysis is reliable. The complete lack of agreement however demonstrates that even qualitative features of the perturbative results are not to be trusted. Let me also point out that the viscosities computed here are the strongly interacting contributions only. For a realistic matter in neutron stars there are additional contributions, among other things, from the electron gas and weak interactions of the matter. For the bulk viscosity, the latter are dominant over the contribution from strong interaction by orders of magnitude (see, e.g.,~\cite{Schmitt:2017efp}).

%%%%%%%%%%%%%%%%%%%%%%%%%%%%%%%%%%%%%
\subsection{Nuclear matter from a homogeneous bulk field}
%%%%%%%%%%%%%%%%%%%%%%%%%%%%%%%%%%%%%

Having established the background, I move to the main topic of this section: the implementation of nuclear matter in V-QCD. As I have reviewed above, there are several ways to attack this problem. Here I will only consider the approach where nuclear matter is modeled through a homogeneous bulk field~\cite{Ishii:2019gta}, which was reviewed in Sec.~\ref{ssec:homog}. This is due to two main reasons: First, we will be interested in neutron stars and their cores where the density of nuclear matter is high, so that treating it as homogeneous matter rather than individual baryons appears natural. Second, the homogeneous approach is much simpler than even computing the solution for a single baryon in the model. And to obtain a reasonable model at high density, one needs to consider an ensemble of baryons with complicated interactions which would be extremely challenging. I will review here the main points of the analysis, see~\cite{Ishii:2019gta} for more details.

\subsubsection{The probe approximation}

The approach for homogeneous nuclear matter follows the ideas outlined in Sec.~\ref{ssec:homog}. I will consider adding nuclear matter in the confined chirally broken phase of V-QCD, i.e., the light blue phase of Fig.~\ref{fig:7aphase}.
In order to simplify the treatment further, I will also work in the probe approximation, i.e., expanding the DBI action to first nontrivial order in the non-Abelian gauge fields and neglecting the backreaction of the baryons to the metric. Replacing the DBI action by a quadratic action may sound like a drastic approximation, but is actually well motivated. First, as I argued in Sec.~\ref{sec:baryons}, the higher order corrections in the gauge fields are suppressed for individual baryons at large coupling in the WSS model. Second, the BPST instanton is self dual, meaning $*F = \pm F$ in terms of differential forms where $*$ is the Hodge dual, and for such self-dual configurations the square root of the determinant in the DBI action can in simple cases be computed explicitly and one finds that the leading order expansion of the action is exact~\cite{Hashimoto:1997px,Brecher:1998su}. In the case of V-QCD the presence of the tachyon however adds an extra complication, and the homogeneous configuration may not follow the properties of individual solitons. However, these observations suggest that dealing with the full non-Abelian DBI action is not worth the effort, since the approach involves rather drastic approximations in any case.

At this point, it is useful to explicitly separate the vectorial Abelian term $\widehat\Phi$ from the non-Abelian terms. In order to do this, we replace, with slight abuse of notation, 
\be A_{L/R}(x_M) \mapsto \widehat \Phi(r) \mathbb{I} dt + A_{L/R}(x_M) \ , \label{eq:AbnonAb}
\ee
where $x_M$ stands for dependence on all coordinates. 
Here, anticipating that the configuration will be homogeneous, we also assumed that $\widehat\Phi$ only depends on the holographic coordinate. Actually, in order for the division of the field into the Abelian and non-Abelian terms to be well-defined, we require
\be
  \int d^4x\, \mathrm{tr}\left(F^{(L)}_{rt}+F^{(R)}_{rt}\right) = 0 \ ,
\ee
i.e., that the averaged Abelian vectorial component of the field strengths for the latter term in~\eqref{eq:AbnonAb} vanish for all values of $r$.
After this, developing the DBI action~\eqref{eq:STDBI} to first nontrivial order in the field strengths $F^{(L/R)}$ and the gauge fields $A_{L/R}$ appearing explicitly through the covariant derivatives of the tachyon yields
\begin{eqnarray} \label{eq:SDBIfinal}
  S_\mathrm{DBI} &=& - M^3 N_c  \int d^5x\, V_{f0}(\l) e^{-\tau^2}
\sqrt{-\det g}\sqrt{R}\\\nonumber
&&\  \times\Bigg[
%\frac{1}{2}R^{-1} e^{-4A} w(\l)^2\Phi'\,\mathrm{tr}\left(F^{(L)}_{rt}+F^{(R)}_{rt}\right)
1+\frac{\kappa(\l)\tau^2}{2}\left( \tilde g^{-1}\right)_s^{MN}\,\mathrm{tr} A_MA_N\\\nonumber
&&\ \ \ \  -\frac{w(\l)^2}{8}\left( \tilde g^{-1}\right)_s^{MN}\left( \tilde g^{-1}\right)_s^{PQ}\\\nonumber
&&\ \ \ \ \times\mathrm{tr}\left(F^{(L)}_{NP}F^{(L)}_{QM}+F^{(R)}_{NP}F^{(R)}_{QM}\right) \Bigg] \,.
\end{eqnarray}
Recall that $T =\tau(r) \mathbb{I}$. We also used a shorthand notation for the expression
\be
 R =1+ e^{-2A}f\kappa(\l)(\tau')^2-e^{-4A}w(\l)^2(\widehat\Phi')^2
\ee
and for the symmetric part of the inverse of the effective TDBI metric
\begin{eqnarray}
 (\tilde g^{-1})_s &=& e^{-2A}\,\mathrm{diag}\big(f R^{-1},\\\nonumber
 && \ \ \ -f^{-1}R^{-1}(1+e^{-2A}f\kappa(\l)(\tau')^2),1,1,1\big)
\end{eqnarray}
with the order of the components $(r,t,x_1,x_2,x_3)$. Notice that here we only developed the non-Abelian terms of the DBI, and kept nonlinear action for the Abelian component $\widehat\Phi'$, which appears in $R$. In principle $\widehat\Phi'$ will be of the same size as the non-Abelian components which suggest that we should also expand the dependence on it as a series. I however think that the full expression of~\eqref{eq:SDBIfinal} is a better approximation than its leading series expansion around $\widehat\Phi'=0$. The arguments on the dependence of the DBI on the soliton discussed above suggest that only the non-Abelian components should be treated as small. Moreover keeping the full dependence on $\widehat\Phi'$ does not violate any principle, and the action in~\eqref{eq:SDBIfinal} is quite compact while the expanded version would be messy.

The Chern-Simons term of~\eqref{eq:CS5} is also important for nuclear matter, because, as we shall see explicitly, it couples the winding number of the soliton to the baryon number (as it is obtained from the dictionary). We use here the expression from~\cite{Casero:2007ae}. There is however, an issue: using their formula as such does not quite work for us. In brief, the details of the complicated CS five form $\Omega_5$ are not important, but it involves a tachyon potential $\propto e^{-\tau^2}$, similar to that appearing in the DBI sector of~\eqref{eq:vqcdfls5} and of~\eqref{eq:SDBIfinal}. As it turns out, for consistent nuclear matter solution the potential of the CS term should vanish faster than the DBI counterpart for large $\tau$. This discrepancy is not a problem since the derivation of this reference involves some assumptions which are satisfied for at least superconformal backgrounds but not necessarily for the V-QCD model. Presumably we should stick to the bottom-up framework also here and treat the potentials of the CS terms as ``free parameters''. Here we will do what is apparently the simplest modification to fix this and replace $\tau \mapsto \sqrt{b} \tau$ with $b>1$ in the CS term, which guarantees the nice IR behavior.

The expression for the baryon number can be computed by following the dictionary. By using the EOM for the Abelian field $\widehat \Phi$, the variation of the on-shell action of the model is given by\footnote{To be precise one should consider here a more general variation of the action since in general changing $\mu$ affects all fields through couplings in the EOMs. However one can show that the other fields do not contribute to the variation of the action unless one also varies their sources, in consistency with the first law of thermodynamics.}
\be 
 \delta S = -\delta\widehat \Phi(r) \frac{\delta S}{\delta \widehat\Phi'(r)}\bigg|_{r=0} = -\delta \mu  \frac{\delta S}{\delta \widehat\Phi'(r)}\bigg|_{r=0}
\ee
where we assumed that the setup is consistent such that no contributions from IR boundary arise. The baryon number is identified as 
\be \label{eq:NBh}
 N_B = -\frac{1}{N_c} \int d^3x\,\frac{\partial \mathcal{L}}{\partial \widehat\Phi'}\bigg|_{r=0} \!= \frac{1}{N_c} \int d^3xdr \frac{\partial  \mathcal{L}_\mathrm{CS}}{\partial \widehat\Phi}
\ee
where the $1/N_c$ factor arises because the source for $\widehat \Phi$ is the quark (rather than baryon) chemical potential, $\mathcal{L}$ is the Lagrangian density of the full action and $\mathcal{L}_\mathrm{CS}$ is the Lagrangian density for the CS term. In order to obtain the last expression in~\eqref{eq:NBh}, we used the $\widehat \Phi$ EOM as well as the fact that only the CS action depends on $\widehat \Phi$ directly whereas other terms depend only on the derivative $\widehat \Phi'(r)$. Inserting here the expression of $\Omega_5$ from~\cite{Casero:2007ae} we obtain
\be \label{eq:NBH4}
 N_B = \frac{1}{24\pi^2} \int d^3x dr H_4
\ee
where
\begin{eqnarray}
H_4 &=&\mathrm{tr}\, d\Big[e^{-b\tau ^2} \big(-3 i A_L\wedge F^{(L)}+3 i A_R\wedge F^{(R)}\\
&&+A_L\wedge A_L\wedge A_L-A_R\wedge A_R\wedge A_R \nonumber\\
  &&\ +b\tau ^2 (A_L-A_R)\wedge (A_L-A_R)\wedge (A_L-A_R)\nonumber\\
  &&+3b i \tau  d\tau\wedge (A_L\wedge A_R-A_R\wedge A_L)\nonumber\\\nonumber
   &&\ -2 ib^2 \tau ^3 d\tau\wedge (A_L\wedge A_R-A_R\wedge A_L)\big)\Big] \,.
\end{eqnarray}

\subsubsection{Homogeneous Ansatz}

We are now ready to insert the homogeneous Ansatz of nuclear matter. As usual with this Ansatz, we restrict to the case of two light flavors, $N_f=2$. This appears to conflict with taking the Veneziano limit, but notice that this was already in practice broken when we decided to treat the nuclear matter as a probe, so taking $N_f=2$ does not lead to any major additional limitations. 

For the case with explicit left and right handed fields, the Ansatz is given by 
\be
 A_L^i = -A_R^i = h(r) \sigma^i \ .
\ee
One can consider higher $N_f$, but this is does not work as nicely as $N_f=2$ because the link between rotations and (full) flavor symmetry is lost. An approach which would still make use of the link would be to consider SU$(2)$ subsectors of the full symmetry, e.g., only a single subgroup or using a block diagonal Ansatz consisting of several SU$(2)$'s (which is most symmetric if $N_f$ is even), but this is not the most general homogeneous Ansatz.

This Ansatz leads to the same issue as in the WSS model discussed above: the baryon number obtained from~\eqref{eq:NBH4} reads
\begin{eqnarray}
 N_B &=& - \frac{2}{\pi^2} \int d^3xdr \\\nonumber
 &&\qquad \times \frac{d}{dr}\left[ e^{- b\, \tau(r)^2}h(r)^3(1-2b\, \tau(r)^2)\right]  \ .
\end{eqnarray}
It is a total derivative (as it should) and will integrate to zero if $h(r)$ is smooth because $h(r)$ must vanish at the boundary in order to avoid turning on sources in the field theory. To cure this in our case we introduce a discontinuity at a generic point $r=r_c$ in the bulk. Actually, we will choose a solution such that $h(r)=0$ for $r>r_c$. The choice in the IR for this field is not unique (there are also two other choices with nonzero $h(r)$) but our results will be insensitive to these choices.

The value of $r_c$ is found by minimizing the action. The interpretation is similar to the case of WSS discussed in Sec.~\ref{ssec:homog}: The soliton centers of the true minimum configuration are expected to lie around $r=r_c$. Away from the centers, at high density, even the true minimum is presumably close to homogeneous. The approximation is therefore drastic only near $r=r_c$.

After inserting the Ansatz, the action reads
\begin{eqnarray} \label{eq:Shdef}
 S_h &=& S_\mathrm{DBI}+S_\mathrm{CS} \\\nonumber
 &=& - 2 M_\mathrm{p}^3 N_c \int d^5x\,V_{f0}(\l) e^{-\tau^2} e^{5A} \sqrt{R}\nonumber\\
 &&\times \bigg[1+6\kappa(\l)\tau^2e^{-2A}h^2+6w(\l)^2e^{-4A}h^4\nonumber\\
 &&\ \ \ +\frac{3}{2} w(\l)^2e^{-4A} f R^{-1} \left(h'\right)^2\bigg] \\
 &&- \frac{2N_c}{\pi^2} \int d^5x\, \widehat\Phi  \frac{d}{dr}\left[ e^{-b\,\tau^2}h^3(1-2b\,\tau^2)\right] \ .
\end{eqnarray}
This actions appears to have a bad singularity at $r=r_c$ since the discontinuity of $h$ leads to a delta function in $h'$. We adopt a prescription where we ignore all localized contributions to the action at $r=r_c$. Near this point our approximation is expected to fail, so we think it is better to ignore the contributions from this region than trying to estimate them within an approach that is dubious. This is however likely to lead to an underestimate of the value of the DBI action.

One can argue that minimizing the action leads to a finite $r_c$ (in the confining phase) as follows. Solving the EOM in terms of the bulk charge 
\be
 \rho = -\frac{\partial \mathcal{L}}{\partial \widehat \Phi'}
\ee
gives 
\be \label{eq:rhosol}
\rho = \left\{\begin{array}{lr}
               \rho_0  + \frac{2}{\pi^2} e^{-b\,\tau^2}h^3(1-2b\,\tau^2) \, , \qquad&(r<r_c) \\
                \frac{2}{\pi^2} e^{-b\,\tau^2}h^3(1-2b\,\tau^2) \, , \qquad&(r>r_c)
              \end{array}\right.
\ee
where 
\begin{eqnarray}
\rho_0 &=&\rho(r=0)  \\\nonumber
 &=& \frac{2}{\pi^2} e^{-b\,\tau(r_c)^2}(1-2b\,\tau(r_c)^2)\left[h(r_c\!+\!\epsilon)^3-h(r_c\!-\!\epsilon)^3\right]
\end{eqnarray}
is the baryon number density. At fixed density we therefore see that the discontinuity of $h(r)$ is roughly proportional to $e^{b\,\tau(r_c)^2}$, which gives a rough lower bound for the size of $h$ near $r=r_c$. If $r_c \to 0$, due to the UV asymptotics~\eqref{eq:tauUV} we have $\tau(r_c) \to 0$. Therefore one finds that $h$ is roughly constant, in which case the action~\eqref{eq:Shdef} diverges due to the factors $e^A\sim 1/r$ arising from the metric. In the IR, for confining backgrounds the tachyon diverges, see~\eqref{eq:tauIR}, and the exponential $\exp(b\tau^2)$ blows up fast. Consequently $h$ will grow exponentially if we take $r_c \to \infty$, and the $~h^4$ term in the action will cause it to diverge. since the action diverges (with the same sign) at both $r_c \to 0$ and at $r_c \to \infty$, its extremum is indeed at finite positive $r_c$. Let me also remark that it is the coupling of the nuclear matter to the tachyon which prevents it from falling in the IR. There is a rough similarity to the WSS model: In that case, the baryon is stopped from falling to the IR by the flavor branes, as is seen clearest in the non-antipodal case for which the flavor branes do not extend down to the tip of the geometry (see~\cite{Aharony:2006da,Horigome:2006xu,Li:2015uea}). The IR fusion of the flavor branes in the WSS model corresponds to the blow up of the tachyon in V-QCD as I discussed in Sec.~\ref{sec:vqcd}, which stops the baryons from falling in the IR in V-QCD.

\subsubsection{Complete model and phase diagram}

We are now ready to write down the complete V-QCD model for dense QCD which contains both (homogeneous) nuclear and quark matter. To avoid double counting we first subtract the background term 
\begin{eqnarray}
 S_{h0} &=& - M^3 N_c  \int d^5x\, V_{f0}(\l) e^{-\tau^2} \\\nonumber
&&\qquad \times \sqrt{-\det g}\sqrt{1+e^{-2A}f\kappa(\l)(\tau')^2} 
\end{eqnarray}
from the nuclear matter action. The full action then reads
\be
 S_\mathrm{V-QCD} =  S_\mathrm{V-QCD}^\mathrm{bg} + c_b(S_h-S_{h0}) \ ,
\ee
with $S_\mathrm{V-QCD}^\mathrm{bg}$ given in~\eqref{eq:vqcdgls5} and~\eqref{eq:vqcdfls5}. The nuclear matter term is treated in the probe limit (only) in the sense the backreaction to the metric is neglected, and the metric is solely solved from the first term. That is, we do not require that the second term is small: it may  in principle contribute to the free energy at the same level as the first term even though typically the contribution of the second term is numerically suppressed. We also included~\cite{Ecker:2019xrw,Jokela:2020piw} the coefficient $c_b$, which will be determined by comparing to data, for two reasons: First, for the quark matter contribution (arising from $S_\mathrm{V-QCD}^\mathrm{bg}$) we will in effect set $N_c=N_f=3$, i.e., consider also the strange quarks, which makes sense as they are active for the temperatures and chemical potentials of the quark matter phase. But for the nuclear matter we used $N_f=2$ to avoid extra complications, so in order to scale the nuclear matter result to a higher number of active quarks, one can simply multiply by the factor $c_b$. Second, the implementation with the homogeneous Ansatz is a somewhat drastic approximation, in particular near $r=r_c$, which may underestimate the action for nuclear matter. So $c_b$ may also correct issues due to the approximation.

Finally, I have some technical remarks. The U$(1)$ gauge fields $\Phi$ in $S_\mathrm{V-QCD}^\mathrm{bg}$ and $\widehat\Phi$ in $S_h$ are actually the same field, but I used different notation because in the case of nuclear matter we neglect the backreaction of the charge to the metric. Notice that they are not turned on simultaneously, but $\Phi$ ($\widehat\Phi$) is nonzero only in quark (nuclear) matter phase so our setup is consistent. Also, there is a constant of integration which we have not specified in the nuclear matter phase. Namely $\widehat\Phi$ is obtained from $\rho$ of~\eqref{eq:rhosol} by direct integration, but determining the constant of integration is somewhat involved (see~\cite{Ishii:2019gta} for details). We actually first Legendre transform the probe action from the grand canonical to the canonical ensemble where the action is a function of $\rho$, minimize the action in this ensemble, and transform back to the grand canonical which ``automatically'' fixes the constant of integration such the it is consistent with the first law of thermodynamics.

\begin{figure}
\centering  \includegraphics[width=0.48\textwidth]{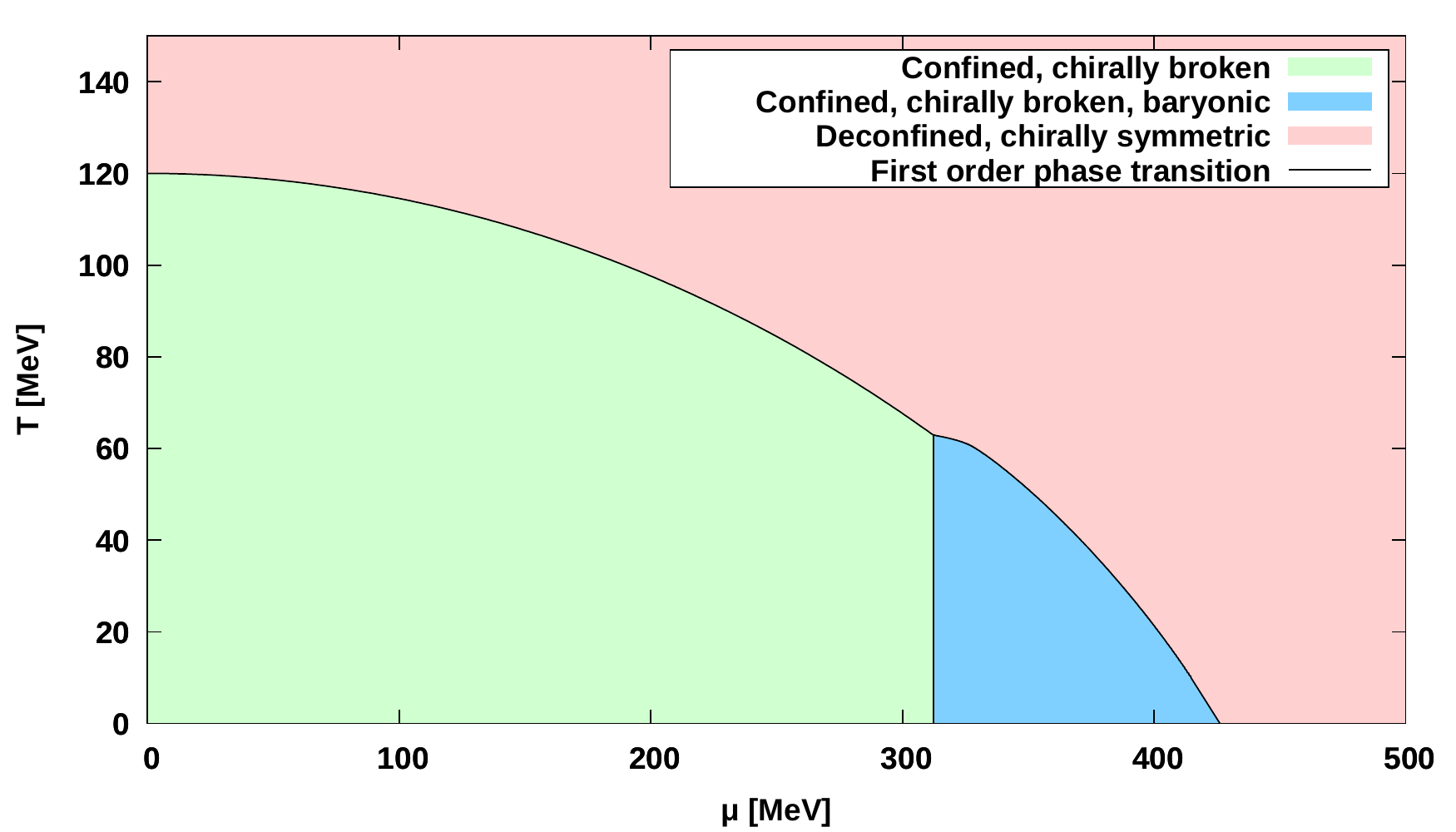}
\caption{The phase diagram including nuclear matter (blue region) for a choice of potentials fitted to lattice data (potentials 7a) and for parameter choices $b=10$ and $c_b=1$. Figure taken from~\protect\cite{Ishii:2019gta}.}
\label{fig:NMphases}       % Give a unique label
\end{figure}

Let me then analyze the phase diagram. Apart from the phases of Fig.~\ref{fig:7aphase}, we now have a third phase:
\begin{itemize}
 \item Horizonless thermal gas solution with probe $h$ condensate in the bulk. This geometry is still independent of the temperature. It is dual to a confined phase with nuclear matter.
\end{itemize}

The phase diagram may be determined numerically by carefully following the holographic dictionary. I show the result in Fig.~\ref{fig:NMphases} using potentials 7a defined in Appendix~B, i.e., the same potentials as in Fig.~\ref{fig:7aphase}. I set here simply $c_b=1$ and chose $b=10$ for which the vacuum to nuclear matter transition is around the correct value of the (quark) chemical potential, i.e., one third of the proton mass (minus small binding energy). The transition from the confined vacuum to the nuclear matter phase, as well as all the other transitions, is of first order, which is as expected for QCD at zero temperature. Recall that there is no temperature dependence in the confined phases which is reflected by the transition line between the thermal gas and nuclear matter phases being a vertical line.

\begin{figure}
\centering  \includegraphics[width=0.48\textwidth]{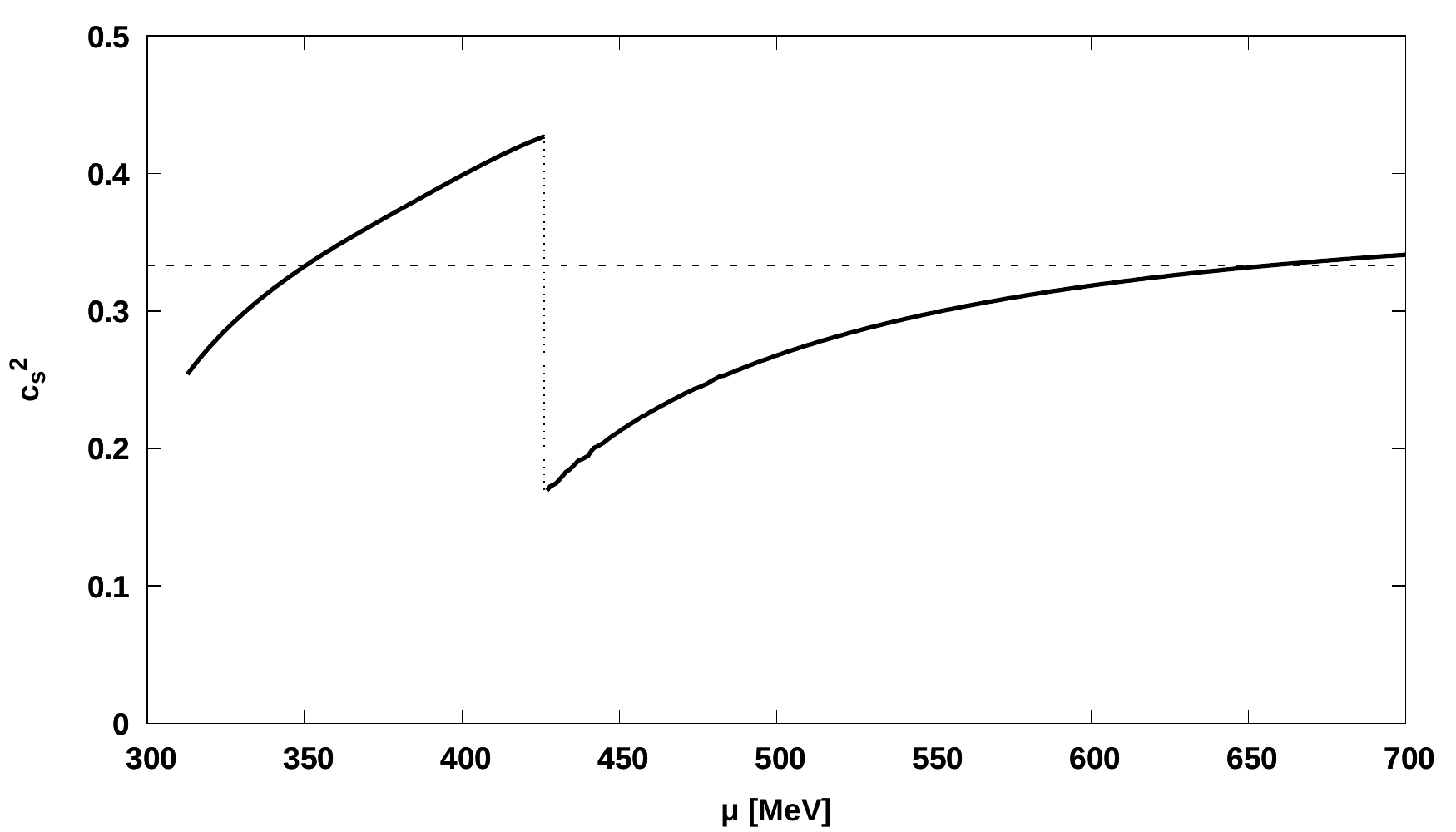}
\caption{The speed of sound at zero temperature as a function of the chemical potentials for potentials 7a, $b=10$, and $c_b=1$. Figure taken from~\protect\cite{Ishii:2019gta}.}
\label{fig:NMcs2}       % Give a unique label
\end{figure}

In Fig.~\ref{fig:NMcs2} I show the speed of sound at zero temperature in the nuclear and quark matter phases. The result is nontrivial and differs significantly from the value of $c_s^2=1/3$ in conformal theories, which is shown as the dashed horizontal line. Interestingly, in the dense nuclear matter phase below the nuclear to quark matter transition, there is clear excess over the conformal value. That is, the EOS is stiff. This is important because, as I discussed in Sec.~\ref{ssec:QCDNS}, a stiff EOS in this regime is required in order to satisfy the lower bound for $M_\mathrm{TOV}$ arising from the neutron star mass measurements. Notice that the conformal value is also exceeded in the quark matter phase.

In conclusion, despite using rough approximations, the results from the model with nuclear matter are encouraging: the phase diagram has the expected structure and the EOS appears feasible for neutron star applications. I will demonstrate below that this is indeed the case.

Before discussing the results for the homogeneous approach further, let me however comment on other possible approaches for nuclear matter in V-QCD.  We also considered an even simpler approach, where nucleons were treated as pointlike sources (basically arising from localized D0 branes) for the $\widehat\Phi$ field, in~\cite{Ishii:2019gta}. We were not able to estimate the coupling of the baryons to the tachyon and tachyon potential in this case, so we left it out, and we were forced to stabilize the nuclear matter by making a nonstandard choice for the gauge field coupling $w(\lambda)$, in some tension with the comparison to lattice data. Interestingly, in this case, we obtained phases were the charge could arise both from the baryons and from behind a black hole horizon, that could be interpreted as quarkyonic phases. Moreover, the low density thermal gas and high density quark matter phases were separated by several second order phase transitions rather then the first order transitions of the homogeneous approach, which is suggestive of the quark-hadron continuity~\cite{Schafer:1998ef}. Similar results were obtained in the WSS model in~\cite{BitaghsirFadafan:2018uzs,Kovensky:2020xif}. Let me stress however, that we had to choose potentials which were in slight conflict with lattice data to obtain the same in V-QCD.

Another approach for baryons and nuclear matter is, naturally, the construction of the solitons dual to individual baryons in V-QCD. This is work in progress.

%%%%%%%%%%%%%%%%%%%%%%%%%%%%%%%%%%%%%
\subsection{Hybrid equations of state} \label{ssec:hybrid}
%%%%%%%%%%%%%%%%%%%%%%%%%%%%%%%%%%%%%

Using a homogeneous bulk field to model nuclear matter is expected to work best at high densities, where nucleons are close and their wave functions start to overlap. At lower densities, in particular densities well below the nuclear saturation density, it is better to treat nuclear matter as a fluid of individual baryons. As I pointed out above, this is somewhat challenging in holography, even though there are some encouraging results for the WSS model. But this is not a problem for us, since the results at low density are known to quite a good accuracy by using other, more traditional methods, such as effective field theory. Therefore we take the approach where we abandon gau\-ge/gra\-vity duality at low densities, and use various nuclear theory models for the EOS in this region instead. We use the V-QCD model only at higher densities, i.e., well above the nuclear saturation density $n_s$. Combining the models like this, we aim to create hybrid EOSs where we use the potentially best available modeling in all regions~\cite{Ecker:2019xrw,Jokela:2020piw}. 

To make this approach concrete, we choose the following nuclear theory models at the lowest densities, ordered roughly from soft to stiff: the soft variation of the Hebeler-Lattimer-Pethick-Schwenk (HLPS) EOSs~\cite{Hebeler:2013nza}, the Akmal-Pandharipande-Ravenhall (APR) model~\cite{Akmal:1998cf}, Skyrme Lyon (SLy) model~\cite{Haensel:1993zw,Douchin:2001sv}, HLPS intermediate, as well as the IUF ~\cite{Fattoyev:2010mx}, and DD2~\cite{Typel:2009sy} variations of the Hempel-Schaffnerr-Bielich model~\cite{Hempel:2009mc}. For V-QCD, we use the variants defined by the three potentials from Appendix~B, which are in order of stiffness 5b, 7a, and 8b. 
At the density $n_\mathrm{tr}$ ranging from 1.2$n_s$ to 2.2$n_s$ we match the nuclear theory models with the V-QCD (nuclear matter) EOS. We require continuity of the pressure and baryon number density at the transition, which fixes the parameters $b$ and $c_b$ of the homogeneous nuclear matter. 

In summary, the resulting hybrid EOS depends on three things:
\begin{enumerate}
 \item Choice of nuclear theory model at low density, which naturally affects mostly the low density nuclear matter regime (densities around and below the saturation density).
 \item Choice of the transition density, which mostly affects the EOS in the intermediate regime, slightly above the transition density, i.e., approximately in the range from $n=1.5\, n_s$ to $2 n_s$).
 \item Choice of the V-QCD potentials: 5b, 7a, or 8b. This choice mostly affects the EOS in the high density regime, i.e., around $2n_s$ and above, which contains both dense nuclear and quark matter as well as the transition between them.
\end{enumerate}

As the uncertainties are largest in the high density region, where the holographic model is used, it is natural to choose samples of the EOSs which represent this uncertainty. For this we use the APR EOS, the transition density $n_\mathrm{tr}=1.6n_s$, and only vary the V-QCD potential. This gives rise to the soft, intermediate, and stiff V-QCD(APR) EOSs, obtained with the potentials 5b, 7a, and 8b, respectively. These EOSs have been published in the CompOSE database~\cite{Typel:2013rza}, see, e.g., \href{https://compose.obspm.fr/eos/198}{https://compose.obspm.fr/eos/198}.

\begin{figure}
\centering  \includegraphics[width=0.48\textwidth]{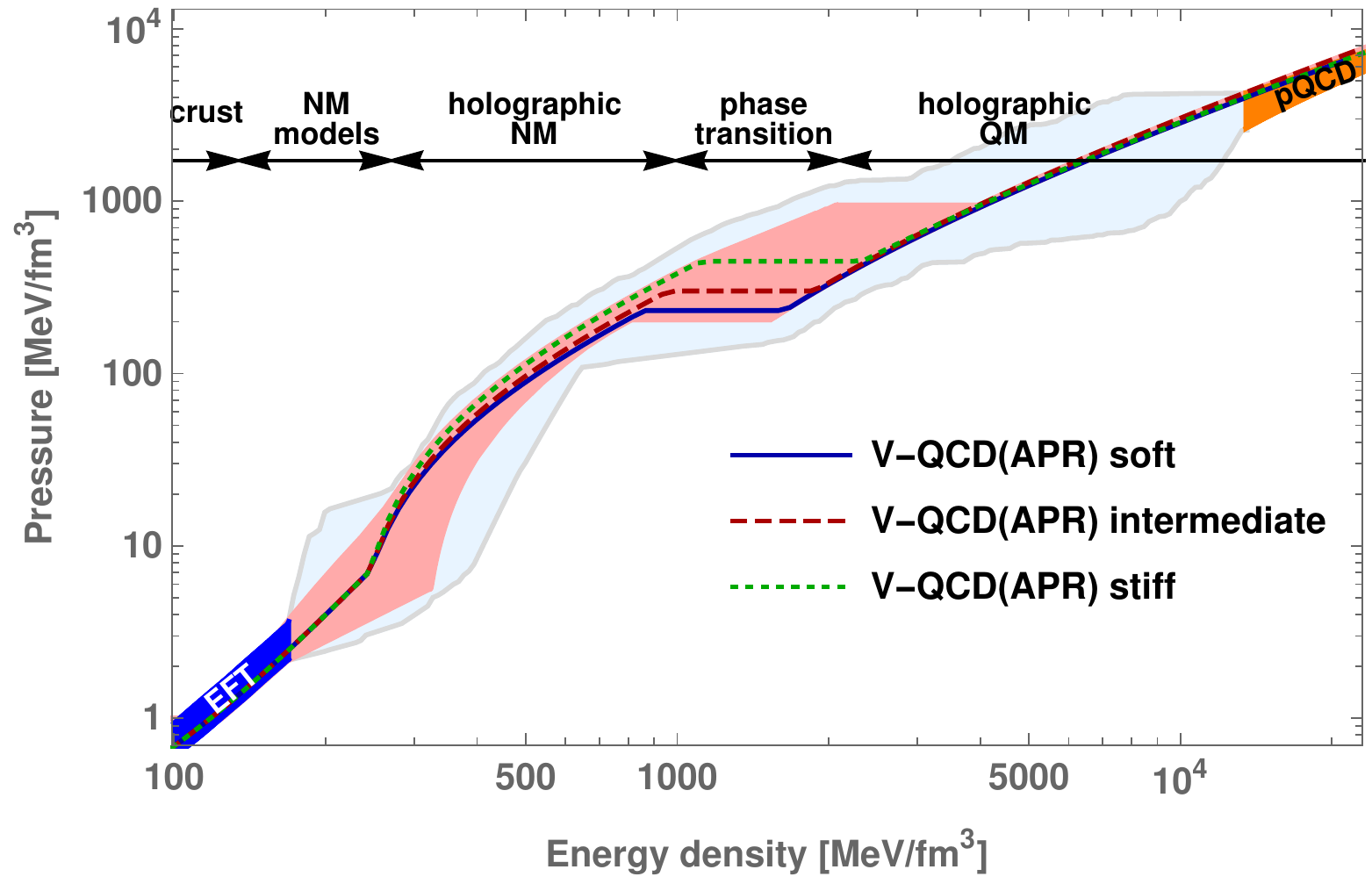}
\caption{Predictions of the hybrid EOS setup. The light blue band is spanned by all interpolating EOSs satisfying the astrophysical bounds, and the light red band is spanned by the hybrid V-QCD EOSs satisfying the same bounds. The curves show the three variants of the V-QCD(APR) EOSs as indicated by the legend. The horizontal lines with arrows show roughly the different regimes of the hybrid EOSs.}
\label{fig:hybridEOS}       % Give a unique label
\end{figure}

\begin{figure}
\centering  \includegraphics[width=0.48\textwidth]{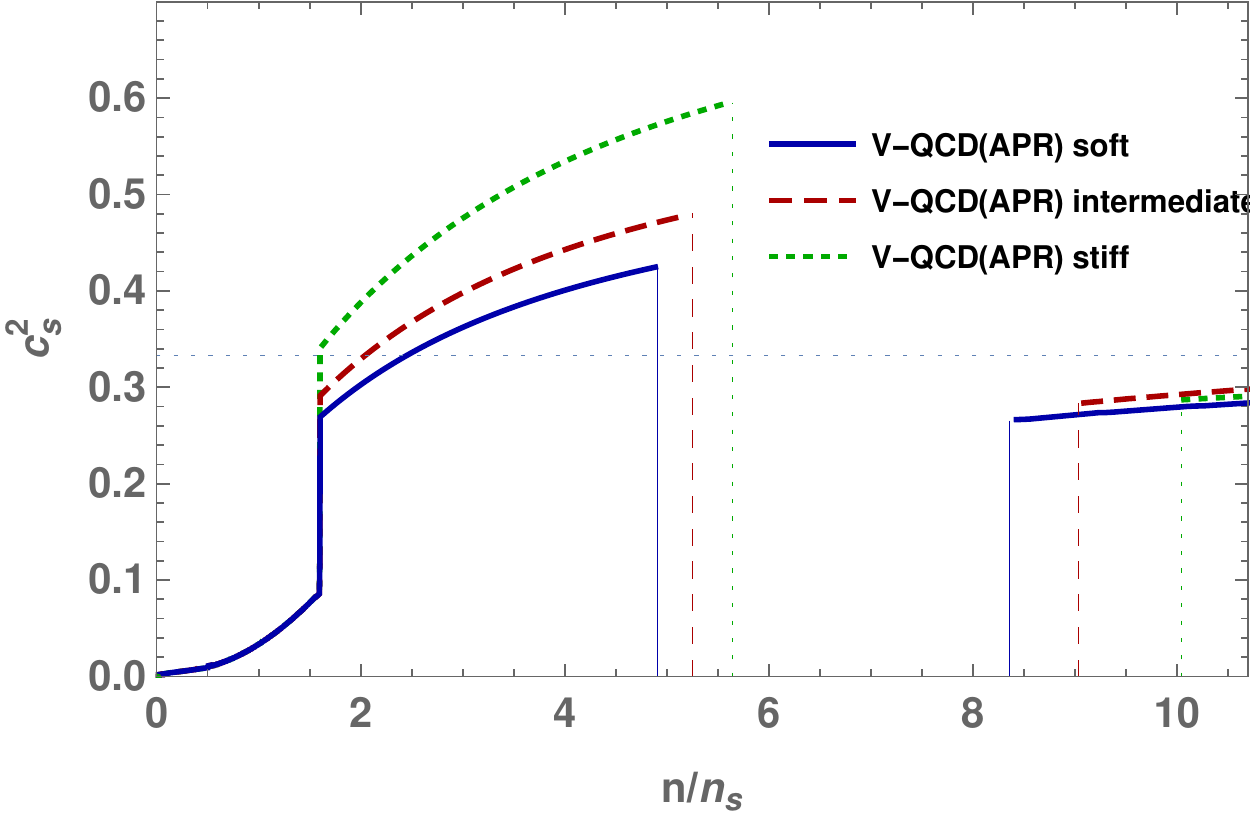}
\caption{Speed of sound for the three variants of hybrid V-QCD(APR) EOSs.}
\label{fig:hybridEOScs2}       % Give a unique label
\end{figure}

I show the predictions from the hybrid EOS setup in Fig.~\ref{fig:hybridEOS}. As in Fig.~\ref{fig:QMEOS}, the light blue band is spanned by all viable EOSs, obtained by using polytropic interpolations between the low density nuclear models and high density perturbative QCD results. I also imposed the observational constraints that $M_\mathrm{TOV}>2.0M_\odot$ and $\Lambda_{1.4}<580$. We notice that the holographic hybrid EOSs, in particular the three V-QCD(APR) variants, lie easily within this band, and are consistent with the known low and high density behaviors. The light red band is spanned by all the holographic hybrid EOSs, i.e., allowing variation of the transition density $n_\mathrm{tr}$ and low density models, and also requiring agreement with the astrophysical observations. Since the red band is significantly narrower than the blue band, imposing the constraints from the hybrid construction narrows down the band considerably. This is mostly due to using the holographic model, but also due to using a sample of nuclear theory models rather than completely general EOS at low density, which excludes EOS with somewhat ``exotic'' behavior at low density.

The speed of sound for representative choices V-QCD(APR) soft, intermediate, and stiff is shown in Fig.~\ref{fig:hybridEOScs2}. Notice that at $n=n_\mathrm{tr}$ the speed of sound is discontinuous. Some level of discontinuity is expected because we only require continuity of the pressure and baryon number density at matching, i.e., a second order transition. But the low density models matches poorly with the holographic model so that the discontinuity is quite sizable (also for other hybrids than those using the APR low density EOS), which might mean that neither the nuclear theory nor the holographic model work well in this region on intermediate density. The uncertainty following from this is however accounted for by the variation of $n_\mathrm{tr}$ in the setup. We also note that there is a gap where the speed of sound is not plotted: this arises because the density jumps at the first order phase transition. On the nuclear matter side the speed of sound is remarkably high, even up to $c_s^2 \approx 0.6$ for the stiff variant. As I noted above this means relative stiffness of the EOSs and makes it easy to construct physically feasible EOSs. Interestingly, the values of the speed of sound both for dense nuclear matter and for quark matter above the transition are close to the predictions of another nonperturbative framework, the functional renormalization group approach~\cite{Drews:2014spa,Drews:2016wpi,Otto:2019zjy}.

Finally, notice that the phase transition from nuclear to quark matter is strongly of first order for all hybrid EOS that we have constructed. We find that the latent heat at the transition satisfies 700~MeV/fm$^3 \lesssim \Delta \epsilon \lesssim 2000$~MeV/fm$^3$. Note that this prediction is insensitive to the astrophysical bounds but we note that the highest values are obtained by combining the extremely soft HLPS soft variant at low densities to the stiffest 8b version of V-QCD, which leads to EOSs that appear somewhat unnatural and have an extreme jump in the speed of sound at the matching density $n_\mathrm{tr}$. With EOSs having more natural behavior we find values of $\Delta \epsilon$ below 1500~MeV/fm$^3$.

For the baryon number density in the nuclear matter phase at the transition we find that $4.2 \lesssim n_b/n_s \lesssim 9.4$. Again the largest values are obtained by combining HLPS soft with the 8b version of V-QCD. For more natural setups nuclear matter densities stay below $8n_s$.

%%%%%%%%%%%%%%%%%%%%%%%%%%%%%%%%%%%%%
%%%%%%%%%%%%%%%%%%%%%%%%%%%%%%%%%%%%%
%%%%%%%%%%%%%%%%%%%%%%%%%%%%%%%%%%%%%
%%%%%%%%%%%%%%%%%%%%%%%%%%%%%%%%%%%%%
%%%%%%%%%%%%%%%%%%%%%%%%%%%%%%%%%%%%%
%%%%%%%%%%%%%%%%%%%%%%%%%%%%%%%%%%%%%
%%%%%%%%%%%%%%%%%%%%%%%%%%%%%%%%%%%%%
%%%%%%%%%%%%%%%%%%%%%%%%%%%%%%%%%%%%%
%%%%%%%%%%%%%%%%%%%%%%%%%%%%%%%%%%%%%
%%%%%%%%%%%%%%%%%%%%%%%%%%%%%%%%%%%%%
%%%%%%%%%%%%%%%%%%%%%%%%%%%%%%%%%%%%%
%%%%%%%%%%%%%%%%%%%%%%%%%%%%%%%%%%%%%
%%%%%%%%%%%%%%%%%%%%%%%%%%%%%%%%%%%%%
%%%%%%%%%%%%%%%%%%%%%%%%%%%%%%%%%%%%%
%%%%%%%%%%%%%%%%%%%%%%%%%%%%%%%%%%%%%
%%%%%%%%%%%%%%%%%%%%%%%%%%%%%%%%%%%%%
%%%%%%%%%%%%%%%%%%%%%%%%%%%%%%%%%%%%%
%%%%%%%%%%%%%%%%%%%%%%%%%%%%%%%%%%%%%
%%%%%%%%%%%%%%%%%%%%%%%%%%%%%%%%%%%%%
%%%%%%%%%%%%%%%%%%%%%%%%%%%%%%%%%%%%%
%%%%%%%%%%%%%%%%%%%%%%%%%%%%%%%%%%%%%
%%%%%%%%%%%%%%%%%%%%%%%%%%%%%%%%%%%%%
\section{Applications of holography to neutron stars}\label{sec:ns}
%%%%%%%%%%%%%%%%%%%%%%%%%%%%%%%%%%%%%

I then discuss the applications of gau\-ge/gra\-vity duality to neutron stars. I will 
mostly focus on results obtained from the V-QCD model. However, before discussing these results, I will give a brief review of the various other approaches discussed in the literature.

%%%%%%%%%%%%%%%%%%%%%%%%%%%%%%%%%%%%%
\subsection{Overview}
%%%%%%%%%%%%%%%%%%%%%%%%%%%%%%%%%%%%%

There is already a high number of papers in the literature discussing the applications of gau\-ge/gra\-vity duality to neutron star physics. Many of the early studies concentrate on the WSS model. However as it turns out, EOSs for nuclear matter made of point-like noninteracting solitons~\cite{Bergman:2007wp,Rozali:2007rx} do not lead to stable neutron stars at least for reasonable ranges of radii~\cite{Kim:2011da} (see however~\cite{Zhang:2019tqd}). In the closely related D4-D6 model~\cite{Kim:2009ey} (i.e., the Witten background of~\eqref{eq:Wittengeom} with different probe flavor brane setup than in WSS) solutions have been found~\cite{Kim:2011da} but radii of the stars were larger than expected for realistic stars. Flavors with different masses in this setup were studied in~\cite{Kim:2014pva}. Neutron stars from instanton gas in the WSS model (taking into account effects due to soliton widths) were considered in~\cite{Ghoroku:2013gja}, and seen to lead to more realistic stars, in particular if pressure from an external crust was added. Neutron stars containing matter made of ``multiquark'' states in the WSS model were considered in~\cite{Burikham:2010sw,Pinkanjanarod:2020mgi,Pinkanjanarod:2021qto}.

There has been also work in other backgrounds. Recently, instanton gas in six dimensional AdS soliton background, together with a model for color superconducting phase~\cite{Ghoroku:2019trx} was considered in~\cite{Ghoroku:2021fos} and shown to lead to potentially more realistic EOSs and mass radius relations.

A somewhat orthogonal direction is that of~\cite{deBoer:2009wk,Arsiwalla:2010bt} where it was argued that a degenerate fermion state in CFT can be dual to a higher dimensional ``neutron star'' in the bulk, and the gravitational collapse of the star could be interpreted as a transition from the high density degenerate state into a thermal state, i.e., similar to the nuclear to quark matter transition in QCD. The neutron star was however found to be unstable in the type IIB AdS$_5\times S^5$ background, but it was argued that it can be reliably embedded in M-theory on AdS$_4\times S^7$.

More recently EOSs where one combines low density nuclear matter result from effective theory to predictions of holographic models at high density have been considered. This was done by combining the soft, intermediate, and stiff HLPS EOSs~\cite{Hebeler:2013nza} with the prediction of the D3-D7 model~\cite{Karch:2007br} in~\cite{Hoyos:2016zke}. The first order transition between the nuclear matter and holographic quark matter was seen to be strong. Consequently, the stable stars only contained regular nuclear matter and stars with holographic quark matter cores were unstable. The estimated location of the transition was reasonable: the density of nuclear matter at the transition was found to be between 2 and 7 times nuclear saturation density.
This study was generalized by varying the quark mass parameter in~\cite{Annala:2017tqz}. The results included stars with exotic structure with quark matter near or at the crust and nuclear matter in the core, or stars made completely of quark matter (following the Bodmer-Witten assumption of stability of quark matter at low density~\cite{Bodmer:1971we,Witten:1984rs}). The stars passed astrophysical constraints, but interestingly violated the universal I-Love-Q relations~\cite{Yagi:2013awa}. Similar approach was also considered in a phenomenologically adjusted D3-D7 setup~\cite{Fadafa:2019euu} containing an intermediate chirally broken deconfined ``massive quark'' phase~\cite{Evans:2011eu}. Color superconducting phases~\cite{BitaghsirFadafan:2018iqr}  were included recently in \cite{BitaghsirFadafan:2020otb}, where it was also found that the intermediate massive quark phase can be made stiff enough to support holographic quark cores. Apart from D3-D7 models, HLPS EOSs have been used with quark matter from a holographic Einstein-Maxwell-dilaton model in~\cite{Mamani:2020pks}. In this case holographic quark cores were unstable.

The key to forming stable compact stars with holographic matter, as one can learn from the examples discussed above, is sufficiently stiff EOS (i.e., high speed of sound) in some of the phases. Indeed, it is difficult to construct viable EOSs without exceeding the conformal ``bound'' of $c_s^2=1/3$ for the speed of sound~\cite{Bedaque:2014sqa}.  This motivated the study of stiff phases in holographic models. Examples were found in~\cite{Hoyos:2016cob}, where top-down ($\mathcal{N}=4$ SYM at finite R-charge density) and bottom-up (Einstein-Maxwell-dilaton) models were considered and seen to have speeds of sound in excess of the conformal value. Such models were shown to produce speeds of sound arbitrarily close to the speed of light in~\cite{Ecker:2017fyh}. Other examples of stiff phases, which have also led to explicit models of neutron stars with holographic matter, are the stiff massive quark phase of~\cite{Evans:2011eu,BitaghsirFadafan:2020otb} and the homogeneous nuclear matter phase in V-QCD~\cite{Ishii:2019gta}, which I discussed above in Sec.~\ref{sec:VQCDNM}. Next I will discuss the results from this latter approach in more detail.

\begin{figure}
\centering  \includegraphics[width=0.45\textwidth]{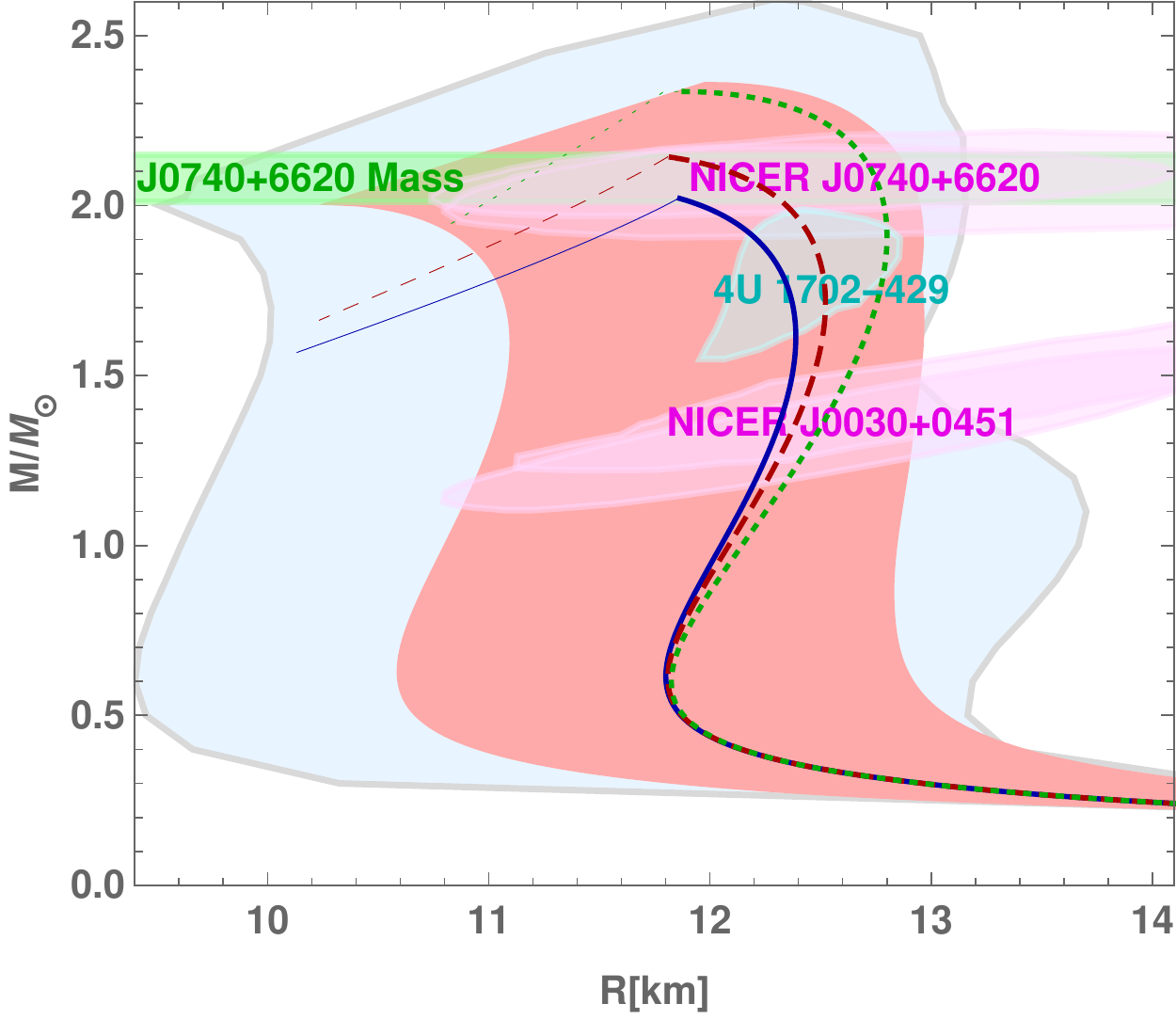}
\caption{Mass-radius relations for the hybrid EOSs. Notation is in Fig.~\protect\ref{fig:hybridEOS}: the light blue band is determined by all interpolating EOSs satisfying the astrophysical bounds, and the light red band is determined by the hybrid V-QCD EOSs satisfying the same bounds. The blue, dashed red, and dotted green curves are the  mass-radius relations for soft, intermediate, and stiff V-QCD(APR) EOSs, respectively. I also show experimental results for neutron star masses and radii, see text for details.}
\label{fig:hybridMR}       % Give a unique label
\end{figure}

%%%%%%%%%%%%%%%%%%%%%%%%%%%%%%%%%%%%%
\subsection{Properties of static neutron stars}
%%%%%%%%%%%%%%%%%%%%%%%%%%%%%%%%%%%%%

I start the discussion of the results from applying the hybrid V-QCD EOS to neutron stars by considering nonrotating stars. The first task is to solve the TOV equation, using the hybrid EOSs from Sec.~\ref{ssec:hybrid} as an input. The results are shown in Fig.~\ref{fig:hybridMR}, following the notation of the EOS plot in Fig.~\ref{fig:hybridEOS}~\cite{Ecker:2019xrw,Jokela:2020piw}. I also compare to some of the available radius measurements from the NICER experiment for the pulsars J0030+0451~\cite{Miller:2019cac,Riley:2019yda} and J0740+6620~\cite{Miller:2021qha,Riley:2021pdl}, which is among the most massive known pulsars, and the result from the analysis of X-ray burst cooling tail spectra for the source 4U 1702-429~\cite{Nattila:2017wtj}. 

Because the holographic EOSs are stiff, the obtained radii of the neutron stars are larger than in approaches based on extrapolating the results of chiral perturbation theory~\cite{Capano:2019eae}. This holds at its clearest at large masses, but even at $M=1.4M_\odot$ we obtain that
\be
 10.9~\mathrm{km} \lesssim R_{1.4} \lesssim 12.8~\mathrm{km} \ ,
\ee
whereas~\cite{Capano:2019eae} obtained $11.0
^{+0.9}_
{-0.6}$~km (90\% credible interval).
Moreover, the lowest radii are obtained by using the soft HLPS hybrids which include a dramatic change in stiffness at the HLPS/V-QCD matching density. Excluding such EOSs lifts to lower bound from 10.9~km to 11.7~km. That is, the remaining possible variation in the radius is roughly 1~km. Interestingly, our values of radii are also in good agreement with the direct radius measurements shown in Fig.~\ref{fig:hybridMR}.

\begin{figure}
\centering  \includegraphics[width=0.45\textwidth]{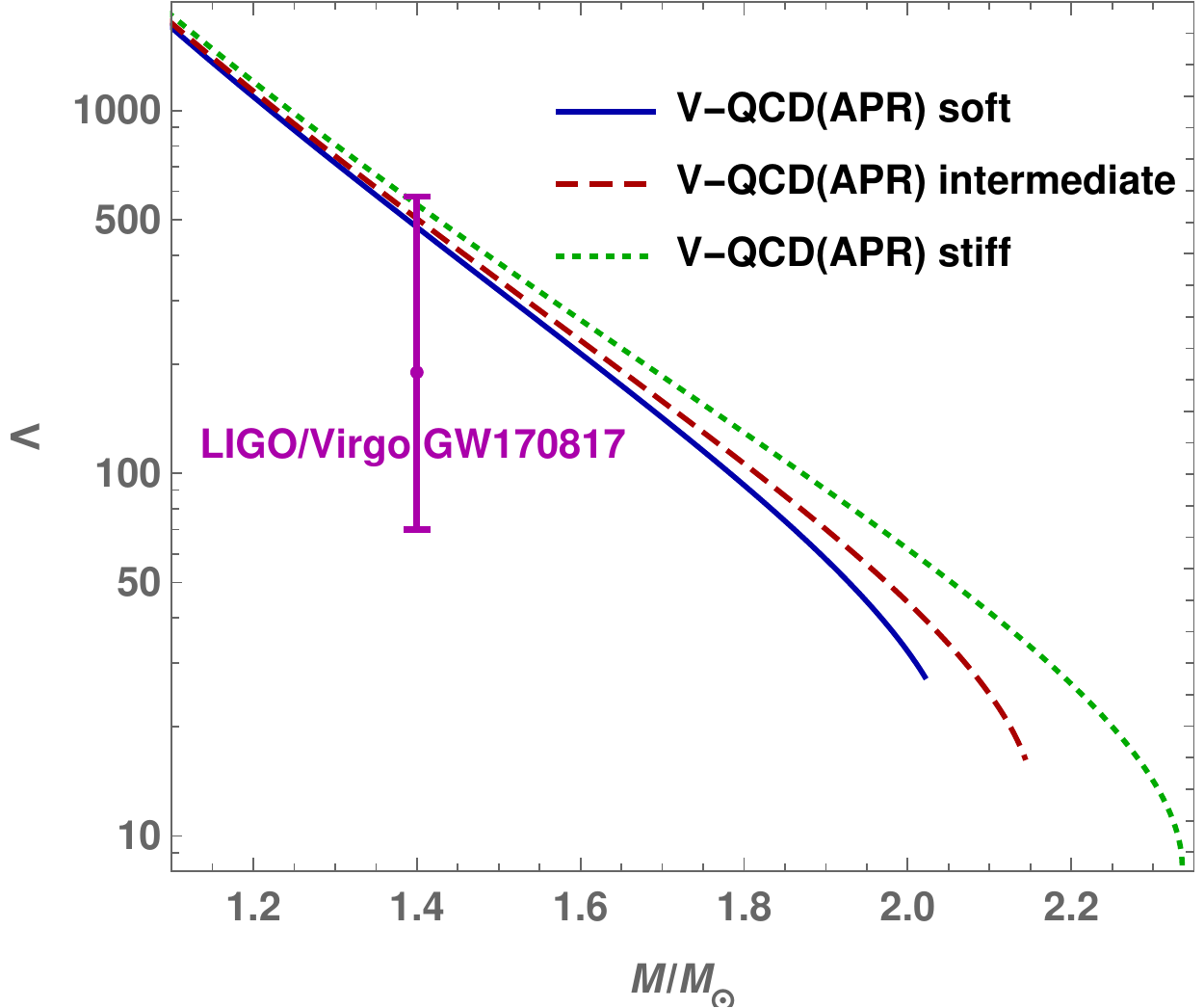}
\caption{Tidal deformability for the V-QCD(APR) hybrid EOSs, as a function of the neutron star mass, compared to the LIGO/Virgo result for the tidal deformability $\Lambda_{1.4}$ at $M=1.4M_\odot$.}
\label{fig:hybridLambda}       % Give a unique label
\end{figure}

In Fig.~\ref{fig:hybridLambda} I show the results for the other key parameter, the tidal deformability $\Lambda$, using the three variants of the V-QCD(APR) EOSs. I also compare to the LIGO/Virgo result $\Lambda_{1.4} = 190^{+390}_{-120}$. At low masses, the variation between the variants is small because we keep the low density nuclear matter model as well as the transition density $n_\mathrm{tr}$ fixed, and even at $M=1.4M_\odot$ the variation is moderate. Varying the choice for the low density model, i.e., taking into account all hybrid EOSs that we have constructed, leads to more variation in the tidal deformability. But our results still predict higher value for $\Lambda_{1.4}$ than the measurement:
\be
 \Lambda_{1.4} \gtrsim 232\ ,
\ee
i.e., the lower bound is higher than the central value reported by LIGO/Virgo. And again the lowest numbers for $\Lambda_{1.4}$ among the hybrids arise from somewhat dubious constructions with HLPS soft involving a drastic jump in stiffness. Other hybrids produce higher values, and using only them would give $\Lambda_{1.4} \gtrsim 326$. For comparison, the soft V-QCD(APR) variant has $\Lambda_{1.4} \approx 478$.

%%%%%%%%%%%%%%%%%%%%%%%%%%%%%%%%%%%%%
\subsection{Properties of rotating neutron stars}
%%%%%%%%%%%%%%%%%%%%%%%%%%%%%%%%%%%%% 

Let me then consider the effects of rotation in neutron stars. First one can start with slow rotation, which is characterized in terms of the moment of inertia $I$ and the quadrupole moment $Q$. These can also be estimated from the results for the tidal deformability $\Lambda$ (which is related to the leading nontrivial Love number $k_2$) by using the approximately universal ``I-Love-Q relations''~\cite{Yagi:2013awa}. We have computed $I$ and $Q$ and checked that these relations hold for our hybrid EOSs within the precision of $\lesssim0.5$\%~\cite{Jokela:2020piw}.

But neutron stars can also be in rapid, relativistic rotation. Rapidly rotating neutron stars are created in neutron star mergers, and even though the speed of rotation slows down rapidly after the merger due to various instabilities, this is still a slow process with respect to the typical timescales of the actual merger event, which range from one millisecond to one second~\cite{Baiotti:2016qnr}. There is however also other topical motivation: the recent observation GW190814 of gravitational waves by LIGO/Virgo from a merger of a black hole with a compact object having a mass of  $2.59^{+0.08}_{-0.09}M_\odot$~\cite{Abbott:2020khf}. The mass of the secondary component falls into the mass gap: it is not clear whether it was a neutron star or a black hole. While it is possible that it was a slowly rotating neutron star, this requires a rather extreme EOS~\cite{Godzieba:2020tjn,Tews:2020ylw}. The mass result is also at odds with the maximum bound of the nonrotating mass $M_\mathrm{TOV}$ inferred from GW170817~\cite{Margalit:2017dij,Rezzolla:2017aly,Shibata:2019ctb}. But it is also possible that this object was a rapidly rotating neutron star, since rotation leads to enhanced bound for the maximum mass (see, e.g.,~\cite{Most:2020bba,Dexheimer:2020rlp,Zhang:2020zsc}).

\begin{figure}
\centering  \includegraphics[width=0.45\textwidth]{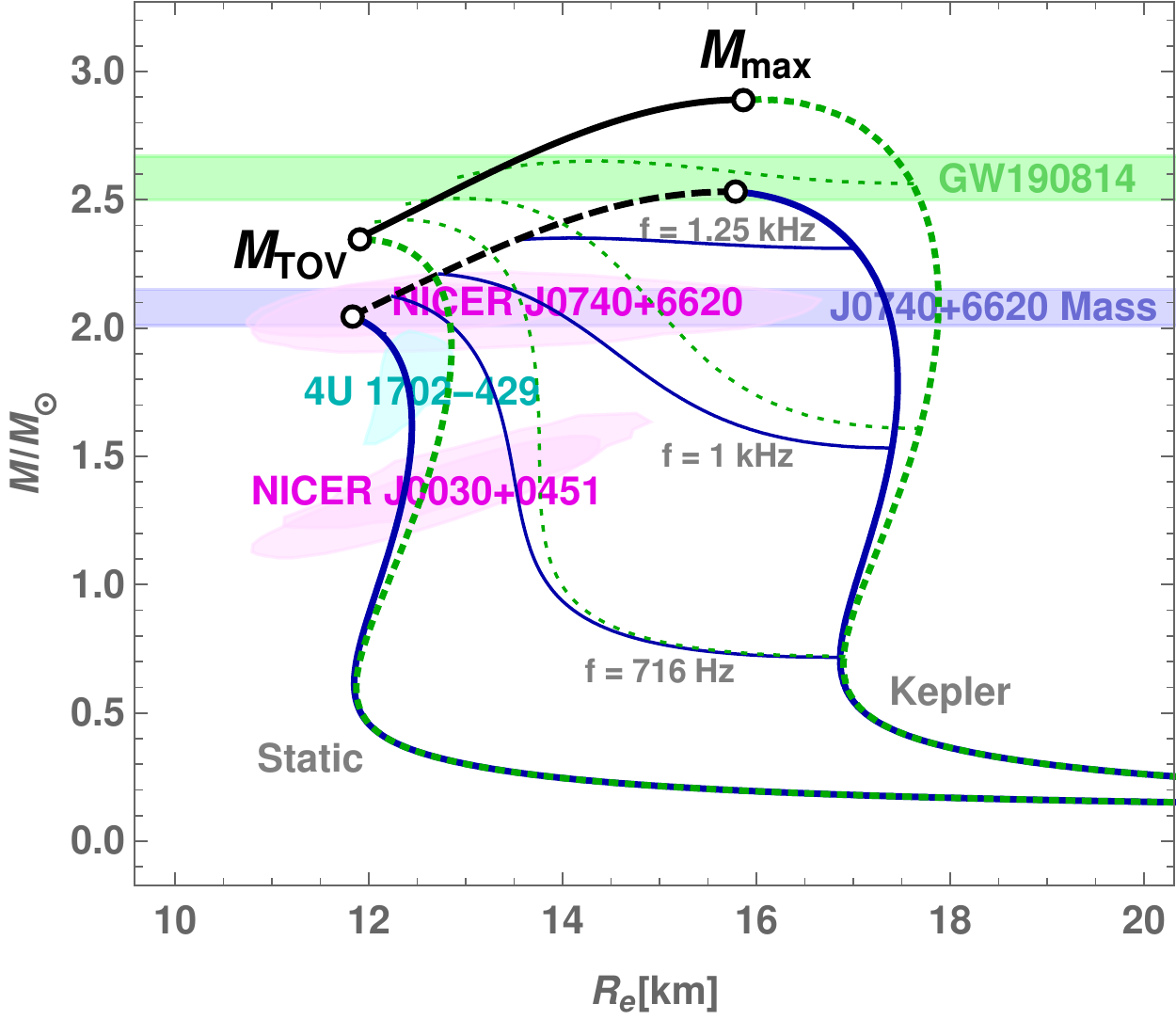}
\caption{Mass-radius curves for the V-QCD(APR) EOSs including rotating stars. Blue (green dotted) curves are for the soft (stiff) V-QCD(APR) EOS. The (dashed and solid) black curves mark lines where the maximal central density is reached. The circles show the maximal nonrotating $M_\mathrm{TOV}$ and rotating $M_\mathrm{max}$ masses. Shown is also experimental data for masses and radii of compact objects. See text for detailed explanation.}
\label{fig:hybridRNS}       % Give a unique label
\end{figure}

In order to study this scenario with the hybrid V-QCD EOSs, we solved the properties of rapidly rotating neutron stars~\cite{Demircik:2020jkc}. The results for the masses and equatorial radii $R_e$ are shown in Fig.~\ref{fig:hybridRNS}. We included the curves for the soft and stiff (but not intermediate) variants. The thick curves are the extremal cases, given by nonrotating stars at low radius, by maximally rotating stars (mass shredding, Kepler limit) at large radius, and the onset of instability at large mass. For the soft EOS, the instability (thick black dashed curve) is due to reaching the nuclear to quark matter phase transition. For the stiff EOS, the maximum mass is set by the secular instability which can be estimated by the stability condition
\be
 \frac{\partial M(n_c,J)}{\partial n_c}>0 \ ,
\ee
where $n_c$ is the central baryon number density and $J$ is the total angular momentum. The thin lines show the curves at three choices of constant rotation frequency. Here 716~Hz is the highest rotation frequency observed in any pulsar to date, set by the pulsar PSR J1748-2446ad~\cite{Hessels:2006ze}. I also show experimental data for masses and radii, including the same neutron star measurements as in Fig.~\ref{fig:hybridMR} and the mass result for the secondary component of GW190814 from LIGO/Virgo. 

The interpretation that the $M=2.59^{+0.08}_{-0.09}M_\odot$ object of GW190814 is a slowly rotating neutron star is inconsistent with the hybrid EOSs, but it may have been a rapidly rotating neutron star. The required rotation frequencies are however high: pretty much the maximal rotation frequency is required for the soft variant, and frequencies above 1~kHz are needed for the stiff variant, which are well above the highest observed rotation frequency of any known pulsar.

\begin{figure}
\centering  \includegraphics[width=0.48\textwidth]{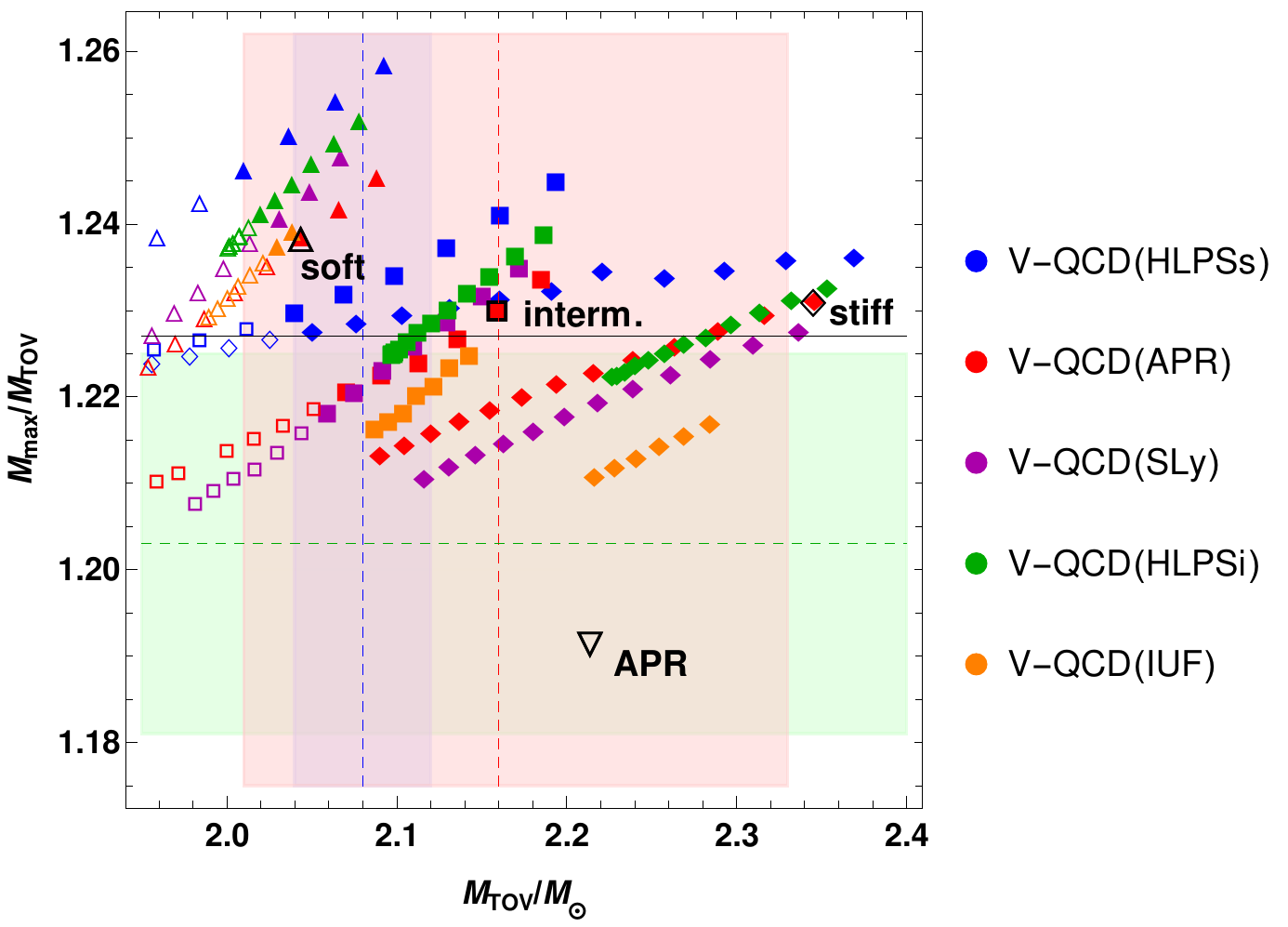}
\caption{Mass ratios $M_\mathrm{max}/M_\mathrm{TOV}$ for the hybrid EOSs. The colors refer to different low density models as indicated by the legend. Triangles, squares, and diamonds stand for soft, intermediate, and
stiff versions of V-QCD (potentials 5b, 7a, and 8b), respectively, with varying $n_\mathrm{tr}$. For the open markers the maximal mass is below the GW190814 band of Fig.~\protect\ref{fig:hybridRNS}. The large black markers show the results for the standard soft, intermediate, and stiff V-QCD(APR) variants with $n_\mathrm{tr}=1.6n_s$, as well as the original APR EOS. The green, red, and blue bands show earlier estimates for the ratio $M_\mathrm{max}/M_\mathrm{TOV}$~\protect\cite{Breu:2016ufb}, for the maximum of $M_\mathrm{TOV}$~\protect\cite{Rezzolla:2017aly}, and the minimum of $M_\mathrm{TOV}$~\protect\cite{Most:2020bba}, respectively. }
\label{fig:hybridmassrat}       % Give a unique label
\end{figure}

A key observable for rotating neutron stars is the ratio of the highest mass $M_\mathrm{max}$ over the maximum mass of static nonrotating stars $M_\mathrm{TOV}$. I show the results for this ratio for all hybrids that pass the astrophysical bounds in Fig.~\ref{fig:hybridmassrat}.\footnote{Notice that the intermediate V-QCD(APR) EOS discussed in this review slightly differs from the intermediate EOS of Ref.~\cite{Demircik:2020jkc}.} We find that the ratio is
\be\label{eq:massratio}
 \frac{M_\mathrm{max}}{M_\mathrm{TOV}} = 1.227^{+0.031}_{-0.016}
\ee
where the central value (horizontal black line in Fig.~\ref{fig:hybridmassrat}) was obtained by polynomially fitting the spin dependence of the maximum mass for all hybrids. The error band was obtained as the maximal deviation from the with among all hybrid EOSs. Our results are somewhat higher than the earlier estimate $\frac{M_\mathrm{max}}{M_\mathrm{TOV}} = 1.203\pm 0.022$~\cite{Breu:2016ufb}, shown as the green band and line, which was based on EOSs without phase transitions. That is, even though all neutron stars are fully hadronic with our EOSs, the presence of the phase transition still affects the ratio considerably, which is possible as the instability due to the onset of the quark-hadron affects both $M_\mathrm{max}$ and $M_\mathrm{TOV}$ even for fully hadronic stars.  In particular the soft variants, where the maximum mass is determined by the nuclear to quark matter phase transition, tend to lie above the earlier estimate.

\begin{figure*}
\centering  \includegraphics[width=0.8\textwidth]{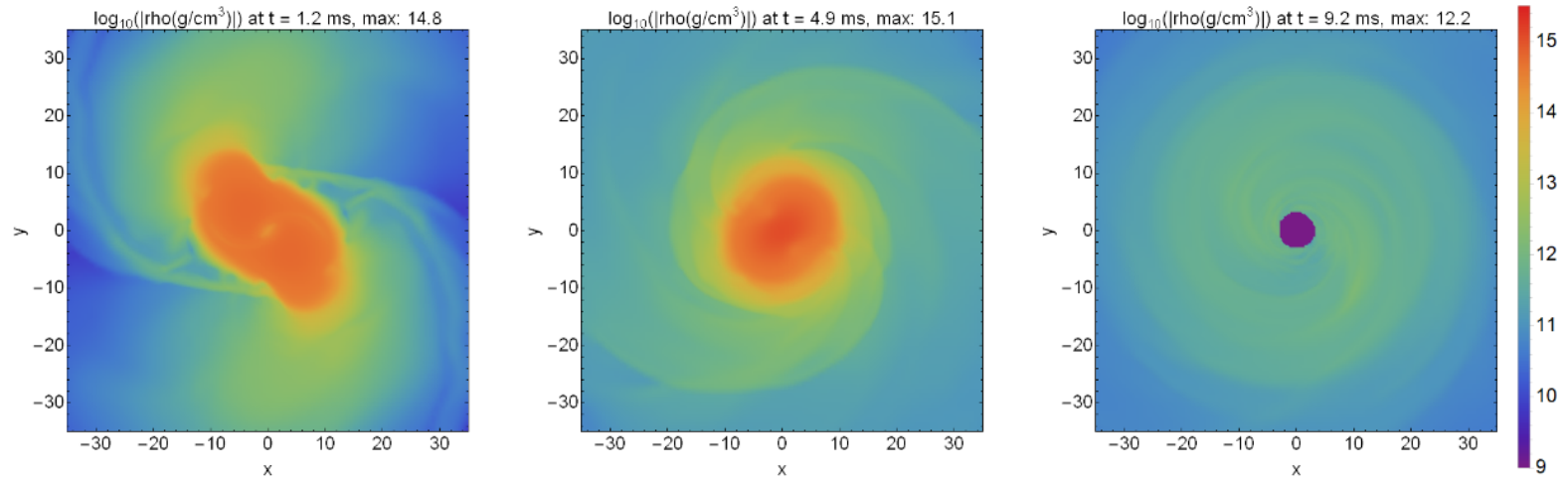}
\caption{Snapshots of the rest mass density profile of the merging neutron stars (both having the mass $M=1.4M_\odot$) in the plane of rotation. Left: density right after merger. Middle: hypermassive neutron star about 5 ms after the merger. Right: mass distribution right after black hole formation. }
\label{fig:snapshots}       % Give a unique label
\end{figure*}

%%%%%%%%%%%%%%%%%%%%%%%%%%%%%%%%%%%%%
\subsection{Neutron star mergers with holographic EOS}
%%%%%%%%%%%%%%%%%%%%%%%%%%%%%%%%%%%%%

The EOS is the basic input for the time evolution of neutron star mergers. We have analyzed the neutron star mergers by using the hybrid holographic EOSs as an input in full four dimensional simulations of mergers in~\cite{Ecker:2019xrw}.  We used publicly available codes in our simulations: the LORENE code for initial data, and Einstein toolkit for the evolution in general relativity with the WhiskyTHC code for relativistic hydrodynamics. Here I will both review briefly the general properties of the merger and the observables, and illustrate by using the results obtained with the holographic EOSs. See~\cite{Baiotti:2016qnr} for an extensive review.

For the EOSs we picked the intermediate (potentials 7a) V-QCD(SLy) variants with $n_\mathrm{tr}=1.61n_s$ and $n_\mathrm{tr}=1.94n_s$, and studied equal mass mergers of neutron stars with the masses ranging from 1.3$M_\odot$ to 1.5$M_\odot$. The high mass simulations ($M=1.5M_\odot$) lead to a collapse to a black hole immediately after the merger. In the low mass simulations ($M=1.3M_\odot$) a rapidly rotating neutron star remnant is formed, which is stable at least within the timescale of the simulation. For the simulation at $M=1.4M_\odot$ and with $n_\mathrm{tr}=1.61n_s$, a differentially rotating hypermassive neutron star is first formed, which then collapses into a black hole about 7.8~ms after the merger. These three cases therefore exhaust the possible basic scenarios expected in the merger~\cite{Baiotti:2016qnr}. As an example, I show snapshots of the latter simulation with $M=1.4M_\odot$ and with $n_\mathrm{tr}=1.61n_s$ with the intermediate hypermassive neutron star in Fig.~\ref{fig:snapshots}. These plots show the rest mass density $\rho = m_n n$, where $m_n$ is the neutron mass and $n$ is the baryon number density, on the plane of rotation.

\begin{figure}
\centering \includegraphics[width=0.455\textwidth]{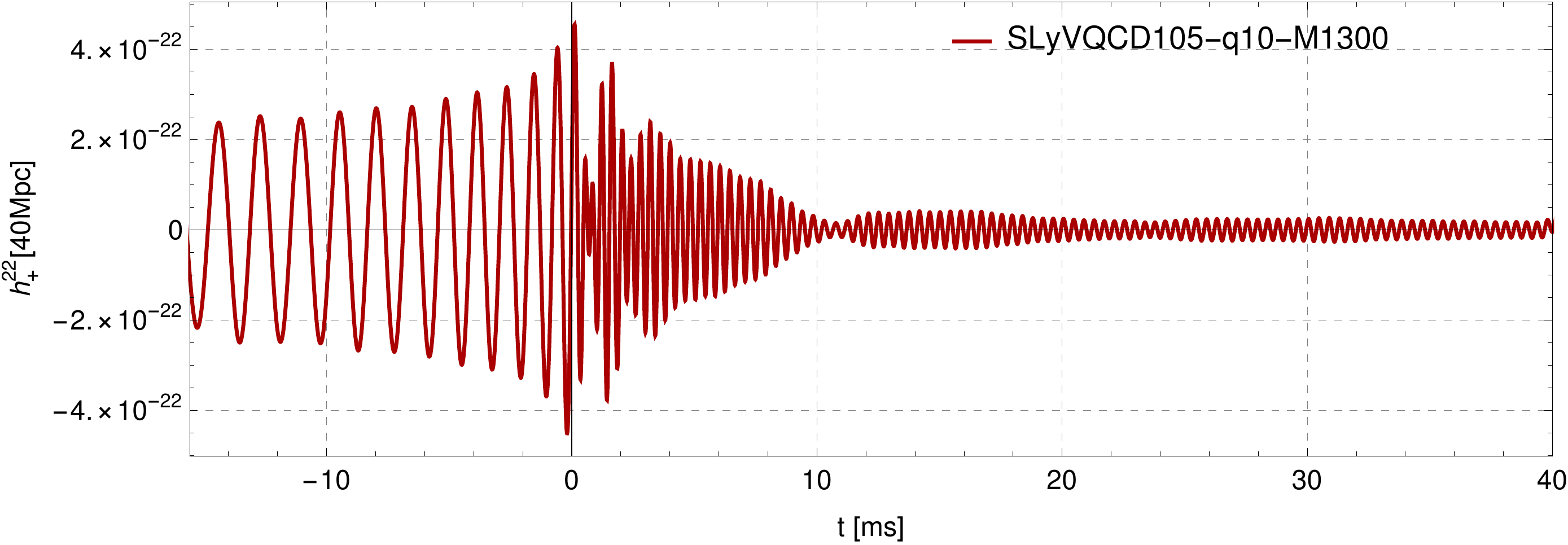}
\includegraphics[width=0.45\textwidth]{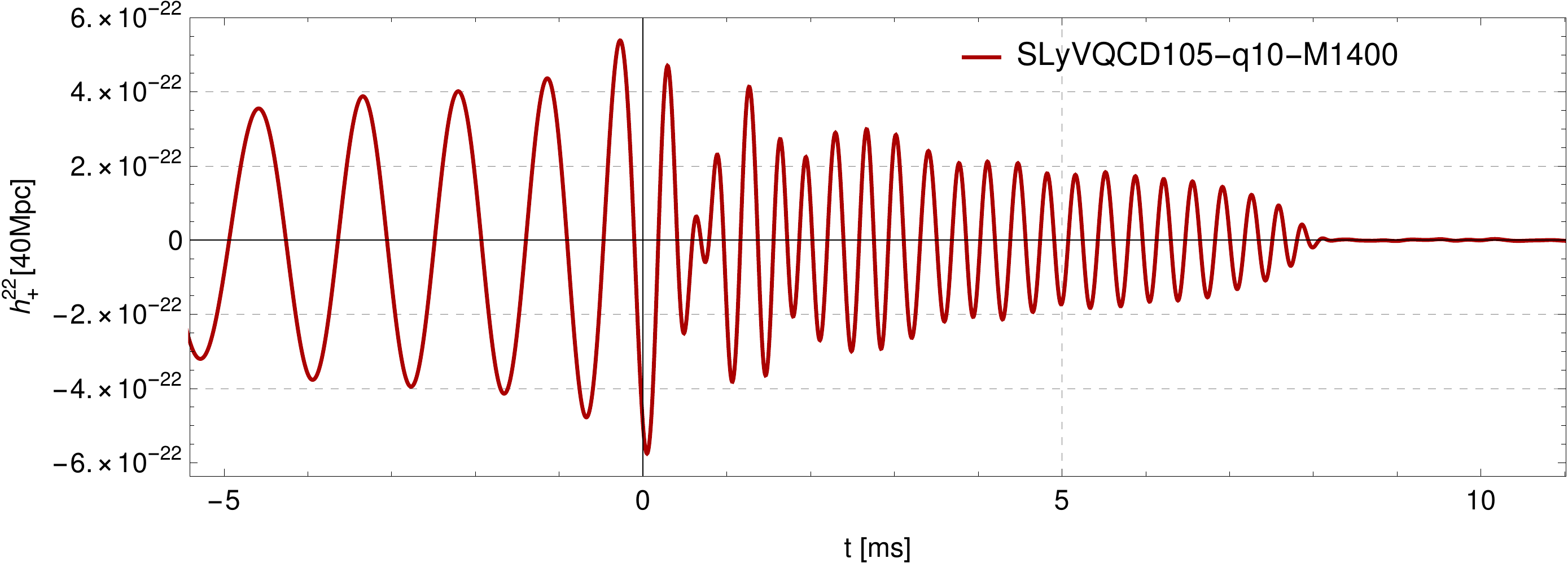}
\includegraphics[width=0.45\textwidth]{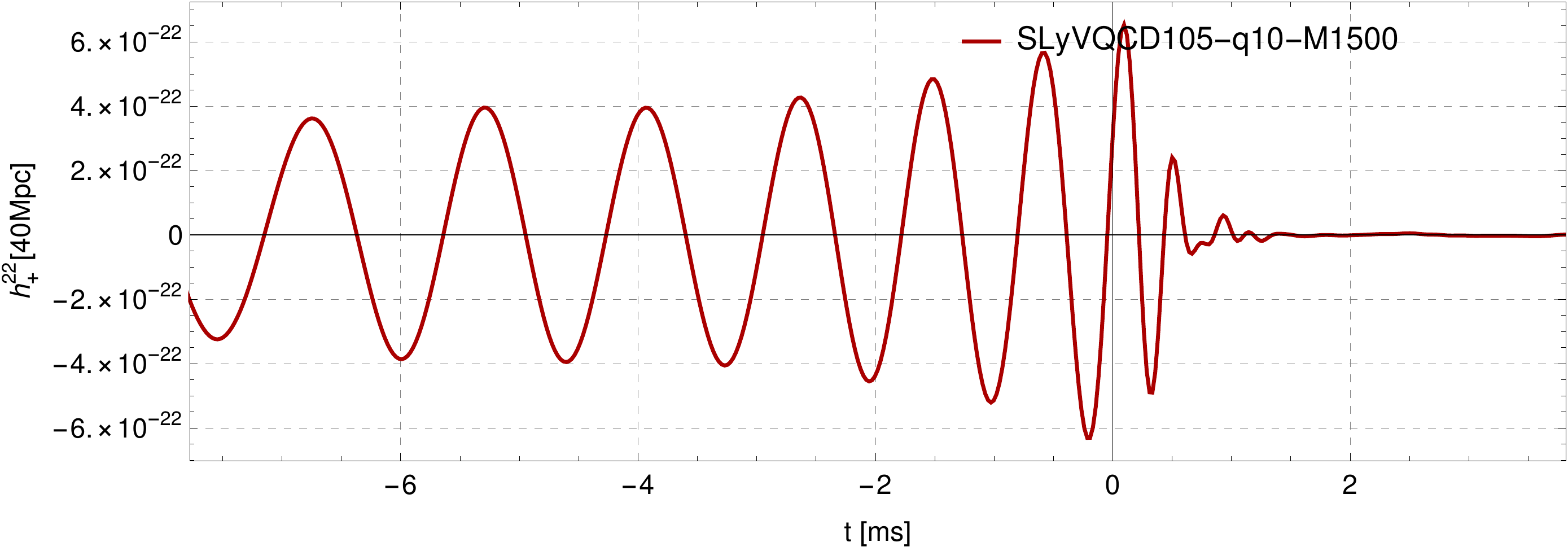}
\caption{Gravitational waves ($h_+^{22}$ at the distance of 40 Mpc) obtained from numerical simulations with the intermediate V-QCD(SLy) EOS with $n_\mathrm{tr}=1.61$. The top, middle, and bottom rows show the signal from simulations with $M=1.3M_\odot$, $1.4M_\odot$, and $1.5M_\odot$, respectively.  }
\label{fig:GW}       % Give a unique label
\end{figure}

Arguably the most important observable that can be extracted from the simulations is the outgoing gravitational wave. As it turns out, a relevant quantity which can be extracted from the asymptotics of the simulated metric is the Newman-Penrose or Weyl scalar $\psi_4$~\cite{Bishop:2016lgv}. It is related to the perturbation of the metric for the two polarizations of the waves in the standard transverse-traceless gauge as 
\be \label{eq:psi4def}
 \psi_4 = \partial_t^2\left( h_+ -i h_\times\right) \ .
\ee
Both $\psi_4$ and $h_{+,\times}$ may be expressed as expansions in the spin-weighted spherical harmonics ${}_sY_{\ell m}(\theta,\phi)$ with $s=-2$. One may then extract from the numerical solution the value of the coefficient of the  dominant mode with $\ell=m=2$, denoted by $\psi_4^{22}$, and use~\eqref{eq:psi4def} to convert this to the coefficients $h_+^{22}$ and $h_\times^{22}$. 

I show the coefficient $h_+^{22}$, extracted from the simulations and extrapolated to the distance of 40~Mpc, i.e., the estimated distance to GW170817~\cite{TheLIGOScientific:2017qsa}, in Fig.~\ref{fig:GW}. The three curves are for different masses so that in the top diagram the remnant is a neutron star, in the middle diagram a short-lived hypermassive neutron star is formed, and in the bottom diagram the system immediately collapse into a black hole after the merger. In these plots, the merger takes place at $t=0$. The gravitational wave signal before that arises from the inspiral phases and has quite regular form. Notice that the time scales of the three curves are different. 

After the merger, in the top and middle curves there is a nontrivial signal which has higher frequency than the inspiral. This part carries nontrivial information on the structure of the merging stars and therefore the EOS. In the bottom curve (as well as in the middle curve for $t \gtrsim 7$~ms), the signal ends in a brief ring down as the black hole forms.

\begin{figure}
\centering \includegraphics[width=0.48\textwidth]{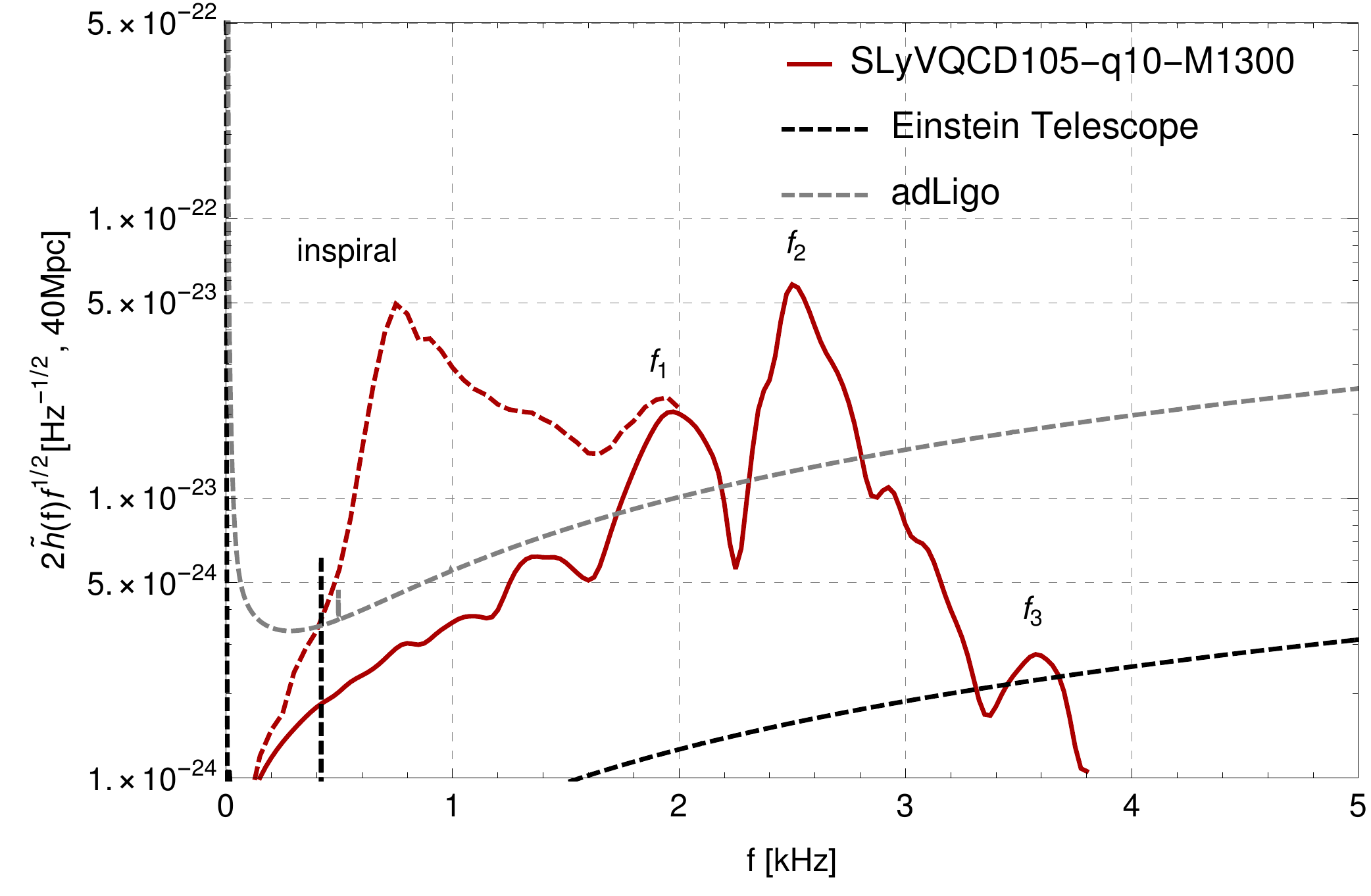}
\caption{The power spectral density  for the merger of two $M=1.3M_\odot$ neutron stars with the V-QCD(SLy) EOS with $n_\mathrm{tr}=1.61$. The dashed and solid curves are with and without including the inspiral part of the signal. I also compare to the sensitivity of advanced LIGO/Virgo and the projected sensitivity of the Einstein telescope assuming the distance 40~Mpc of GW170817.}
\label{fig:PSD}       % Give a unique label
\end{figure}

A standard method to analyze the spectrum of the signal further is to compute the power spectral density. To this end, we first carry out a Fourier transform
\be
\tilde h_{+,\times}(f) = \int dt\ h_{+,\times}^{22}(t)\, e^{-2\pi if t} \ .
\ee
The power spectral density is then obtained by computing the average over polarizations:
\be
 \tilde h(f) \equiv \sqrt{\frac{|\tilde h_{+}(f)|^2+|\tilde h_{\times}(f)|^2}{2}} \ .
\ee
I show the power spectral density for the merger of two $M=1.3M_\odot$ neutron stars in Fig.~\ref{fig:PSD}, i.e., for the signal of Fig.~\ref{fig:GW} (top). I plot $2\tilde h(f)\sqrt{f}$ assuming the estimated distance to GW170817, i.e., 40~Mpc. The dashed (solid) curve is the signal with (without) the inspiral part. We therefore see that the signal for $f\lesssim 2$~kHz arises mostly from the inspiral, whereas the signal from $f \gtrsim 2$~kHz arises mostly from the remnant.

The postmerger part of the signal shows, for mergers with a neutron star remnant, three characteristic peaks at frequencies $f_{1,2,3}$. The origin of these peaks is relatively simple~\cite{Takami:2014tva}: After the merger the cores of the two neutron stars oscillate, causing variation of the mass of inertia of the whole remnant. The extremal rotation frequencies of the system due to the oscillation of the mass of inertia give rise to $f_1$ and $f_3$. After a while the oscillations cease, and the final rotation frequency gives the most prominent $f_2$ peak. Another interesting frequency is the instantaneous frequency of the gravitational signal at the time of the merger $f_\mathrm{mrg}$, which is typically close to $f_1$~\cite{Breschi:2019srl}.

We find that the holographic EOS favor relatively low characteristic frequencies of the power spectral densities. For example, the signal of Fig.~\ref{fig:PSD}, which uses the V-QCD(SLy) EOS with $n_\mathrm{tr}=1.61n_s$, has $f_2 \approx 2.53$~kHz, the same simulation with the V-QCD(SLy) EOS and $n_\mathrm{tr}=1.94n_s$ gives  $f_2 \approx 2.80$~kHz, whereas using the ``pure'' SLy EOS gives  $f_2 \approx 3.19$~kHz. Also the other frequencies are shifted towards zero with respect to the SLy results. These results reflect the stiffness of the holographic EOS at high densities.

I also remark that, whenever a collapse to a black hole happened in the simulations, it was always triggered by the phase transition to quark matter, which may be detectable by analyzing the gravitational wave signal~\cite{Most:2018eaw,Bauswein:2018bma}. Consequently, only limited amounts of quark matter was produced in the simulations. This is expected because the V-QCD quark matter EOS above the transition is quite soft so that formation of a quark matter of mixed phase leads to an immediate collapse.

Notice also that the $f_2$ peak in Fig.~\ref{fig:PSD} lies well above the sensitivity estimate for advanced LIGO/Virgo when the distance to the event is the same as for GW170817, so it is expected that the postmerger signal will be observed for future events. For GW170817 only the inspiral signal was detected. Studies of the electromagnetic signal from the GW170817 kilonova~\cite{GBM:2017lvd} suggest that a (rather long lived) hypermassive neutron star was formed in the event, so that a nontrivial postmerger signal was generated in the event, but missed detection (see, e.g., ~\cite{Granot:2017tbr,Metzger:2018qfl,Gill:2019bvq}). This happened because at the time of the event the sensitivity of the experiment was a bit lower than that marked in the figure.  

The main feature of the electromagnetic signal from GW170817 was its relative brightness, which implies that a large amount (around 0.05$M_\odot$) of matter was ejected during and after the merger~\cite{Drout:2017ijr,Perego:2017wtu,Tanaka:2017qxj,Tanvir:2017pws,J-GEM:2017tyx,Villar:2017wcc}. This is actually the reason, together with the delay of the observed gamma ray burst GRB 170817A \cite{GBM:2017lvd}, that the hypermassive remnant must have been long lived, otherwise not enough material would have been ejected~\cite{Granot:2017tbr,Metzger:2018qfl,Gill:2019bvq}. Notice that the amount of ejected matter also affects the estimates for the maximum of $M_\mathrm{TOV}$ from GW170817~\cite{Margalit:2017dij,Rezzolla:2017aly,Shibata:2019ctb} which I discussed in Sec.~\ref{ssec:expNS}. 

\begin{figure}
\centering \includegraphics[width=0.48\textwidth]{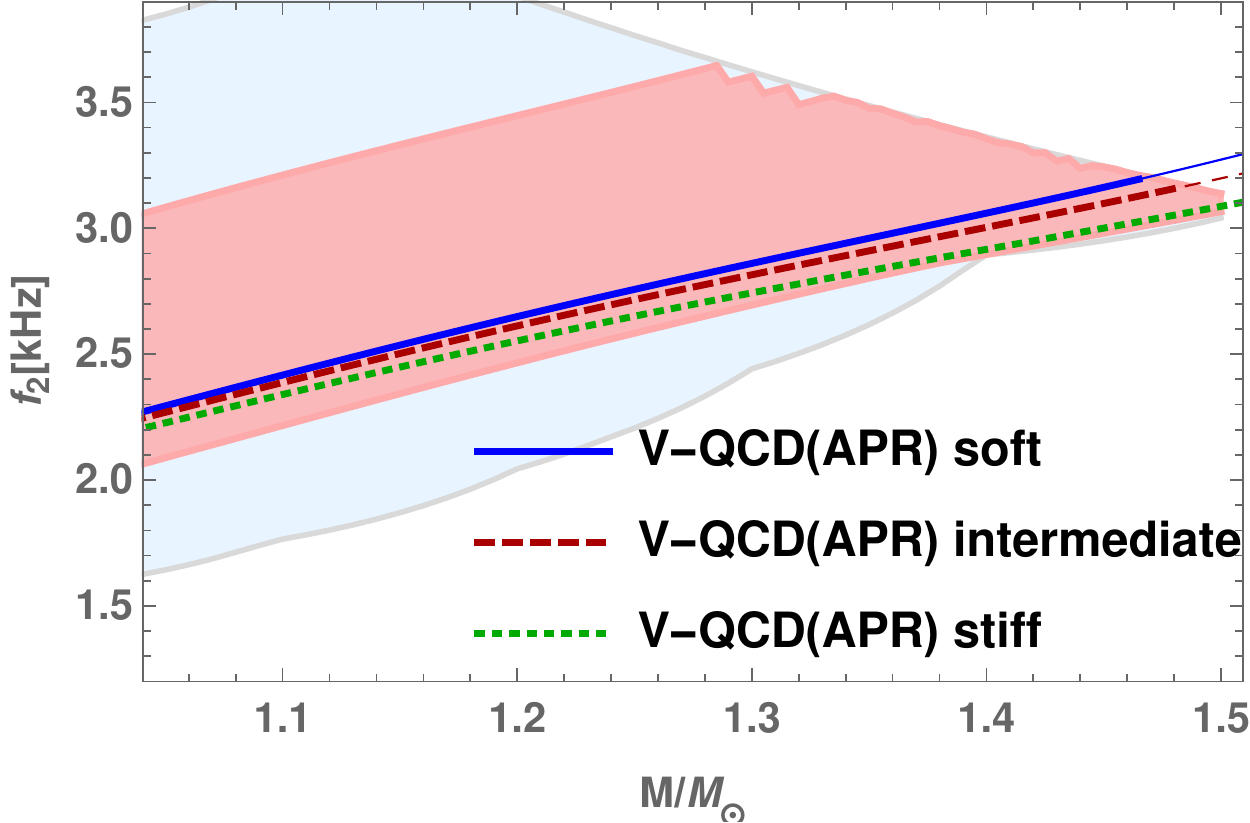}
\caption{The prediction for the peak frequency $f_2$ as a function of neutron star mass for equal mass mergers from the approximate relations derived in~\protect\cite{Breschi:2019srl} using the hybrid EOSs as input. Notation as in Figs.~\protect\ref{fig:hybridEOS} and~\protect\ref{fig:hybridMR}: the light blue (red) band is spanned by all (hybrid holographic) EOSs satisfying the astrophysical bounds. The curves show the predictions for the three V-QCD(APR) variants.  }
\label{fig:f2}       % Give a unique label
\end{figure}

As I discussed above, the signal in Fig.~\ref{fig:PSD} naturally divides into the inspiral and postmerger regions, which naively appear to be essentially independent. However this not true: The inspiral signal contains information about the EOS through its dependence on the tidal deformability $\Lambda$. Both the characteristic frequencies of the postmerger signal and $\Lambda$ depend on the stiffness of the EOS, so there are approximate relations between them~\cite{Takami:2014zpa,Tsang:2019esi,Breschi:2019srl}. Such relations were studied for the hybrid EOSs in~\cite{Jokela:2020piw}, and we also compared our numerical results from the simulations of~\cite{Ecker:2019xrw} to predictions from the relations of~\cite{Breschi:2019srl} and found deviation of at most 7\%. That is, our results with the intermediate V-QCD(SLy) EOS with $n_\mathrm{tr}=1.61n_s$ agreed well with the relations at $M=1.4M_\odot$ (0.3\% deviation) but the deviation was larger at $M=1.35M_\odot$ and $M=1.3M_\odot$ (6.7\% and 6.0\%, respectively) and close to the edge of the 2$\sigma$ confidence band of the fit. This deviation may be due to the fact that our EOS contains a rather rapid change in stiffness at densities right above the saturation density, as seen from the jump in the speed of sound in Fig.~\ref{fig:hybridEOScs2}.

We have also analyzed the results for the characteristic frequencies for all the hybrid EOSs only based on approximate relations and without running any simulations~\cite{Jokela:2020piw}. The results for $f_2$ in equal mass mergers from this analysis are shown in Fig.~\ref{fig:f2}. As in Figs.~\ref{fig:hybridEOS} and~\ref{fig:hybridMR}, the red band is the combined prediction from all the hybrid EOSs, and the curves show the results for the three V-QCD(APR) variants. Similarly to the mass-radius curve of Fig.~\ref{fig:hybridMR} a part of the red band arises from somewhat unlikely configurations with an extreme jump in the speed of sound: in this case it is the part of the band with highest frequencies. The upper left part of the bands is cut off by the threshold for direct collapse into a black hole, in which case the postmerger signal is a ringdown and there is no $f_2$. For equal masses the threshold means that prompt collapse is found roughly when (following~\cite{Breschi:2019srl})
\be
 370 \lesssim \Lambda \ .
\ee
This threshold also gives a rough bound on the EOSs: for GW170817, for which the average mass was slightly below $1.4M_\odot$, studies of the electromagnetic signal discussed above strongly suggest that a hypermassive neutron star was formed instead of a direct collapse. That is, the softer EOSs (with lower $\Lambda$ and therefore higher $f_2$ in Fig.~\ref{fig:f2}) are disfavored also by the electromagnetic signal from GW170817.

%%%%%%%%%%%%%%%%%%%%%%%%%%%%%%%%%%%%%
\section{Conclusion and Outlook}\label{sec:concl}
%%%%%%%%%%%%%%%%%%%%%%%%%%%%%%%%%%%%%

Physics of neutron stars and neutron star mergers is a rapidly evolving field at the moment thanks to the wealth of incoming experimental data from measurements of neutron stars and neutron star mergers. The progress in experiments is complemented by advances in numerical general relativity and theoretical modeling of dense QCD which make it possible to carry out increasingly realistic simulations of neutron star mergers. There has also been increasing interest to fill the holes of theoretical understanding of dense QCD by employing gau\-ge/gra\-vity duality in recent years. Reviewing progress in this subfield was the main purpose of this article.

I reviewed the advances in applications of gau\-ge/gra\-vity duality to dense QCD in neutron stars in various models, concentrating on results from the V-QCD model which is a rich bottom-up model that has been carefully fitted to lattice and other QCD data. Within the V-QCD approach, several details were seen to work particularly well: 
\begin{enumerate}
 \item The model was seen to be able to describe the lattice data for QCD thermodynamics at small densities extremely well (Figs.~\ref{fig:Vgfit},~\ref{fig:Vf0fit}, and~\ref{fig:wfit}).
 \item Extrapolation of the quark matter equation of state from the lattice QCD region towards higher densities, including the regime relevant for neutron star cores also worked (Fig.~\ref{fig:QMEOS}), and led to a feasible model for the quark matter EOS at all temperatures and densities.
 \item Including nuclear matter in V-QCD using a simple approach with a homogeneous bulk field produced a phenomenologically desirable, stiff EOS for dense nuclear matter (Fig.~\ref{fig:NMcs2}).
 \item Combining the V-QCD (nuclear and quark matter) EOS with low density nuclear models led to the construction of a family of feasible hybrid EOSs, many of which pass all known bounds to the QCD EOS (Fig.~\ref{fig:hybridEOS}). 
\end{enumerate}
By using the hybrid EOS, I narrowed down the band of available QCD EOSs in the cold and dense region. Notice in particular the V-QCD is one of the few models where the nuclear and quark matter EOSs, and the transition between these two phases, can be analyzed within a single framework. The model predicts that the transition is strongly of first order at low temperatures. 
I also computed several predictions for static and rotating neutron stars as well as for the gravitational wave signal produced in neutron star mergers. 

There are also several open questions and new directions yet to be explored through gau\-ge/gra\-vity duality. These include
\begin{itemize}
 \item Detailed analysis of the temperature dependence of the EOS, in particular in the nuclear matter phase.
 \item Flavor dependence: quark masses (in particular the strange quark mass) and coupling of the strongly coupled QCD fluid to flavor dependent electroweak currents.
 \item Magnetic field dependence at large densities of nuclear/quark matter, possibly with properly implemented flavor dependence. 
 \item Transport in the presence of magnetic field, both for nuclear and quark matter.
 \item Construction of individual baryons as solitons in modern, realistic bottom-up holographic models, as well as the study of their properties. 
 \item Proper overall simultaneous fit of particle spectra and lattice thermodynamics with holography. 
 \item Analysis of the interfaces between nuclear and quark matter, as well as nuclear matter and vacuum.
 \item Detailed analysis of color superconducting and other paired phases at high density.
\end{itemize}
Many of these questions can be, and will be studied in near future by using V-QCD and other (holographic) models.

\begin{acknowledgements}
I thank J.~Cruz Rojas, T.~Demircik, C.~Ecker, N.~Jokela, J.~Remes, K.~Rigatos, J.~Sonnenschein, and A.~Vuorinen for discussions and comments on the man\-u\-script.
This work has been supported 
by an appointment to the JRG Program at the APCTP through the Science and Technology Promotion Fund and Lottery Fund of the Korean Government. It has also been supported by the Korean Local Governments -- Gyeong\-sang\-buk-do Province and Pohang City -- and by the National Research Foundation of Korea (NRF) funded by the Korean government (MSIT) (grant number 2021R1A2C1010834).
\end{acknowledgements}

\appendix

%%%%%%%%%%%%%%%%%%%%%%%%%%%%%%%%%%%%%
\section*{Appendix A: Precise dictionary of V-QCD} %\label{app:dictionary} 
%%%%%%%%%%%%%%%%%%%%%%%%%%%%%%%%%%%%%
\addcontentsline{toc}{section}{A\ \,  Precise dictionary of V-QCD}

\renewcommand{\theequation}{A.\arabic{equation}}

In this Appendix I write down the precise dictionary for V-QCD (see also Table~\ref{tab:dictionary}). First, to do this, I need to specify the conventions for QCD. I take the QCD action to read
\begin{eqnarray}
\label{eq:QCDS} 
{\cal S}_\mathrm{QCD} &=& \int d^4x\,\bigg [
-{1\over2g^2}\,{\mathbb Tr}\, G_{\mu\nu}\,G^{\mu\nu}+i\bar \psi \slashed D \psi
\\\nonumber
&&\qquad\qquad -\bar \psi_R\,M_q\,\psi_L-\bar \psi_L\,M_q^\dagger\,\psi_R \\\nonumber
&&\qquad\qquad+{\theta \over 32\pi^2}\,\epsilon^{\mu\nu\rho\sigma}\,{\mathbb Tr}\, G_{\mu\nu}\,G_{\rho\sigma}
\bigg]
\end{eqnarray}
where $\mathbb Tr$ is over the color indices, the flavor indices are implicit, $M_q$ is the (possibly complex) mass matrix, and I also included the $\theta$-angle. The dictionary between the operators (${\mathbb Tr}\, G_{\mu\nu}\,G^{\mu\nu}$, $\epsilon^{\mu\nu\rho\sigma}\,{\mathbb Tr}\, G_{\mu\nu}\,G_{\rho\sigma}$, $J^{(L/R)ij}_{\mu} = \bar \psi^i\, \gamma_\mu(1\pm\gamma_5)\, \psi^j/2\,$, and $\bar \psi^i\, (1\pm\gamma_5)\, \psi^j$) and the five dimensional fields ($\lambda$, $a$, $A_{L/R}^{ij}$, and $T^{ij}$) can then be specified by writing down the four dimensional coupling between the operators and the boundary values of the fields~\cite{Arean:2016hcs}:
\begin{eqnarray}
\label{eq:dict}
S_{\delta} &=& -\frac{N_c}{2} \int_{r=\delta} d^4 x \, \frac{1}{\lambda}\,{\mathbb Tr}\, G_{\mu\nu}\,G^{\mu\nu} \\\nonumber
&&+{1 \over 32 \pi^2} \int_{r=\delta} d^4 x \, a \,
\epsilon^{\mu\nu\rho\sigma}{\mathbb Tr} (G_{\mu\nu}\, G_{\rho\sigma})\\\nonumber
&&+ \int_{r=\delta} d^4 x\, J^{(L)ij}_{\mu}{A_L^{\mu\, ij}}
+ \int_{r=\delta} d^4 x\, J^{(R)ij}_{\mu}A_R^{\mu\, ij} \nonumber \\\nonumber
&&- K_T \int_{r=\delta} d^4 x\,\frac{1}{\ell\, \delta}\,\bar \psi_R\,T\,\psi_L\\\nonumber
&&-  K_T \int_{r=\delta} d^4 x\,\frac{1}{\ell\, \delta}\,\bar \psi_L\,T^\dagger\,\psi_R
\end{eqnarray}
where $\delta$ is a UV cutoff and the flavor indices $i,j$ run from 1 to $N_f$. The coefficient $K_T$ is a $\mathcal{O}(1)$ number which determines the normalization of the  quark mass in the tachyon asymptotics but cannot be precisely determined. 
For simplicity, I did not include the sources for the metric. 
The correspondence between the boundary values of the fields and QCD sources can be read off by comparing~\eqref{eq:QCDS} and~\eqref{eq:dict} 
The generating functional QCD can be explicitly written as
\be
 Z_\mathrm{QCD} = \int \mathcal{D}\psi \mathcal{D}G \exp\left[-\int d^4x\,\bar\psi\slashed D\psi + iS_\delta\right]
\ee
and the correspondence says that it equals to the on-shell gravity partition function
\be
 Z_\mathrm{grav} = e^{iS_\mathrm{V-QCD}^{(\mathrm{on-shell})}} \ .
\ee
To be precise, the actions also contain the UV divergences of QCD and needs to be renormalized. The holographic renormalization~\cite{Skenderis:2002wp} for actions in this class has been studied in detail in~\cite{Papadimitriou:2011qb}. 

\begin{table}
\caption{The parameters of the potentials 5b, 7a, and 8b.} \label{tab:thermofit}
\begin{center}
\begin{tabular}{|c||c|c|c|}
\hline
 & \textbf{5b} & \textbf{7a}  & \textbf{8b} \\
\hline
\hline
$W_0$ & 1.0 & 2.5 &  5.886 \\
\hline 
\hline 
$W_\mathrm{IR}$ & 0.85 & 0.9 & 1.0 \\
\hline
$w_0$ & 0.57 & 1.28 & 1.09 \\
\hline
$w_1$ & 3.0 & 0 & 1.0 \\
\hline
$\bar w_0$ & 65 & 18 & 22 \\
\hline
$8 \pi^2/\hat \lambda_0$ & 0.94 & 1.18 & 1.16 \\
\hline
\hline 
$\bar\kappa_0$ & 1.8 & 1.8 & 3.029 \\
\hline
$\bar\kappa_1$ & -0.857 & -0.23 & 0 \\
\hline
\hline 
$\Lambda_\mathrm{UV}$/MeV & 226 & 211 & 157 \\
\hline
$180 \pi^2 M_\mathrm{p}^3\ell^3/11$ & 1.34 & 1.32 & 1.22 \\
\hline

\end{tabular}
\end{center}
\end{table}

%%%%%%%%%%%%%%%%%%%%%%%%%%%%%%%%%%%%%
\section*{Appendix B: Choices of potentials for V-QCD} %\label{app:potentials}
%%%%%%%%%%%%%%%%%%%%%%%%%%%%%%%%%%%%%
\addcontentsline{toc}{section}{B\ \,  Choices of potentials for V-QCD}

\renewcommand{\theequation}{B.\arabic{equation}}

In this Appendix I specify the potential sets 5b, 7a, and 8b, which define the soft, intermediate and stiff variants of the V-QCD EOSs, respectively. 
I take $V_f(\lambda,\tau) = V_{f0}(\lambda) e^{-\tau^2}$ and
\begin{eqnarray}
 V_g(\lambda)&=&12\,\biggl[1+V_1 \l+{V_2\lambda^2
\over 1+\l/\l_0}\\\nonumber
&&+V_\mathrm{IR} e^{-\l_0/\l}(\l/\l_0)^{4/3}\sqrt{\log(1+\lambda/\l_0)}\biggr]\  \\
 V_{f0}(\lambda) &=& W_0 + W_1 \l +\frac{W_2 \l^2}{1+\l/\l_0} \\\nonumber
&&+ W_\mathrm{IR} e^{-\l_0/\l}(\l/\l_0)^{2}   \\
\frac{1}{\kappa(\l)} &=& \kappa_0 \biggl[1+ \kappa_1 \l \\\nonumber
&&+ \bar \kappa_0 \left(1+\frac{\bar \kappa_1 \l_0}{\l} \right) e^{-\l_0/\l }\frac{(\l/\l_0)^{4/3}}{\sqrt{\log(1+\lambda/\l_0)}}\biggr]  \\
\frac{1}{w(\l)} &=&  w_0\biggl[1 + \frac{w_1 \l/\l_0}{1+\l/\l_0} \\\nonumber
&& + 
\bar w_0 
e^{-\hat\l_0/\l}\frac{(\l/\hat\l_0)^{4/3}}{\log(1+\lambda/\hat\l_0)}\biggr] 
\end{eqnarray}
for the potentials appearing in~\eqref{eq:vqcdgls5} and in~\eqref{eq:Shdef}. Most of the UV parameters were determined by comparing to the RG flow of perturbation theory:
\begin{eqnarray}
 V_1 &=& \frac{11}{27\pi^2} \ , \qquad V_2 = \frac{4619}{46656 \pi ^4} \ ,  \\ 
 \kappa_0 &=& \frac{3}{2} - \frac{W_0}{8} \ , \qquad
   \kappa_1 = \frac{11}{24\pi^2} \ , \\ 
   W_1 &=& \frac{8+3\, W_0}{9 \pi ^2} \ , \qquad W_2 = \frac{6488+999\, W_0}{15552 \pi ^4} \ .
\end{eqnarray}
All the other parameters were determined by comparing to lattice data as I explained in Sec.~\ref{ssec:latticefit}. 
All three potentials use the same fit result for the parameters in $V_g$:
\be
  \l_0 = 8\pi^2/3 \ , \qquad V_\mathrm{IR} = 2.05 \ .
\ee
The flavor sector  fit parameters for the three potentials can be found in Table~\ref{tab:thermofit}. The  AdS radius is given by
\be
 \ell = \frac{1}{\sqrt{1-W_0/12}} \ .
\ee

\bibliographystyle{ucsd_mod}  
\bibliography{review-epjc}   % name your BibTeX data base

% % Non-BibTeX users please use
% \begin{thebibliography}{}
% %
% % and use \bibitem to create references. Consult the Instructions
% % for authors for reference list style.
% % Format for Journal Reference
% Author, Article title, Journal, Volume, page numbers (year)
% Author, Book title, page numbers. Publisher, place (year)
% \end{thebibliography}

\end{document}